# THE PALOMAR DISTANT CLUSTER SURVEY:
# I. THE CLUSTER CATALOG

Marc Postman[1]
Space Telescope Science Institute[2], 3700 San Martin Dr., Baltimore, MD 21218

Lori M. Lubin
Princeton University Observatory, Princeton NJ 08544

James E. Gunn[3]
Princeton University Observatory, Princeton, NJ 08544

J. B. Oke
Palomar Observatory, California Institute of Technology, Pasadena, CA 91125
and
Dominion Astrophysical Observatory, Victoria, BC V8X 4M6, Canada

John G. Hoessel[1]
Washburn Observatory, University of Wisconsin — Madison
475 N. Charter St., Madison, WI 53706

Donald P. Schneider[1]
Department of Astronomy and Astrophysics, The Pennsylvania State University,
525 Davey Lab, University Park, PA 16802
and
Institute for Advanced Study, School of Natural Sciences, Princeton, NJ 08540

Jennifer A. Christensen
Space Telescope Science Institute[2], 3700 San Martin Dr., Baltimore, MD 21218

---

[1]Guest Observer, Palomar Observatory

[2]Space Telescope Science Institute is operated by the Association of Universities for Research in Astronomy, Inc., under contract to the National Aeronautics and Space Administration.

[3]Visiting Associate, Palomar Observatory

# ABSTRACT

We present an optical/near IR selected catalog of 79 clusters distributed over an area of 5.1 square degrees. The catalog was constructed from images obtained with the 4-Shooter CCD mosaic camera on the Hale 5m telescope operating in "scan" mode. The survey, hereafter known as the Palomar Distant Cluster Survey (PDCS), was conducted in two broad band filters that closely resemble $V$ and $I$. The $4\sigma$ limiting magnitudes for our 300 s exposures are $\sim 23.8$ ($V$) and 22.5 ($I$). A matched filter algorithm was developed and employed to identify the cluster candidates by using positional and photometric data simultaneously. The clusters cover the range $0.2 \lesssim z \lesssim 1.2$, based on the redshift estimates derived in the cluster detection procedure. An accurate selection function is generated from extensive simulations. We find that the cumulative surface density of clusters with richness class R $\geq 1$ is about a factor of 5 higher than the extrapolated density of R $\geq 1$ Abell clusters. The PDCS results are consistent with a constant comoving density of clusters to $z \lesssim 0.6$, albeit at the above high density level. Constraints on cluster abundances at $z > 0.6$ will be possible with the acquisition of spectroscopic redshifts for a large subset of these cluster candidates. We also present a supplemental catalog of 28 clusters that do not satisfy all our selection criteria but which include some of the most distant systems detected in the survey. Finding charts for all 107 cluster candidates are provided. (Astro-ph readers: due to their large size, the finding charts are not included in this version of the paper).

*Subject headings:* galaxies: clustering; cosmology: observations

# 1. INTRODUCTION

The evolution of the space density of clusters, of their internal kinematics and spatial structure, of their large scale clustering properties, and of their host galaxy populations are fundamental predictions of all viable galaxy and cluster formation theories. The accuracy of the observational constraints on these parameters depends on the availability of statistically complete cluster catalogs over a large redshift range ($0 < z \lesssim 1$ and beyond). At redshifts less than 0.2, the available catalogs (Abell 1958; Zwicky $et$ $al.$ 1968; Shectman 1985; Abell, Olowin, & Corwin 1989, hereafter ACO; Lumsden $et$ $al.$ 1992 (Edinburgh-Durham); Dalton $et$ $al.$ 1994 (APM)) have all been based on the large photographic sky surveys done with the 1.2m Schmidt at Palomar Mountain and at the UK Schmidt at Siding Spring. These surveys have only recently become available in digital format and, consequently, only the Shectman, APM, and Edinburgh-Durham catalogs have been generated using automated and objective cluster detection algorithms. At higher redshifts, the catalogs are based on deep prime focus optical/near IR photographic plates taken on 4-5m class telescopes (Gunn, Hoessel, & Oke 1986, hereafter GHO; Couch $et$ $al.$ 1991). The low quantum efficiency of photographic material limits the completeness of these surveys beyond $z \sim 0.5$. These catalogs have yielded important information on galaxy evolution ($e.g.$, Butcher & Oemler 1984; Gunn & Dressler 1988; Rakos & Schombert 1995) and on the clustering properties and space densities of low redshift clusters as a function of richness ($e.g.$, Bahcall & Soneira 1983; Dalton $et$ $al.$ 1992; Nichol $et$ $al.$ 1992; Postman, Huchra, & Geller 1992). The Abell/ACO and Zwicky $et$ $al.$ cluster catalogs and the previously published optically selected cluster catalogs at higher ($z \gtrsim 0.3$) redshifts were constructed via visual inspection, limiting their utility. For a cluster catalog to be statistically useful, it must be homogeneous and its selection biases must be accurately quantified. Furthermore, in order to compare the predictions of models to the observations reliably, the same cluster selection process that is used on the observations should be applicable to large scale simulations.

The 4-Shooter CCD mosaic camera on the Hale 5m (installed in 1983) was designed, in part, to conduct digital, two color surveys for distant ($z \lesssim 1$) clusters over a moderate area of sky. We present here the cluster catalog resulting from the first such survey, hereafter referred to as the Palomar distant cluster survey (PDCS). We have developed a matched filter cluster selection algorithm which uses both positional and photometric data simultaneously to search for overdensities in the galaxy distribution. The algorithm works well over a large dynamic range in redshift and is optimal in the regime where one is looking for small signals in large backgrounds, as is the case for distant clusters. We have applied this algorithm to the uniform galaxy catalogs derived from our CCD data.

Section 2 of this paper describes the observations and their calibration. The construction of galaxy catalogs from the CCD data is discussed in section 3. Our cluster selection algorithm, the cluster selection criteria, the cluster catalog, the catalog completeness and the false positive rate are discussed in full detail in section 4. Preliminary constraints on the space density of optically selected clusters are discussed in section 5. The properties of the clusters in this catalog will be discussed in subsequent papers.

## 2. OBSERVATIONS AND DATA CALIBRATION

### 2.1. Observational Technique

The PDCS consists of five separate fields. The shape of each field is roughly a square that is 1 degree on a side. The fields are all located within or near the original high–latitude Gunn & Oke (1975) survey areas. In addition, field centers were chosen to minimize the number of bright stars in the survey regions and to avoid very low-redshift ($z < 0.1$) clusters and regions with high reddening. There is one R = 1 Abell cluster, A 1739 ($z_{est} \approx 0.17$), located in our survey area. The coordinates of the field centers are given in Table 1; the five regions will be denoted in this paper as the $0^h$, $2^h$, $9^h$, $13^h$, and $16^h$ fields. The $2^h$ CCD survey field is not included in the original Gunn & Oke survey; this field is used in lieu of the original $3^h$ field which suffers from high galactic extinction.

The PDCS was performed with the 4-Shooter CCD camera (Gunn *et al.* 1987) mounted at the cassegrain focus of the Hale telescope. The focal plane of the telescope is coincident with a reflective four-faceted pyramid in the 4-Shooter. The pyramid directs the incoming light into four reimaging cameras. The 4-Shooter detectors are four $800 \times 800$ Texas Instruments CCDs. The cameras produce an image scale of approximately $0.335''$ pixel$^{-1}$; the instrument's entire field of view is $8.9'$ on a side. A diagram of the 4-Shooter focal plane is given in Figure 1.

The CCDs are positioned so that there should be no gaps in sky coverage along the pyramid facets (there are typically about 20 rows/columns of pixels that do not see the sky due to pyramid shadowing), but the few arcseconds of sky located where the CCDs join are not usable because of imperfections in the pyramid surface along the facets. The effective size of the field for this project is $\approx 2 \times 780$ pixels, or a width of $520''$. Images of each field are obtained in two filters, the F555W and F785LP of HST's Wide Field/Planetary Camera (Griffiths 1990). To minimize confusion with the HST photometric systems, we will denote the 4-Shooter passbands as $V_4$ (F555W) and $I_4$ (F785LP); see §2.2 for a description of the PDCS photometric system. The limiting magnitude of the $V_4$ data is fainter than that of the $I_4$ observations (due to the fainter sky and higher detector sensitivity at 5500 Å), but the $I_4$ images are in some respects just as important; at $z \gtrsim 0.6$ the 4000 Å "break" lies redward of the $V_4$ band, so the $I_4$ data provides the wavelength coverage needed to detect clusters out to $z \sim 1$.

The data are acquired in Time-Delay-and-Integrate, or TDI, mode. (This is frequently called "scan" mode in the literature.) Details of how TDI observations are taken with the 4-Shooter, as well as the initial data processing of the data, are given in Schneider, Schmidt, & Gunn (1994). In TDI mode the sky is moved along the columns of the CCD at a constant rate; the CCDs are read so that the charge packet shifts one row in the time that it takes the sky to traverse one pixel. TDI mode was chosen for the PDCS for three reasons: 1) This is a very efficient method to image large sections of sky to moderate depth (there is no overhead associated with CCD readout; one is exposing essentially 100% of the time), 2) Data taken in TDI mode can be extremely well flat-fielded (see Schneider, Schmidt, & Gunn 1994 and §2.3 of this paper). This property is especially important for these survey images, as the Texas Instruments CCDs have very poor fringing properties at near-infrared wavelengths. 3) The configuration of the 4-Shooter focal plane (Figure 1) is well-suited for two-filter observations in TDI mode.

The PDCS data were obtained by moving the telescope north along lines of constant Right Ascension (equinox 1950.0) at a rate of $0\rlap{.}''893$ s$^{-1}$; this yields a CCD transit time of 300 seconds. Because of the aforementioned pyramid shadowing in the 4-Shooter, the effective exposure time is 293 seconds. The data are acquired in two colors, nearly simultaneously, by making use of the fact that the 4-Shooter can be equipped with one set of filters for the leading two cameras (CCDs 1 and 4, see Figure 1) in the scan and another set for the trailing ones (CCDs 2 and 3).

A survey field consists of eight adjacent scans. Each scan requires approximately 80 minutes to complete and overlaps the adjacent scan by $\approx 30''$. The total area imaged in each field is usually slightly larger than a square degree (see Table 1).

Imaging of the five fields began in October 1986 and required approximately three years to complete. The data were all obtained under photometric conditions. In order to maintain accurate star–galaxy classification and uniform completeness limits, only data with FWHM seeing better than $1\rlap{.}''5$ are used in this analysis. The median FWHM seeing in all slow-scan data obtained during the course of the observations is $1\rlap{.}''40$; the FWHM median seeing in data accepted for use in our cluster survey is $1\rlap{.}''23$.

## 2.2. The PDCS Photometric System

The system response as a function of wavelength for the filter+4-Shooter+telescope are shown in Figure 2. The lower curve for each filter is the quantum efficiency through 1.2 airmasses from Palomar; this is the response curve used to calculate filter properties (*e.g.*, effective wavelengths) and transformations to other photometric systems.

The effective wavelengths and widths of the filters, following the definitions in Schneider, Gunn, & Hoessel (1983) and Griffiths (1990), are 5430 Å and 1230 Å for $V_4$ and 8690 Å and 1400 Å for $I_4$. The effective wavelength of the $V_4$ band is $\approx 100$ Å bluer than that of $V$, and $V_4$ is about 50% wider than $V$. $I_4$ has nearly the same width as the Kron-Cousins $I$ band (Cousins 1976; this band will be denoted as $I_{\mathrm{KC}}$), but the effective wavelength of $I_4$ is about 500 Å redder than that of $I_{\mathrm{KC}}$. The $I_4$ band has some sensitivity at wavelengths longward of 10,500 Å.

The zero points of the $V_4, I_4$ magnitudes are based on the $AB$ magnitude system of Oke & Gunn (1983). Objects whose spectral energy distribution are constant in terms of ergs s$^{-1}$ cm$^{-1}$ Hz$^{-1}$ have neutral color in an $AB$-based system. The magnitudes of Vega are $V_4 = +0.03$ and $I_4 = +0.46$. The $V_4, I_4$ magnitudes for the fundamental standard of the Gunn system, the F subdwarf BD +17 4708 (defined as magnitude 9.500 in all bands), are 9.45 and 9.20, respectively. Magnitudes in the $V_4, I_4$ system for the six standard stars that provided the photometric calibration for the PDCS are listed in Table 2. (Kent 1985 gives coordinates and $g, r$ magnitudes for these stars.)

Since the $V_4$ and $I_4$ filters are quite broad, it is impossible to define precise transformations to other photometric systems when dealing with galaxies with a wide range of redshifts and spectral energy distributions. We have calculated approximate transformations from $V_4$ and $I_4$ measurements to a number of other photometric systems that are frequently used in studies of faint galaxies and/or clusters. In addition to $V$ and $I_{KC}$, we present the transformations to $B_J$ ($\lambda_{eff} = 4390$ Å, Gullixson *et al.* 1994), and the $g$ ($\lambda_{eff} = 4910$ Å), $r$ ($\lambda_{eff} = 6530$ Å), and $i$ ($\lambda_{eff} = 7860$ Å), bands of Thuan & Gunn (1976) and Wade *et al.* (1979). (Also see Kent 1985 for further information on the $g$ and $r$ photometric systems.) The rough transformations were derived by synthesizing magnitudes from the spectrophotometric atlas of stars of Gunn & Stryker (1983), and restricting the fits to objects with ($V_4 - I_4 < +3$).

$$B_J = V_4 + 0.29 + 0.676\ (V_4 - I_4) - 0.083\ (V_4 - I_4)^2 \tag{1}$$

$$g = V_4 + 0.04 + 0.295\ (V_4 - I_4) \tag{2}$$

$$V = V_4 - 0.02 - 0.056\ (V_4 - I_4) + 0.012\ (V_4 - I_4)^2 \tag{3}$$

$$r = V_4 + 0.13 - 0.514\ (V_4 - I_4) + 0.055\ (V_4 - I_4)^2 \tag{4}$$

$$i = I_4 + 0.27 + 0.145\ (V_4 - I_4) \tag{5}$$

$$I_{KC} = I_4 - 0.43 + 0.089\ (V_4 - I_4) \tag{6}$$

The average rms scatter about the above relations is a few hundredths of a magnitude. The correction for Galactic extinction for the $V_4$ and $I_4$ filters is

$$A_{V_4} = A_V = 3.05\ E(B - V) \tag{7}$$

$$A_{I_4} = 0.497\ A_{V_4} = 1.52\ E(B - V) \tag{8}$$

where $E(B-V)$ values are derived from the procedure presented in Burstein & Heiles (1982). Reddening corrections for the five survey fields are given in Table 1.

Figures 3a–c show the $V_4$ and $I_4$ k-corrections and expected $V_4 - I_4$ colors as functions of redshift, respectively, for five morphological types: E, Sab, Sbc, Scd, and Im. The spectral energy distributions are from Coleman *et al.* (1980) and no spectral evolution is assumed.

## 2.3. TDI Image Data Processing

The raw data from a 4-Shooter TDI observation form a continuous pixel stream 816 columns wide and approximately 51,000 rows long (800 rows are generated every 300 seconds for each CCD). To take advantage of the large amount of existing picture-based image processing software, the data from each CCD are subdivided into $816 \times 800$ frames with an 80 pixel ($26''.8$) overlap introduced in the scan direction. The overlap allows reliable photometric and astrometric measurements of essentially all of the objects that fall near the boundaries of the image frames. The size of the overlap was chosen as a compromise between processing efficiency (larger overlaps mean more redundant processing) and the typical diameters of galaxies in moderate to high redshift clusters ($\lesssim 15''$). Analysis software may not accurately measure the parameters of galaxies with diameters larger than 80 pixels that are located in the overlap area. The number of such cases is extremely small ($< 10$ per square degree), and in any case they are not the targets of study of this program.

In TDI mode, the data can be used directly to construct a one-dimensional "skyflat" vector that will produce extremely accurate measurements of the variation in pixel-to-pixel sensitivity perpendicular to the scan direction. No adjustment for pixel-to-pixel sensitivity variations parallel to the scan direction is required in TDI mode. The median sky values in the $V_4$ and $I_4$ bands are 2400 DN pixel$^{-1}$ ($\sim$21.3 mag per arcsec$^2$) and 6000 DN pixel$^{-1}$ ($\sim$19.3 mag per arcsec$^2$), respectively; thus there is sufficient signal to generate accurate flat-field vectors.

To correct for the relative zeropoint offsets between the two CCDs assigned to a given passband (*e.g.*, CCDs 1 and 4 for the $V_4$ band), a sequence of raw frames from each CCD are debiased and then flattened by multiplying the data by their corresponding skyflat vector. A "balance factor" for each CCD, $BF_i$, is computed using the expression

$$BF_i = \frac{\frac{1}{2} \sum_{k=1}^{N} (\mathrm{Sky}_{k,i} + \mathrm{Sky}_{k,j})}{\sum_{k=1}^{N} \mathrm{Sky}_{k,i}} \tag{9}$$

where $\mathrm{Sky}_{k,i}$ is the sky level in the $k^{\mathrm{th}}$ frame from CCD $i$, $\mathrm{Sky}_{k,j}$ is the sky level in the $k^{\mathrm{th}}$ frame from CCD $j$, where $j$ is the CCD assigned the same filter as CCD $i$, and $N$ is the total number of frames processed for this calibration (usually four frames per CCD). The initial skyflat vector is then multiplied by the balance factor to produce the final, corrected skyflat vector. This method of computing relative offsets distributes the offset correction between the two CCDs as opposed to applying the correction all to one CCD. The assumption underlying this approach is that sky level differences between corresponding frames from CCDs with a given passband are due to different relative CCD zeropoints rather than actual sky variations on the scale of a few arcminutes. This quite reasonable assumption is supported both by the insensitivity of the derived balance factor values to the specific frames used in the calculation and by the excellent agreement between the balance factors derived from the scans and those derived from $V_4$ and $I_4$ dome flat exposures. The typical zeropoint offsets are $\approx 0^{\mathrm{m}}_{\cdot}36$ and $0^{\mathrm{m}}_{\cdot}12$ for the $V_4$ and $I_4$ CCDs, respectively (*i.e.*, CCD 4 is 1.39 times more sensitive than CCD 1 in the $V_4$ band, and CCD 3 is 1.12 times more sensitive than CCD 2 in the $I_4$ band).

The CCD signature is removed from the raw data and, as a final step, the median sky level as a function of row (scan direction) is fit to a cubic polynomial. The appropriate linear, quadratic, and cubic components are subtracted from each image row. This process removes any medium-term temporal variations in sky brightness and produces frames with extremely flat sky levels (sky brightness levels measured in different regions in any given frame agree to within 0.002 mag per arcsec$^2$). Most frames are quite flat even before applying the final sky correction. The largest temporal variations in sky brightness across a frame seen in the data have amplitudes less than 2% of the mean sky level. One consequence of this sky correction is that the mean sky level will change in a non-continuous manner from frame to frame. Measurements of objects in the 80 pixel overlap region between frames shows that this has no influence on number counts or object photometry and yields very stable object detection performance.

## 3. GALAXY CATALOG CONSTRUCTION

### 3.1. Object Detection and Classification

We used a modified version of the Faint Object Classification and Analysis System (Jarvis & Tyson 1981, Valdes 1982; hereafter FOCAS) to detect, measure, and classify objects in the calibrated CCD images. The modifications include new software to perform photometric classification of galaxies (discussed later in this section), the addition of object catalog parameters related to this photometric classification as well as a new object compactness parameter, and a variety of new image statistics, catalog editing, and quality assurance programs. The basic FOCAS object detection, point spread function (PSF) measurement, deblending, and star–galaxy–noise classification algorithms have not been altered.

As in the image calibration process, the frames are processed individually. The object isophotal detection threshold in each frame is set to be $3\sigma_{sky}$ above sky (typically 25.7 mag per arcsec$^2$ in the $V_4$ band and 24.8 mag per arcsec$^2$ in the $I_4$ band) and a minimum object area requirement of 15 pixels (1.68 arcsec$^2$) is imposed. Because the sky level is lower and the system response is higher in the $V_4$ band than in the $I_4$ band, the $V_4$ images are deeper. Consequently, about 40% of the objects detected in a typical $V_4$ image are undetected by FOCAS in the corresponding $I_4$ image. Nearly all the objects which are only detected in the $V_4$ band are faint ($V_4 \gtrsim 23$). About 95% of the objects with $V_4 \leq 23$, however, are detected in both passbands.

After object detection, merged objects are deblended using the FOCAS *splits* routine. Approximately 15% of all detected objects with $V_4 \leq 24$ required deblending. Object parameters are evaluated (*e.g.*, isophotal, core, and total magnitudes, axial ratio, intensity-weighted and unweighted moments, etc.) and the FOCAS *resolution* object classifier is run on all of the objects. The PSF used by the *resolution* classifier is computed separately for each frame to compensate for variations in seeing that can occur during the course of a scan. Occasionally there is an insufficient number of suitable PSF stars in a frame for an accurate

PSF to be measured; in such cases, the PSF from an adjacent frame (in the scan direction) is used for object classification.

FOCAS classifies objects based on the results of fitting a series of templates derived from the observed PSF to each object. Specifically, the templates are parameterized by

$$t(r_i) = f \times p(r_i/s) + (1 - f) \times p(r_i) \tag{10}$$

where $r_i$ is the position of pixel $i$, $p(r_i)$ is the observed PSF, $s$ is the FOCAS *resolution* scale parameter, and $f$ is the FOCAS *fraction* parameter. The object classification is determined from the values of $s$ and $f$ which yield the best fit to the two-dimensional data. We adopted the following classification rules:

| | | |
|---|---|---|
| Noise | $0.1 \leq s \leq 0.7$ | $0 < f \leq 1$ |
| Star | $0.7 < s \leq 1.3$ | $0 < f \leq 1$ |
| Fuzzy Star | $1.3 < s \leq 10$ | $0 < f < 0.25$ |
| Galaxy | $1.3 < s \leq 10$ | $0.25 \leq f \leq 1$ |
| Diffuse | $10 < s \leq 100$ | $0 < f \leq 1$ |

Practically all "Fuzzy stars" are stellar objects and every diffuse object brighter than $V_4 = 23$ is a galaxy. The range of the scale parameter used to select stars was chosen to be broad enough to account for reasonable PSF variations that occur across a single 4-Shooter CCD but narrow enough to minimize galaxy misclassifications. The 4-Shooter re-imaging cameras are relatively fast ($f/1.84$) and the PSF varies in shape from center to edge. In TDI mode, PSF variations due to the camera optics are a function of CCD column only. FOCAS currently does not allow one to use a variable PSF for object classification on a given CCD frame. The scale parameter range used to select stars was, therefore, chosen to be broad enough to account for reasonable PSF variations that occur across a single 4-Shooter CCD but narrow enough to minimize galaxy misclassifications. We find that there are position dependent star classification errors due to the PSF variations but the galaxy classifications are not significantly affected. This is primarily because galaxies dramatically outnumber stars at the magnitude limit of the present survey. A detailed discussion of the accuracy of the object classifications is presented in §3.3.

Finally, we perform a crude morphological classification of the galaxies based on their radial surface brightness profile. For galaxies with azimuthally averaged radial profiles extending for $1\overset{''}{.}34$ or more, a comparison is made between the observed profile and a library of seeing-convolved reference profiles for exponential disk and de Vaucouleurs (1948) $r^{-\frac{1}{4}}$ models. The reference profiles span a wide range of effective radii and axial ratios. A maximum likelihood technique is used to select the best fit model. If the galaxy has an axial ratio greater than 2.5, only the disk models are used in the comparisons. A new

"total" magnitude is then computed by integrating the profile of the best fit model. This total magnitude, the effective radius of the best fit model, the $\chi^2_\nu$ value of the fit, and the profile type (disk or de Vaucouleurs) are stored along with the other object parameters. The morphological classification algorithm reliably classifies pure disk or pure bulge systems down to $V_4 = 22$. At the survey limit, however, only about 50% of simulated $r^{-\frac{1}{4}}$ galaxies are correctly classified. There is a classification bias in the present algorithm that results in galaxies with $V_4 > 22.5$ being preferentially classified as exponential disks.

## 3.2. Photometric Calibration

Although all of the data were obtained under photometric conditions, very few observations of photometric standards were made on the nights that the scans were performed. Consequently, a bootstrap procedure was used for the final photometric calibration of the PDCS.

All eight scans of a given field were placed on a uniform flux scale by comparing the instrumental magnitudes of the objects located in the $\sim 30''$ overlap region between adjacent scans. This procedure is done for both the $V_4$ and $I_4$ data independently. The observation dates for some adjacent scans are separated by as much as two years; temporal variations in the reflectivity/transmission of the telescope and 4-Shooter optics will alter the photometric zero points. If the data are indeed photometric, then a single constant for each scan will suffice to place the data on a given instrumental system. If the observations were taken in nonphotometric conditions, this offset will vary with time.

Figure 4 shows the photometric offset as a function of time (for the PDCS observations, the declination is linearly related to the time elapsed since the start of the scan) for four pairs of adjacent scans. The upper two panels show scans taken on the same night; as expected, there is no significant offset between scans. The lower two panels compare data for scans taken at times separated by more than a month; while there is a definite change in the flux scale between adjacent scans, there is no evidence for any temporal variations. Objects with $V_4 \leq 22$ were used to measure the offset in the $V_4$ band data ($I_4 \leq 21$ in the $I_4$ band). The observed scatter is consistent with the estimated uncertainties in the FOCAS photometry ($\pm 0.12$ mag for a galaxy with $V_4 = 22$). The handful of points with discrepant offsets are from objects which straddle the overlap zones and, hence, their magnitudes cannot be accurately compared. These objects are excluded from zeropoint offset computations. The typical photometric offsets seen in the data are $\pm 0\overset{m}{.}09$ and $\pm 0\overset{m}{.}10$ in the $V_4$ and $I_4$ passbands, respectively. There are some cases where the offsets are as large as $\pm 0\overset{m}{.}25$. There is little indication of zeropoint drifts which would occur if the photometric quality of the sky varied during the course of a scan. The photometric offsets in each band are normalized to the scan with the best seeing; following this adjustment the data in each field are on the same flux scale.

To place the observations on an absolute flux scale, we acquired three "fast" scans for each of the five fields. The fast scans were taken in a direction perpendicular to the motion of the survey scans. The fast scans have an effective exposure time of 98 seconds, span the entire width of the field, and are centered on the northern, central, and southern segments of each field. The fast scan data were obtained on three photometric nights (27 September 1987 and 12 and 13 March 1988) on which we also obtained observations of photometric standards (see Table 2 for a list of standards). In addition to providing the absolute photometric calibration, the data produced by the fast scans are used to tie the strips together in the astrometric calibration (§3.4).

The accuracy of the photometric calibration of the data can be measured by comparing number counts from the five independent survey fields. Figures 5a and 5b show the differential galaxy surface density as a function of $V_4$ and $I_4$ magnitude, respectively. Figure 6 shows the residuals between the mean galaxy surface density for all five fields and the galaxy surface density in each individual field as a function of $V_4$ and $I_4$ magnitude, respectively. All magnitudes have been corrected for Galactic extinction using the values in Table 1. The typical field-to-field number fluctuations are $\pm 9\%$ over the range $19 \leq V_4 \leq 23.5$ and $18.5 \leq I_4 \leq 22$. At $V_4 < 19$ ($I_4 < 18.5$), the number of galaxies is too small to yield reliable statistics; at $V_4 > 23.8$ ($I_4 > 22.5$) incompleteness sets in. The solid curve shows the observed rms fluctuations averaged over all 5 survey fields. For comparison, the dashed curve shows the expected fluctuations between fields of this area, given the observed angular two point correlation function derived from the galaxy survey of Neushaefer, Windhorst, & Dressler (1991). The observed fluctuations are in very good agreement with the clustered distribution prediction, indicating that, on average, zeropoint fluctuations between survey fields must be $\lesssim \pm 0\overset{m}{.}09$. This suggests that the galaxy catalogs are not seriously contaminated by misclassified stars, a conclusion confirmed by our checks on classification accuracy described in the next section. The systematic excess of galaxies in the $0^h$ field is most likely due to a photometric zeropoint error (there is no particularly rich cluster of bright galaxies in this field). A zeropoint error of $-0\overset{m}{.}20$ (roughly a $2\sigma$ deviation) in both passbands for the $0^h$ data is sufficient to explain the excess at faint magnitudes ($V_4 \geq 20$) and to make the differences at $V_4 < 20$ statistically insignificant. Our cluster identification algorithm is insensitive to zeropoint errors of this magnitude. The good agreement in the faint galaxy counts between fields demonstrates that on very large scales ($\gtrsim 600 h^{-1}$ Mpc) the optical universe does appear to be homogeneous.

### 3.3. Object Classification Accuracy

The uniformity of the log $N_{gal}$ - magnitude relationship from field to field is certainly a necessary but not sufficient condition for confirming good classification accuracy. An additional check is to compare object classifications from the $V_4$ and $I_4$ images. We obtain identical classifications for 95% of the objects detected in both passbands when the object $V_4$ magnitude is 20.5 or brighter. For objects with $V_4 \leq 22.5$, identical classifications are obtained 86% of the time. For the statistically complete survey used for cluster detection ($V_4 \leq 23.8$), identical classifications are obtained for 75% of the objects.

There are two reasons why object classification accuracy is not perfect. The first is that the FOCAS classifier (and, indeed, most any classifier) breaks down for objects fainter than about 1 magnitude above the completeness limit. Near the limit, the signal-to-noise ratio is low (S/N $\sim$ 4 at $V_4 = 23.8$) and the mean object area is $\lesssim 5\times$ FWHM$^2$, making reliable classification difficult. The second is that the PSF varies in shape from center to edge (*cf.*, §3.1). Since we are using a single PSF per frame, classification accuracy should degrade somewhat near the east/west edges of every scan (in TDI mode, the PSF is a function of column only). We have attempted to minimize the latter effect by allowing a 30% tolerance in the fit to the mean PSF for objects classified as stars (*c.f.* §3.1); however, the effect is nonetheless still present. Fortunately, the net result is to introduce only minor contamination of the galaxy catalogs, as shown in Figures 7 and 8. The star catalogs are more seriously contaminated. Statistical corrections are made to compensate for the contamination, as discussed below.

We now demonstrate the nature and amplitude of the misclassifications. Figure 7a and 7b show the galaxy-to-star ratio (GSR) for the 13$^{\mathrm{h}}$ field as a function of $V_4$ and $I_4$ magnitude, respectively (results for the other fields are very similar). The solid points represent the data for all objects regardless of CCD column; the open points represent data for objects with $200 \leq$ CCD Column $\leq 600$. The breakdown of the classifier at low signal-to-noise levels is the source of the turnover in the GSR in Figures 7a and 7b. The probability of a faint star being classified as a galaxy is higher than the probability of a faint galaxy being classified as a star. This has been verified by running the classifier on simulated CCD images. Galaxies outnumber stars by about 7 to 1 at $V_4 \sim 23$ and, consequently, even a small error in galaxy classification results in a big change in the faint star counts. For the same reason, the galaxy counts are not significantly affected. For $V_4 \lesssim 21$ the GSR is essentially independent of CCD column. At fainter magnitudes, the GSR is systematically lower for objects located within the central 400 columns. This is precisely the trend expected from a PSF variation - the assumed PSF is rounder than the actual PSF near the east/west edges and hence relatively more stars are misclassified as galaxies in those areas. This is shown directly in Figures 8a and 8b which depict the number of stars and galaxies as a function CCD column number. There is an obvious drop in star counts in the 150 outermost columns. Again, the galaxy counts are not as noticeably affected because galaxies vastly outnumber stars at $V_4 \sim 23$ and because the broadened PSF is still narrower than the typical galaxy profile.

Using both the observed data and results from simulations, we have adopted effective weighting factors to correct the galaxy counts depending on position and magnitude. At $V_4 = 23.8$ and $I_4 = 22.5$ (the limits adopted for the purpose of finding clusters) and near the east/west CCD edges the galaxy count correction is typically 15%, only a modest effect. At the same magnitude limits but near the CCD center, the galaxy count correction is only $\lesssim 10\%$. Fortunately, our final galaxy cluster catalogs are insensitive to the application of these corrections (*i.e.*, the same clusters are detected whether or not we apply these corrections).

### 3.4. Astrometric Calibration

A precise astrometric calibration is required to obtain accurate celestial coordinates for subsequent spectroscopic programs and to perform accurate catalog matching in order to generate $V_4 - I_4$ colors. To perform this calibration, we first construct a global cartesian coordinate system for each color in each survey field by determining the relative offsets between the CCDs in the 4-Shooter as well as the offsets between adjacent scans. The global coordinate system provides the $x$-$y$ position relative to the origin of the first scan in a survey field. The $x$-$y$ offsets between the $V_4$ band CCDs (CCDs 1 and 4) and between the $I_4$ band CCDs (CCDs 2 and 3) are determined by using the photometric calibrations, or "fast" scans (see §3.2), which run perpendicular to the survey scans. The relative positions of objects on the fast scans give their true separations which, when combined with their apparent separations on CCDs 1 and 4 or CCDs 2 and 3, allow the inter-CCD offsets to be computed quite accurately. The inter-CCD offsets remain constant for any given scan but can vary from scan to scan as the 4-Shooter CCDs are periodically adjusted. Offsets between adjacent survey scans are determined by using the objects in the scan overlap regions. The precise $x$-$y$ offsets between overlapping frames in adjacent scans also depends on the declination (since the scans converge as one moves towards the 1950.0 pole). To account for this, the observed offsets at the start, middle, and end of the adjacent scans are measured; the offset for each overlapping pair of frames is then calculated by a linear interpolation of the measured offsets. When the offsets are determined, the global $x$-$y$ position for each object is computed and stored in the catalogs. Because the offsets are all derived using 4-Shooter data directly, the angular scale per unit position in the global system is identical to the CCD image scale. Indeed, we recover the published 4-Shooter CCD scale with great accuracy when our astrometric solution is computed. Our mean best fit 4-Shooter image scale (in arcseconds per pixel) is $0.33501 \pm 0.00011$ ($V_4$) and $0.33503 \pm 0.00021$ ($I_4$).

The astrometric reference objects are obtained by selecting 50–60 unsaturated ($V_4 \gtrsim 17$) stars per survey field brighter than $V_4 = 19$. The J2000 celestial positions of these stars are then measured from the digitized "Quick V" (epoch 1983) Palomar Observatory Schmidt plates (Lasker *et al.* 1990). The stellar centroid is measured from the digitized scan using the density-weighted position and the celestial coordinates are determined using the plate solution generated during construction of the HST Guide Star Catalog V1.0 (see Lasker *et al.* 1990 and Russell *et al.* 1990 for details). An astrometric solution for each 4-Shooter survey field is then computed using a 6-term per coordinate polynomial fitting procedure, with one iteration to exclude reference stars with large ($> 1\farcs5$) residuals. The typical rms deviations in the final solution are $0\farcs48$ in right ascension and $0\farcs36$ in declination. The scatter in the astrometric solution is due to two roughly equal components — uncertainties in measuring the relative scan and CCD offsets and uncertainties in the measuring the centroids of reference stars from the digitized Schmidt plates.

## 4. CLUSTER CATALOG CONSTRUCTION

## 4.1.  The Cluster Finding Algorithm

The detection of a rich cluster at low redshift is successfully accomplished by many different methods (*e.g.*, visual, percolation, wavelets, matched filters). For statistical studies of distant clusters, though, the important properties of the algorithm are the detection rate for small amplitude fluctuations (*e.g.*, poor or distant cluster completeness), the false positive detection rate (spurious detections due to chance projections on the sky), and the accuracy with which the selection function can be determined. The latter is particularly important because the stability and accuracy of the cluster selection function are direct measures of the objectivity of a given cluster detection procedure. Our approach to cluster detection is to make use of both positional and photometric information simultaneously by using a matched filter algorithm which optimally enhances the contrast of a cluster galaxy distribution with respect to the foreground and background galaxy distributions. Since the shape of such a filter will naturally depend on the redshift of the clusters being searched for, the amplitude of the filtered signal provides a useful redshift estimate.

The matched filter method has many attractive features — 1) it is optimal for identifying weak signals in a noise dominated background (exactly the case for distant clusters), 2) photometric information is incorporated in an optimal way, 3) the contrast of overdensities which approximate the filter shape is greatly enhanced, 4) improved suppression of false detections, 5) redshift estimates are produced as a byproduct, and 6) good performance is seen over a large dynamic range in redshift. There are some disadvantages, in particular, one must assume a form for the cluster luminosity function and radial profile. We find that with sensible choices, however, the filters are broad enough to allow detection of systems with a range of properties.

The objective is to design a filter which preferentially suppresses galaxy fluctuations which are not due to real clusters. We can derive such a filter directly from a maximum likelihood estimator by modeling the spatial and luminosity distribution of the galaxies within a cluster and then determining the likelihood $\mathcal{L}$ that our model fits the actual data. We begin with the distribution of the *number* of galaxies per unit area per magnitude in any given patch of sky. This distribution is represented as

$$
\begin{aligned}
D(r, m) &= \text{background} + \text{cluster} \\
&\equiv b(m) + \Lambda_{cl} P(r/r_c) \phi(m - m^*)
\end{aligned} \tag{11}
$$

where $D(r, m)$ is the total number of galaxies per magnitude per arcsec$^2$ at a given magnitude $m$ and a given distant $r$ from a cluster center, $b(m)$ is the background galaxy counts (num/mag/arcsec$^2$), $P(r/r_c)$ is the projected radial profile of the cluster galaxies (num/arcsec$^2$), $\phi(m - m^*)$ is the differential cluster luminosity function (num/mag), and $\Lambda_{cl}$ is proportional to the total number of cluster galaxies and is a measure of the cluster's richness. The parameter $m^*$ is the apparent magnitude corresponding to the characteristic luminosity of the cluster galaxies and $r_c$ is a characteristic cluster scale length (as seen in projection).

The likelihood $\mathcal{L}$ of the data given the model or, equivalently, the likelihood of having a cluster at a given position is given by

$$\ln \mathcal{L} \propto \int_{Area, m} \ln \sigma + \int_{Area, m} \frac{[b(m) + \Lambda_{cl}P(r/r_c)\phi(m - m^*) - D(r, m)]^2}{\sigma^2}. \quad (12)$$

The above equation is valid when $b(m)$ is large enough that the Gaussian approximation holds; therefore, the statistics are dominated by the background galaxy counts and attribute all uncertainty to Poissonian noise in the background. This implies $\sigma^2 = b(m)$. We explicitly ignore the two point galaxy-galaxy correlation function. We do not solve for $b(m)$ but assume it a priori.

The likelihood is, thus, a function of $m^*$, $r_c$, and $\Lambda_{cl}$, i.e. $\mathcal{L} = \mathcal{L}(m^*, r_c, \Lambda_{cl})$. The dependence on $m^*$ and $r_c$ allows one to obtain a redshift estimate for each cluster. We choose to accomplish this by maximizing the likelihood with respect to the characteristic $m^*$. The characteristic magnitude, $m^*$, is much more sensitive to the redshift than the cluster scale, $r_c$.

Expanding the right hand side of the likelihood equation (Eq. 12), we get

$$\ln \mathcal{L} \propto \int \left[ \frac{2\Lambda_{cl}P(r/r_c)\phi(m - m^*)D(r, m)}{b(m)} - 2\Lambda_{cl}P(r/r_c)\phi(m - m^*) \right] d^2r \ dm$$
$$- \int \left[ \frac{\Lambda_{cl}^2 P^2(r/r_c)\phi^2(m - m^*)}{b(m)} + C \right] d^2r \ dm \quad (13)$$

The expression for $\Lambda_{cl}$, determined by setting $\frac{\partial \ln \mathcal{L}}{\partial \Lambda_{cl}} = 0$, is

$$\Lambda_{cl} = \frac{\int \frac{P(r/r_c)\phi(m - m^*)D(r, m)}{b(m)} d^2r \ dm - \int P(r/r_c)\phi(m - m^*) \ d^2r \ dm}{\int P^2(r/r_c)\frac{\phi^2(m - m^*)}{b(m)} d^2r \ dm} \quad (14)$$

Inserting $\Lambda_{cl}$ (Eq. 14) into the expression for the likelihood (Eq. 13), we find

$$\ln \mathcal{L} \propto \frac{\left[ \int \frac{P(r/r_c)\phi(m - m^*)D(r, m)}{b(m)} d^2r \ dm - \int P(r/r_c)\phi(m - m^*) \ d^2r \ dm \right]^2}{\int P^2(r/r_c)\frac{\phi^2(m - m^*)}{b(m)} d^2r \ dm} \quad (15)$$

The expression in the numerator is simply the square of the *excess* above the background, as the background contribution is removed by the second term in Eq. 15. Since this term is a constant (if the integrals converge; see below), we could simply maximize the first term if we needed only to maximize the numerator. The denominator is *not* a constant; however, it is monotonic function which is proportional to some multiple of $1/b(m^*)$, which to a good approximation is just a power-law in the flux. Thus, maximizing the expression on the RHS of Eq. 15 is roughly equivalent to maximizing

$$\ln \mathcal{L} \propto \int P(r/r_c) \ \frac{\phi(m - m^*)}{b(m)} \ D(r, m) \ d^2r \ dm \quad (16)$$

except for a shift $\Delta m^*$ in the location of the maximum which is constant if $b(m)$ can be accurately modeled as a power-law in $m$. We choose to use this expression not primarily because of its simpler form but because it depends only on the *shape* of the background, while Eq. 15 is very sensitive to the correct *normalization* of the background. In practice, there are no quantitative differences in the cluster detections (and their estimated properties) when maximizing the formal likelihood equation (Eq. 15), compared to maximizing Eq. 16.

The application of the matched filter to an input galaxy catalog is accomplished by evaluating the integral in Eq. 16 at every point in the survey and over a range of $m^*$ and $r_c$ values (or equivalently, over a range of redshifts). The characteristic luminosity, $L^*$, and the intrinsic cluster scale length are assumed to remain fixed in physical units. The observables, $m^*$ and $r_c$, are assumed to vary with redshift as prescribed by a Friedman-Robertson-Walker cosmology with $H_\circ = 75$ km s$^{-1}$ Mpc$^{-1}$ and $q_\circ = 0.5$. The redshift dependence of $m^*$ also includes a k-correction discussed in §4.2. Because the galaxy distribution is a set of discrete coordinates and fluxes, one can represent $D(r, m)$ as a series of $\delta$-functions at the observed positions and magnitudes. The integral in Eq. 16 is, therefore, equivalent to the sum over all galaxies in a field

$$S(i,j) \;=\; \sum_{k=1}^{N_g} P(r_k(i,j)) L(m_k) \equiv \int P(r/r_c)\,\frac{\phi(m-m^*)}{b(m)}\,D(r,m)\,d^2r\,dm \qquad (17)$$

The "filtered" galaxy catalog is created by evaluating this sum for each element in a two-dimensional array, $S(i,j)$. The elements correspond to a series of grid points covering the entire survey area. The array, $S(i,j)$, can be treated as an "image" and we shall hereafter refer to its elements as pixels. In Eq. (17), $N_g$ is the total number of galaxies in the catalog, $P(r_k(i,j))$ is the angular weighting function (hereafter the "radial" filter), $r_k(i,j)$ is the separation between the $k^{\text{th}}$ galaxy and the $(i,j)$ pixel center, $m_k$ is the magnitude of the $k$th galaxy, and $L(m_k)$ is the luminosity weighting function (hereafter the flux filter). The dimensions of the array $S(i,j)$ are dependent on both the angular size of the survey and the redshift to which the filters are tuned. The best choice of pixel size is to maintain a constant metric width that is approximately half a cluster core radius ($r_c$), although the results are quite similar with pixels ranging from $0.25r_c$ to $2r_c$. While the above summation runs over all galaxies, in practice the radial filter will have a cutoff radius and, thus, the execution time is greatly reduced for a large catalog.

The *optimal* flux filter is $\frac{\phi(m-m^*)}{b(m)}$; however, the integral of this function diverges for Schechter (1978) luminosity functions with $\alpha < -1$ as one sums to magnitudes fainter than $m^*$. We can correct for this and, at the same time, place most of the weight of the filter near $m^*$ (where the contrast of the cluster against the background is the largest) by using the modified flux filter

$$L(m) = \frac{\Phi(m-m^*)}{b(m)} \equiv \frac{\phi(m-m^*)\,10^{-0.4(m-m^*)}}{b(m)} \qquad (18)$$

Here we have introduced a power-law cutoff of the dimensionless form $10^{-\beta(m-m^*)}$ to keep the function integrable. We choose $\beta = 0.4$ as this would represent weighing by the *flux* of the galaxy.

We choose as the radial filter

$$P(r/r_c) = \begin{array}{ll} \frac{1}{\sqrt{1+(r/r_c)^2}} - \frac{1}{\sqrt{1+(r_{co}/r_c)^2}} & \text{if } r < r_{co} \\ 0 & \text{otherwise} \end{array} \tag{19}$$

where $r_c$ is the cluster core radius and $r_{co}$ is an arbitrary cutoff radius. The function $P(r/r_c)$ is circularly symmetric. However, since the diameter of the filter extends well beyond a core radius, systems which are not circularly symmetric are still easily detected, albeit at a different significance level.

We choose the following normalizations for the radial and flux filters

$$\int_0^\infty P(r/r_c) \, 2\pi r \, dr = 1 \tag{20}$$

$$\int_0^{m_{lim}} \Phi(m - m^*) \, dm \equiv \int_0^{m_{lim}} \phi(m - m^*) \, 10^{-0.4(m-m^*)} \, dm = 1 \tag{21}$$

The limits of integration for $P(r/r_c)$ are effectively $[0, r_{co}]$. The limits for $L(m)$ are the magnitude range of the sample $[0, m_{lim}]$. Figures 9 and 10 show the final radial and flux filters as a function of filter redshift.

With the above normalizations, the "background level" of a filtered image is $S(i,j) = S_{bg} = 1$. We expect the distribution of $S(i,j)$ values for a given field (and a given filter redshift) should be a roughly Gaussian distribution with a standard deviation characterized by the Poissonian error in the background distribution, a mode of 1, and a tail toward high values of $S(i,j)$ representing significant cluster detections. In practice, the mode ($S_{bg}$) is not strictly 1 but varies slightly with the filter redshift because we have estimated the background counts, $b(m)$, as the *total* galaxy counts in the field (background + clusters); we have no way of separating the two components a priori. Figure 11 shows the distribution of $S(i,j)$ values for the $0^h$ field with the matched filter "tuned" to a redshift of $z = 0.6$.

### 4.1.1. Effect of the Normalization on the Redshift Estimate

The normalization of the flux filter attempts to compensate for the portion of the flux filter that extends beyond the magnitude limit of the sample. This correction is *strictly* appropriate for a pure background distribution $b(m)$, as it assumes that the cluster galaxy counts are small compared to the field galaxy counts. However, when a significant section of the flux filter is truncated by the survey magnitude limit, a small but correctable bias is introduced in the estimated redshifts derived from the matched filter for the most distant clusters. The redshift estimate for a cluster candidate is determined by finding the filter redshift at which the signal, $S(i,j)$, is maximized. The bias is essentially insensitive to the radial filter normalization and we will ignore the radial filter $P(r/r_c)$ in the computations here. The flux filter normalization then implies the following for the data $D = b + \phi_{cl}$

$$
\begin{aligned}
S(i,j) &= \frac{\int_0^{m_{lim}} \frac{D\Phi}{b} \, dm}{\int_0^{m_{lim}} \Phi \, dm} = \frac{\int_0^{m_{lim}} \frac{[\phi_{cl}+b]\Phi}{b} \, dm}{\int_0^{m_{lim}} \Phi \, dm} \\
&= \frac{\int_0^{m_{lim}} \frac{\phi_{cl}\Phi}{b} \, dm}{\int_0^{m_{lim}} \Phi \, dm} + 1
\end{aligned}
\tag{22}
$$

This is explicitly correct for background only ($\phi_{cl} = 0$), as $S(i,j) = S_{bg} = 1$. For real data ($\phi_{cl} > 0$), the above normalization results in an *overcorrection* of the cluster signal at high filter redshifts; that is,

$$
\frac{\int_0^{m_{lim}} \frac{\phi_{cl}\Phi}{b} \, dm}{\int_0^{m_{lim}} \Phi \, dm} > \frac{\int_0^{\infty} \frac{\phi_{cl}\Phi}{b} \, dm}{\int_0^{\infty} \Phi \, dm}
\tag{23}
$$

This overcorrection becomes important only when a significant portion of the flux filter extends beyond the magnitude limit of the sample ($m^* \to m_{lim}$).

We can quantify the effect by *analytically* calculating the cluster signal, $S_{cl}$, as a function of both the filter $m^*$ (or equivalently filter redshift $z_{est}$) and the cluster redshift. We calculate $S_{cl}$ from the expression

$$
S_{cl} = \int_0^{m_{lim}} \frac{\phi(m - m^*_{cl})\Phi(m - m^*)}{b(m)} dm \int_0^{\infty} P_{cl}(r)P(r/r_c)d^2r
\tag{24}
$$

where $\phi(m - m^*_{cl})$ and $P_{cl}(r)$ are the characteristic luminosity function and radial profile of the input cluster. We have assumed a cluster luminosity function which has the same form and k-correction as the flux filter and a cluster radial profile which has the same form as the radial filter (see §4.2, §4.3, and Table 3 for explicit information on the filter parameters used). $P(r/r_c)$ and $\Phi(m - m^*)$ are normalized according to Eq. (20) and (21).

The dependence of $S_{cl}$ on filter redshift is affected by the assumed functional form of the background galaxy distribution $b(m)$ and the magnitude limit of the sample. Figure 12 shows $S_{cl}$ as a function $m^*$ (the equivalent filter redshift, $z_{est}$, is listed along the top axis) for the $V_4$ and $I_4$ bands. Each curve represents a cluster at a redshift from 0.2 to 1.2 in intervals of 0.1 (from top to bottom). The filter is run at each $z_{est}$ (or $m^*$), and the resulting values of $S_{cl}$ are plotted. The survey magnitude limits are indicated by dashed lines.

For clusters at lower redshifts, $S_{cl}$ reaches a maximum at or close to the actual redshift and, hence, a reliable redshift estimate is obtained. However, for a cluster at $z > 0.6$, *i.e.*, as $m^* \to m_{lim}$, the maximum of $S_{cl}$ cannot be used to gauge the redshift of the cluster since $S_{cl}$ is always maximized at the highest filter redshift. This behavior is the direct result of the overcorrection of the cluster signal by the adopted flux filter normalization. When $m^* << m_{lim}$, the problem is negligible.

We can correct for this bias by applying an appropriate, redshift–dependent cluster signal correction ($CSC$) to the detected cluster signal, $S(i,j) - S_{bg}$. We derive this correction by making Eq. (23) into an equality, which is the case when the flux filter is not truncated by the magnitude limit, *i.e.*,

$$CSC = \frac{\int_0^{m_{lim}} \frac{\phi_{cl}\Phi}{b} \, dm}{\int_0^{m_{lim}} \Phi \, dm} \bigg/ \frac{\int_0^{\infty} \frac{\phi_{cl}\Phi}{b} \, dm}{\int_0^{\infty} \Phi \, dm} \qquad (25)$$

The corrected signal is then

$$S_{cl}^{cor} = \frac{(S(i,j) - S_{bg})}{CSC} \qquad (26)$$

The correction factor is dependent on the actual cluster redshift and the assumed luminosity function of the cluster $\phi_{cl}$ (here we have assumed that the clusters have the same luminosity function and k-correction as the flux filter), but *not* on the cluster richness. In deriving the $CSC$, we have assumed that the clusters have the same luminosity function, k-correction, and radial profile as our matched filter. The fiducial redshift we have chosen for computing the $CSC$ is $z = 0.7$. The $CSC$ is well approximated by the expression $(1 + z_{est}^n)^\gamma$. The advantage of this approximation is strictly for computational convenience. The cluster redshift estimates do not change significantly if we use the full expression given in Eq. 25. The best-fit functions are

$$CSC(V_4) = (1 + z_{est}^{4.4})^{2.1} \qquad (27)$$
$$CSC(I_4) = (1 + z_{est}^{3.0})^{1.0} \qquad (28)$$

Note that these corrections are only *estimates* to the cluster signal as they explicitly depend on the actual redshift of the cluster, its luminosity function, and its k-correction. Figure 13 shows $S_{cl}$ as a function $m^*$ (and $z_{est}$; top axis) after application of the $CSC$. The signal now peaks at or near the actual redshift (up to $z = 1.2$) and, thus, is a better redshift estimator (see §4.2.1 for a discussion of the redshift accuracy). Nonetheless, estimated redshifts of clusters at $z \gtrsim 0.7$ in the $V_4$ band and $z \gtrsim 1.0$ in the $I_4$ band are still quite uncertain as the matched filter is fitting only the bright end of the luminosity function.

## 4.2. Cluster Detection Criteria

The algorithm is run on each of the ten galaxy catalogs (five survey fields in two passbands), tuned to redshifts in the range $0.2 \leq z \leq 1.2$ in 0.1 intervals. The required inputs are lists of galaxy positions and magnitudes. At each redshift a filtered image of the galaxy catalog is produced according to the prescription in the previous section. Use of a matched filter does not remove the need to select a detection threshold. Detection is done on the filtered galaxy maps where the cluster contrast is improved. Clusters are identified by searching for the local maxima within a moving box which is $1.25h^{-1}$ Mpc across. We center the box on each "pixel" of the filtered galaxy map array, $S(i,j)$. A candidate cluster is registered when the central pixel in the box is the local maximum and it lies above a prescribed threshold. The detection threshold is taken to be the 95th percentile (approximately the $2\sigma$ level).

The resulting $V_4$ and $I_4$ cluster candidates are matched with each other to locate those systems which are detected in both bands. We then impose the additional criteria that the peak signal (highest valued pixel in a given detection) must exceed the $3\sigma$ threshold in at least one passband, if the cluster candidate is detected in both passbands, or must exceed the $4\sigma$ threshold if it is only detected in a single passband [note, we use the $S_{cl}^{cor}$ (Eq. 26) distribution to determine the peak signal and significance of all detections]. Detection in a single bandpass is often explained by the fact that the $V_4$ scans cover a bit more area than the $I_4$ scans or by the presence of a bright red or blue star which reduces the effective $I_4$ or $V_4$ survey areas differently. There are, none the less, some interesting cluster candidates detected in only one of the two bands. These are listed or described in Tables 4 and 5. A cluster candidate must satisfy the above criteria in at least two of the matched filter redshift intervals to be included in our main catalog. This last criterion provides additional rejection of spurious detections.

To compute the $95^{\text{th}}$ percentile detection threshold and the significance of the peak signal in a cluster candidate, we examine the binned distribution of the values in the filtered galaxy map array, $S(i, j)$. The distribution is typically Gaussian with a tail toward higher values, representing the cluster detections (see Figure 11). We characterize this distribution by calculating the background level, $S_{bg}$, as the mode and the dispersion around the background, $\sigma_{bg}$, as $0.741 \times (Q_3 - Q_1)$ where $Q_1$ and $Q_3$ are the first and third quartiles, respectively. The coefficient is correct for computing the standard deviation from the interquartile range of a normal distribution. The significance of a given peak signal, $S_p$, is then simply $(S_p - S_{bg})/\sigma_{bg}$.

The redshift estimate for a given cluster candidate is determined by finding the filter redshift at which the peak signal is maximized (*e.g.*, see Figure 15). The choice of the k-correction and of the Schechter (1976) luminosity function parameters used to define $L(m)$ strongly affect the redshift estimate but have little effect on cluster detection reliability providing extreme values are not used and that the luminosity functions of distant clusters are closely approximated by the Schechter form. We quantify this last statement in §4.3.

### 4.2.1. Redshift Estimation Accuracy

Figure 14 shows the offset between the estimated and true redshift as a function of true redshift for $\sim 400$ simulated clusters. The luminosity functions of the simulated clusters exactly match the luminosity function used to compute the flux filter function. The offset is small and is independent of the cluster redshift as long as the assumed k-correction and form of the cluster luminosity function are accurate and the $CSC$ has been applied.

The real test of the accuracy of our redshift estimates, of course, is to compare them with the observed redshifts of real clusters in our survey. Unfortunately, there are only six clusters in the survey which have reliably determined redshifts, although they span a reasonable range in redshift, $0.3 < z_{obs} < 0.8$. Figures 15a and 15b show the offset, $z_{est} - z_{obs}$, as functions of the matched filter signal ($S_{cl}^{cor}$) and passband for these six clusters.

In all cases, the matched filter signal reaches its peak within $z_{obs} \pm 0.2$. A subregion of the filtered $13^{\mathrm{h}}$ galaxy catalog (with the $CSC$ applied) centered on the GHO cluster 1322+3028 ($z_{obs} = 0.70$) is shown in Figure 16 at the 11 filter redshift intervals ($0.2 \rightarrow 1.2$) used. The GHO cluster 1322+3027 ($z_{obs} = 0.75$) also lies in this subregion. The most distant clusters appear to have estimated redshifts which are systematically lower (by $0.1 — 0.2$) than their observed redshifts. Such a trend is expected if we have underestimated the amount of spectral evolution (*i.e.*, overestimated the k-correction). Since our nominal flux filter is constructed assuming no evolution (see §4.3), this trend is not surprising. However, given that precise evolutionary corrections are not well known, we simply publish our redshift estimates uncorrected for such systematic trends and present this result as a caveat to users of this catalog.

The $V_4$ and $I_4$ redshift estimates are quite consistent. Of the 74 matched clusters in our main catalog, 58 clusters (78%) have $|z_V - z_I| \leq 0.2$. Of the 16 matched clusters with $|z_V - z_I| > 0.2$, half are cases of distant clusters located near clusters at moderate redshifts where the detection algorithm locked onto the distant cluster in one passband and the foreground cluster in the other passband. The remaining differences are consistent with the inherent errors in the redshift estimates.

### *4.2.2. Richness Measure $\Lambda_{cl}$*

Once we have determined the estimated redshift of the cluster, the cluster richness is given by calculating $\Lambda_{cl}$ at $z_{est}$. From Eqs. 11 and 14 and the adopted filter normalizations (Eqs. 20 and 21), $\Lambda_{cl}$ is given by

$$
\begin{aligned}
\Lambda_{cl} &= \frac{S(i,j) - 1}{\int_0^\infty P^2(r/r_c)\, 2\pi r\, dr \int_0^{m_{lim}} \frac{\phi(m-m^*)\Phi(m-m^*)}{b(m)}\, dm} \\
&= \frac{S(i,j) - 1}{\int_0^\infty P^2(r/r_c)\, 2\pi r\, dr \int_0^{m_{lim}} \frac{\phi^2(m-m^*)f(m)}{b(m)f(m^*)}\, dm}
\end{aligned}
\tag{29}
$$

where $f(m) = 10^{-0.4m}$. We use $S(i,j)$ prior to application of the $CSC$ (the $CSC$ is only needed for redshift estimation). The normalization of the differential cluster luminosity function $\phi(m - m^*)$ is $\int \phi(m - m^*)\, f(m)\, dm = L^*$ (which is exactly equivalent to Eq. 21); this normalization implies that $\Lambda_{cl}$ is a measure of the effective number of $L^*$ galaxies in a cluster. We can easily see this by determining the total flux of the cluster ($L_{tot}$). Integrating Eq. (11) over luminosity ($L = 10^{-0.4m}$), we have

$$
\begin{aligned}
L_{tot} &= \int 10^{-0.4m}\, D(r,m)\, d^2r\, dm \\
&= \int 10^{-0.4m}\, b(m)\, dm \int d^2r + \Lambda_{cl} \int \phi(m - m^*)\, 10^{-0.4m}\, dm \int P(r/r_c)\, d^2r \\
&= \text{constant} + \Lambda_{cl} 10^{-0.4m^*}
\end{aligned}
\tag{30}
$$

Therefore, the cluster luminosity is $L_{cl} = \Lambda_{cl} 10^{-0.4m^*} = \Lambda_{cl} L^*$, implying that $\Lambda_{cl}$ is the equivalent number of $L^*$ galaxies in the cluster.

### 4.2.3. Abell Richness Measure

Since estimates of cluster richness based on parameters from the matched filter cluster algorithm, such as the peak signal or the detection area, depend on the cluster profile and estimated redshift, we make an independent estimate of richness from the cluster galaxy counts in each of the two passbands. We follow the specification of Abell (1958) and calculate richness ($N_R$) as the number of member galaxies (above background) that are brighter than $m_3 + 2^m$ (where $m_3$ is the magnitude of the third brightest galaxy) and located within a projected radius of $1.0h^{-1}$ Mpc from the cluster center. We use a radius of $1.0h^{-1}$ Mpc, rather than Abell's historical $1.5h^{-1}$ Mpc, because of the growing uncertainty in background subtraction at larger radii. Even at $1.0h^{-1}$ Mpc, it is still difficult to calculate $m_3$ accurately from the cluster number counts; therefore, we first calculate the central density ($N_0$; Bahcall 1981). That is, we determine $m_3$ from the background subtracted galaxy counts within the smaller radius of $0.25h^{-1}$ Mpc.

For each field, we calculate the differential background galaxy surface density (galaxies/deg$^2$/0.20 mag) by using regions which contain no detected clusters. For each cluster in the field, we bin all galaxies within the central $0.25h^{-1}$ Mpc (at the estimated cluster redshift) in the same 0.20 magnitude bins as the background galaxies. The expected number of background galaxies in each magnitude bin is subtracted. From the final cluster number counts, we determine the magnitude bin which contains the third brightest galaxy. The sum of all the galaxies in bins with $m \leq m_3 + 2^m$ is defined as the central density of the cluster. We then perform the same procedure at a radius of $1.0h^{-1}$ Mpc, except that we fix $m_3$ at the value determined in the central density calculation.

With the above method for estimating $m_3$, we find that there is a good correlation between $N_R$ computed within a radius of $1.0h^{-1}$ Mpc and $N_R$ computed within a radius of $1.5h^{-1}$ Mpc. Specifically, $N_R(1.0)/N_R(1.5) = 0.72 \pm 0.05$. We use this relationship to estimate the conventional Abell richness of a distant cluster candidate in order to make some comparisons with the space density of low redshift clusters in §5.

Figure 17 shows the relation between $N_R(1.0)$ and $\Lambda_{cl}$ for both the real cluster candidates and for cluster detected using the algorithm in simulated galaxy catalogs. The correlation between the two richness parameters is significant at a greater than 99.2% level (using the Spearman rank correlation function), indicating that these richness measures are meaningful. The rms scatter, however, is large with $sigma_{\Lambda_{cl}} \sim 23$ for a given $N_R$. The scatter is, in part, due to the dependence of $\Lambda_{cl}$ on the shape of both the matched filter and the actual cluster, as well as background subtraction and redshift errors in the calculation of $N_R$. When $m_3 + 2$ is fainter than the survey magnitude limit, the estimated cluster richness count is highly uncertain and is often too small. Consequently, for the highest redshift clusters (indicated by stars in Figure 17), the richness estimate $N_R$ systematically underestimated the actual cluster richness.

## 4.3. The Cluster Catalog

The matched filter and cluster detection parameters used to generate the cluster catalog are listed in Table 3. These parameters were chosen because they provide a reasonable compromise between minimizing spurious detections and maximizing completeness at high redshift. We choose $100h^{-1}$ kpc and $1h^{-1}$ Mpc as the nominal values for the core radius and the cutoff radius, respectively, in the definition of the radial filter (Eq. 19). This choice was motivated by several studies on the profiles of nearby ($z < 0.2$) Abell clusters (*e.g.*, Beers & Tonry 1986; Oegerle *et al.* 1987; Postman, Geller & Huchra 1988; Broadhurst 1995; Smail *et al.* 1995; Squires *et al.* 1995; Tyson & Fischer 1995). Quantitative descriptions of how the selection of radial filter parameters affects catalog membership are given in §4.4.2 and §4.4.3. A detailed analysis of the radial profiles of the distant clusters found in the present survey will be presented in paper II (Lubin & Postman 1995). Briefly, we find that the typical surface density profile of our cluster candidates has $r_c \lesssim 140h^{-1}$ kpc and a power law profile on scales out to at least $1h^{-1}$ Mpc with slope $-1.4 \pm 0.4$.

The luminosity function parameters used are based on observations of low redshift clusters and field galaxies (Kirshner *et al.* 1979; Lugger 1986; Postman *et al.* 1988; Colless 1989; Marzke *et al.* 1994). We assume that the shape of the luminosity function is independent of redshift and adopt a k-correction appropriate for a non-evolving elliptical galaxy (see Figures 3a and 3b). While there is evidence to suggest that the galaxy luminosity function does depend on redshift (Eales 1993; Lilly 1993; Lonsdale & Chokshi 1993), our choice to ignore this primarily affects the accuracy of the redshift estimates (see §4.2.1) and does not significantly affect the catalog membership. This is because catalog membership is more sensitive to the shape of the flux filter than the location of its peak. The redshift estimate, on the other hand, is more sensitive to the location of the peak of the filter. Plausible changes in the shape of the cluster luminosity function with redshift result in changes in the flux filter shape which are significantly less than its width. For example, if we vary the slope of faint end of a Schechter luminosity function from $\alpha = -0.6$ to $\alpha = -1.6$ the FWHM of the flux filter varies by $\lesssim 15\%$ relative to the FWHM of our nominal ($\alpha = -1.1$) filter (the flux filter becomes broader as we steepen the faint end slope). The peak of the flux filter shifts from 0.2 mag brighter than the peak of our nominal filter to 0.3 mag fainter than the peak of the nominal filter as we vary $\alpha$ over the above range (the filter peaks at fainter magnitudes as we steepen the faint end slope). For comparison, the FWHM of our nominal flux filter is 1.87 mag, which is significantly larger than the above shift. To be sure, we ran the algorithm on the 13[h] field and varied the slope of the luminosity from $-0.6$ to $-1.6$ in 0.2 intervals. The resulting catalogs did not vary by more than 10% in size (relative to the 13[h] catalog published here) and the detailed agreement is excellent (17 of the 19 13[h] cluster candidates were found at all the $\alpha$ values explored). The width of the flux filter is insensitive to the characteristic luminosity, $L^*$, or the k-correction. The location of the peak of the flux filter, however, is quite sensitive to these parameters. If the amount of spectral evolution is underestimated (*i.e.*, the k-correction is overestimated) then the redshift estimate will be underestimated and visa versa. For example, if $z \gtrsim 0.5$ cluster galaxies have spectral energy distributions similar to non-evolving Sbc galaxies then using a non-evolving elliptical galaxy k-correction in the construction of the flux filter will result in $z_{est} - z_{obs} \approx -0.1$ at $z_{obs} = 0.5$ and $z_{est} - z_{obs} \approx -0.2$ at $z_{obs} = 0.8$.

Table 4 contains the 79 cluster candidates which satisfy the detection criteria with the above matched filter parameters. For each cluster, the first column contains a numeric ID, columns 2 and 3 contain the J2000 equatorial coordinates of the cluster (derived from the weighted average of the locations of the peak signal in the $V_4$ and $I_4$ bands for systems detected in both bands), columns 4 through 10 contain, respectively, an effective $V_4$ band cluster radius in arcseconds, the $V_4$ band redshift estimate, the peak $V_4$ band cluster signal ($S_{cl}^{cor}$), the significance of the peak signal, the richness estimate ($\Lambda_{cl}$) derived from the $V_4$ cluster signal, and two richness estimates derived from the unfiltered $V_4$ band galaxy catalog ($N_o$ is the background subtracted galaxy count within $0.25h^{-1}$ Mpc; $N_R$ is the counts within $1.0h^{-1}$ Mpc; see §4.2.3). Columns 11 through 17 contain the same information for the $I_4$ band. The effective radius is defined as

$$R_{eff} = \sqrt{A_{Det}/\pi} \qquad (31)$$

where $A_{Det}$ is the area of the cluster detection that lies above the $95^{\text{th}}$ percentile threshold. If a cluster is only detected in a single passband, we measure its properties in the undetected band using the position and estimated redshift derived from the actual detection. Such clusters are identified in Table 4 by a missing entry in the effective radius column for the band in which the cluster was not detected. There is one cluster, #028, for which $I_4$ data does not exist. This cluster is a low $z$ ($z_{est} = 0.2$) system which would have been easily detected in the $I_4$ band.

Single band detections with peak signal $< 4\sigma$, $z_{est} \geq 0.4$, and which appear to be real clusters via visual inspection of the CCD images are listed in our supplemental catalog (Table 5). The supplemental catalog is not intended to be complete but contains 28 excellent distant cluster candidates, one of which is a known GHO cluster. The notes to Tables 4 and 5 contain details about some of the cluster candidates in the main and supplemental catalogs, including alternate identifications when available.

Pictures of all 107 cluster candidates are displayed in Plates 1 – 6. The images are of the central $134'' \times 134''$ regions; north is up and west is to the right. The cluster ID is given at the top of each picture (supplemental cluster IDs are followed by the letter S). The clusters are shown in the passband in which the significance of the peak signal is highest. Occasionally there are narrow gaps in the images – these are just the unusable areas between 4-Shooter CCDs.

Although the PDCS fields are not all completely interior to the original Gunn & Oke (1975) survey areas, there are 27 GHO clusters that were imaged by the present CCD survey.

With the adopted matched filter and cluster detection parameters, 24 of the 27 GHO clusters are detected but only 16 of these satisfy all the criteria for inclusion into the main catalog. GHO cross identifications are given in Table 6. The 3 GHO clusters not detected at all are CL0030+0536, CL0030+0512, and CL1605+4106; they are all missed because their peak signals fall below the 95$^{\text{th}}$ percentile detection threshold.

## 4.4. False Positive Rate

The number of spurious clusters detected (*i.e.*, the false positive rate) is a function of the detection threshold and the matched filter parameters. We are able, fortunately, to measure these dependencies with sufficient accuracy that quantitative results can be extracted from our cluster catalogs. In the following sections, we discuss how these factors affect cluster catalog membership.

### 4.4.1. Peak Threshold Dependence

To evaluate the dependence of the false positive rate on the detection threshold, we ran simulations on artificial fields using a simple Monte-Carlo technique to reproduce, as closely as possible, the projected galaxy and magnitude distribution of the actual PDCS fields. In order to reproduce our matching criteria in the cluster detection method, we created ten pairs of background galaxy fields by simulating each field as viewed through the $V_4$ and $I_4$ passbands. We accomplished this by first generating a galaxy catalog in the $V_4$ band by using all galaxies detected in the $V_4$ band of one of the PDCS fields. The simulated field is divided into a grid of square cells of 6 arcmin sides. A background galaxy population is created by distributing these galaxies, with their corresponding magnitudes, at random within each cell with the *observed* variation in number from cell to cell and the average count per field. We then generate a matched simulated field in the $I_4$ band by placing all the galaxies detected in the $I_4$ band of the PDCS field at the positions of the galaxies in the simulated $V_4$ field such that the galaxy at that position would have an appropriate color (see Figure 3b).

The cluster detection algorithm is then run on each simulated background galaxy catalog using the same matched filter parameters as in the real data (see Table 3). We fix the 95$^{\text{th}}$ percentile detection threshold and standard deviation ($\sigma$) at values obtained from the actual galaxy fields. That is, we examine only those spurious detections in the uniform background which would have been considered cluster detections had they been present in the actual fields. Using the same detection criteria as discussed in §4.2, these spurious detections are first identified individually in $V_4$ and $I_4$ and then matched with each other to determine which are detected in both bands. The number of matched spurious detections (those detected in both passbands) as functions of peak signal threshold and radial extent (Eq. (31)) are shown in Figure 18. The histograms for the actual cluster detections are shown for comparison. There are 4.2 and 0.8 matched spurious detections deg$^{-2}$ with peak signals of $\geq 3\sigma$ and $\geq 4\sigma$, respectively. The observed density of matched cluster candidates is 13.4 deg$^{-2}$ and 6.8 deg$^{-2}$ for peak signal thresholds of $\geq 3\sigma$ and $\geq 4\sigma$, respectively. Approximately 88% of the observed cluster candidates are matched detections. This would imply false positive rates of $\sim 31\%$ and $\sim 12\%$ for the matched detections with peak signals $\geq 3\sigma$ and $\geq 4\sigma$, respectively. The observed densities of unmatched $\geq 4\sigma$ detections in the $V_4$ and $I_4$ passbands are 1.1 deg$^{-2}$ and 0.6 deg$^{-2}$, respectively. In the simulations, we find the densities of unmatched $\geq 4\sigma$ detections in the $V_4$ and $I_4$ passbands are 0.7 deg$^{-2}$ and 0.4 deg$^{-2}$, respectively. This would imply a very high false positive rate for the unmatched detections (about 65%). Figure 18 demonstrates, however, that the actual detections are typically larger than the spurious ones. If one restricts the radial extent to be at least 200 arcseconds, the false positive rates drop to $\sim 15\%$ and $\sim 9\%$ for matched detections with peak signals $\geq 3\sigma$ and $\geq 4\sigma$, respectively.

We can use the Abell richness estimates ($N_R$) of our cluster candidates to check these estimated spurious detection rates. We do this by computing the percentage of clusters with richnesses which are within two standard deviations of zero (i.e. consistent with being simply background). If such clusters are all false positive detections then we find spurious detection rates of $\leq 20\%$ for $\geq 3\sigma$ matched detections, $\leq 9\%$ for $\geq 4\sigma$ matched detections, and $\leq 22\%$ for $\geq 4\sigma$ single band detections. These rates are in good agreement with our estimates of the false positive rates in the simulated field galaxy distributions discussed above.

If the peak threshold is lowered below the $3\sigma$ level, the number of detections increases significantly: the detection rate is about 50% higher if a $2.5\sigma$ peak threshold is used. However, many of the additional detections are either superpositions of field galaxies or very poor systems. There are inevitably richer but quite distant clusters which fall below the $3\sigma$ threshold. Some are listed in the supplemental catalog (Table 5). Because the false positive rate at the $2.5\sigma$ peak threshold is comparable with the probable number of real detections, we have adopted the more reasonable $3\sigma$ peak threshold. In any case, the advantage of an objective detection algorithm is that a selection function can be accurately quantified and we do so in §4.7.1.

### 4.4.2. Core Radius Dependence

To quantify the dependence of the catalog size on the core radius used in the radial filter, we ran the cluster finding algorithm on the $0^{\rm h}$ field using three different core radii: $100h^{-1}$ kpc (the nominal value), $33h^{-1}$ kpc (3 times smaller), and $300h^{-1}$ kpc (3 times larger). These tests were done with the cutoff radius held constant at $1h^{-1}$ Mpc. All the candidates found when we use a radial filter with a $300h^{-1}$ kpc core radius are also found when the $100h^{-1}$ kpc core radius filter is used. Conversely, only 70% of the candidates found with the $100h^{-1}$ kpc core radius are found with the $300h^{-1}$ kpc core radius filter. Some of the candidates missed with the larger core radius filter are among our most distant cluster candidates. The number of cluster candidates found when we use a radial filter with a $33h^{-1}$ kpc core radius is twice the number found when a $100h^{-1}$ kpc core radius is used. As one lowers the core radius scale, one picks up smaller systems and consequently the spurious detection rate also increases. Some of the smaller systems detected are poor groups of galaxies. The rest are almost certainly superpositions of field galaxies that are not physical associations. None of the systems detected with the $33h^{-1}$ kpc core radius filter but missed at larger radii are rich clusters. Simulations reveal as the adopted core radius is decreased the false positive rate increases, reaching nearly twice the rates quoted in the previous section when $r_c = 33h^{-1}$ kpc. Conversely, the false positive rate declines to about 65% the rates quoted in the previous section when $r_c = 300h^{-1}$ kpc but then so does the detection probability of compact systems and very distant systems. The selection of a $100h^{-1}$ kpc core radius thus appears to be appropriate for reliable identification of clusters out to $z \approx 1$.

### 4.4.3. Cutoff Radius Dependence

The catalog size also depends on the cutoff radius of the radial filter. We have explored a range of cutoff radii from $700h^{-1}$ kpc to $3h^{-1}$ Mpc (with the core radius held constant at $100h^{-1}$ kpc). The significance of a detection, on average, gradually decreases as one increases the cutoff radius from $0.7h^{-1}$ Mpc to $3h^{-1}$ Mpc, dropping by $20 - 40\%$ over this range. The dependence of the detection significance on cutoff radius is less well defined, however, in the range $700h^{-1}$ kpc $\lesssim r_{co} \lesssim 1h^{-1}$ Mpc. A cutoff radius of $1h^{-1}$ Mpc appears to be about optimal and is indeed what we have chosen. The most robust catalog members are obviously the $\geq 4\sigma$ detections. Their inclusion in the final catalog is insensitive to the cutoff radius in the range explored here. The dependence of detection significance on the cutoff radius is understood. As one increases the cutoff radius, more weight is given to galaxies at larger radii and less to galaxies near the cluster center (a consequence of the radial filter normalization given in Eq. 20). If the cluster dimensions are significantly smaller than the cutoff radius then the matched filter signal will be reduced because of the increased contribution from interlopers at large radii. If the cluster dimensions are significantly larger than the cutoff radius, the signal may be truncated but this depends on the precise shape of the cluster profile (clusters with steeper profiles than the radial filter are not as significantly affected).

### 4.4.4. Cluster Deblending

About 5% of the cluster candidates have a neighboring cluster candidate that lies within a projected separation of $2h^{-1}$ Mpc and that has an estimated redshift that differs by $\leq 0.2$. While it is possible (although unlikely) for clusters to be so close, an alternative explanation is that substructures within a cluster were detected as separate systems. For reference, fewer than 3% of the R $\geq 0$ Abell clusters in the volume limited sample produced by Lauer & Postman (1994) have projected separations less than $2h^{-1}$ Mpc. In the notes to Tables 4 and 5 we attempt to identify possible related systems. We have not explicitly removed such systems because it is possible that they really are separate clusters (visual inspection was inconclusive). Follow-up spectroscopic observations will eventually resolve such cases. In any case, the effect is significantly smaller than the estimated spurious detection rate.

## 4.5. False Negative Rate

Real clusters that are not detected fall into two categories. First are those clusters or groups which fail to satisfy either the 95[th] percentile detection threshold or the peak signal threshold because they are either very poor systems or extremely distant systems. It is inevitable that we will miss such systems. Fortunately, we are able to precisely quantify this selection bias. Second are those clusters which fall below the detection thresholds because they exhibit very flat profiles. If such systems exist, they may be missed because their shape is not well matched to the shape of the radial filter. Poor clusters with very "cuspy" profiles (core radii $\lesssim 30h^{-1}$ kpc) might also be missed as they will be significantly smoothed, although no presently observed clusters have such narrow profiles. An accurate determination of the cluster selection function is essential, therefore, not only for understanding the completeness of our survey but for interpreting the observed cluster abundances in the context of competing cluster formation models (*e.g.*, Frenk *et al.* 1990).

### 4.5.1. The Selection Function

The probability that a cluster will be included in the sample depends on its redshift, richness, profile, the evolutionary history of its galaxies, and the observation passband. In order to derive the selection function which describes all these dependencies, we first need to establish the relationship between the matched filter signal and these quantities. To do this accurately, we simulated a total of 10560 clusters at different redshifts ($0.2 \leq z \leq 1.2$), richnesses (Abell R=1−4), projected radial profiles (from flat to $r^{-2}$), passbands ($V_4$ and $I_4$), and assumed two different k-corrections: one appropriate for a non-evolving elliptical galaxy and one appropriate for a non-evolving Sbc galaxy (see Figures 3a and 3b). The simulated clusters all have Schechter luminosity functions with faint end slopes that are identical to that used in the construction of the flux filter (*i.e.*, $\alpha = -1.1$). Each simulated cluster was then added into each real galaxy catalog and the cluster detection algorithm was run. This allows us to directly compute the matched filter signal and detection frequency as functions of redshift, richness, profile slope, k-correction, and passband.

Figure 19 shows the matched filter richness, $\Lambda_{cl}$, as a function of redshift, richness, k-correction, and passband for simulated clusters with $r^{-1.4}$ profiles (which is the average profile slope of the actual cluster candidates). The matched filter parameters used are identical to those listed in Table 3. This is an important calibration since the peak signal and $\Lambda_{cl}$ are not related to the cluster galaxy overdensity in a simple way. The observed $\Lambda_{cl}$ values of all the cluster candidates (including the supplemental clusters) are plotted for comparison, along with the typical detection threshold level. The richness of detected clusters tends to increase with redshift but as Figure 19 demonstrates the precise trend depends on the degree of spectral evolution of the cluster galaxies.

We also use our simulations to quantify the effect of mismatch between the cluster profile and the radial filter. The range of cluster profiles used in the simulations ($r^0$ to $r^{-2}$) bracket the range of profiles typically seen in Abell clusters. It was found that, as expected, the $r^{-2}$ clusters yield the strongest signal and the flat ($r^0$) clusters yield the weakest signal. The relative matched filter signal ratios for $r^0, r^{-1}$, and $r^{-2}$ clusters *of a given richness class* are $0.20 : 1 : 2.85$. These ratios are independent of richness class and redshift. The uncertainties in these ratios are about $\pm 10\%$. In other words, a richness class 1 cluster with a flat profile (if such a system exists) would yield a peak signal that is about 14 times smaller than that for a richness class 1 cluster with a $r^{-2}$ profile at the same redshift.

The cluster selection function is computed by simply counting the number of simulated clusters that are detected as a function of redshift, richness, profile, k-correction, and the observation passband. The selection functions for clusters with $r^{-1.4}$ profiles are shown in Figure 20. The trends are well understood. Obviously, the richer clusters are detected more frequently than poorer ones at high redshift. Likewise, clusters with the Sbc k-correction are more easily detected at high redshift than those with the E-galaxy k-correction. Distant clusters are also more easily detected in the $I_4$ band. The selection functions for clusters with flat or $r^{-2}$ profiles are qualitatively similar but, of course, differ quantitatively.

### 4.5.2. Merged Clusters

A standard problem with any two-dimensional or quasi two-dimensional cluster detection algorithm is separating 2 or more superposed clusters at different redshifts. Our simulations have shown that in only 10% of the cases where two clusters had a projected separation of $1.25h^{-1}$ Mpc or less, did the algorithm successfully detect both systems. This is largely a consequence of the $1.25h^{-1}$ Mpc detection box used to locate local maxima in the filtered galaxy maps. All clusters with projected separations greater than $1.25h^{-1}$ Mpc are resolved. By selecting a smaller detection box one will resolve finer scale structures but at the cost of introducing many more multiply detected clusters (see §4.4.4). Indeed, if we reduce the size of the detection box by 50% the number of multiply detected clusters increases almost tenfold. The frequency of occurance of superposed systems is a function of the spatial distribution of distant clusters. Since this is not yet known, it is difficult to estimate how many systems are missed or have richness overestimated because of this effect. Spectroscopic follow-up surveys of these systems are required to definitively answer this question.

## 5. DISCUSSION

One of the prime motivations for conducting this survey is to constrain the comoving density of clusters as a function of redshift. This will be accurately accomplished only when redshifts exist for a substantial number of these distant clusters. At the present time we have spectroscopic redshifts for only a handful and, thus, we present preliminary constraints based upon our estimated redshift data.

In Figure 21 we plot the cumulative number of clusters per square degree as a function of redshift and richness ($\Lambda_{cl}$ and $N_R(1.0h^{-1}$ Mpc)) for the PDCS. The $N(<z)$ for the GHO catalog, the Abell/ACO catalog, the APM cluster catalog (Dalton *et al.* 1992), and the Edinburgh-Durham-Milano (EDM) cluster redshift catalog (Nichol *et al.* 1992) are also shown. The Abell cluster curves are extrapolations normalized to the $z \leq 0.07$, R $\geq 0$ Abell cluster space density and assume this density remains constant in comoving coordinates. For reference, a richness class 1 cluster should have $35 \leq N_R < 56$. If the same cluster has $z \lesssim 0.7$ and a $r^{-1.4}$ profile, then it would have $25 \lesssim \Lambda_{cl} \lesssim 60$ in the $V_4$ band and $30 \lesssim \Lambda_{cl} \lesssim 65$ in the $I_4$ band (see Figure 19). No corrections have been made for false positive detections in Figure 21.

Figure 22 shows the observed and expected differential redshift histograms for the clusters in our main catalog with richness class $\geq 1$. Figure 23 shows the cumulative histograms. The observed histograms are based solely on our estimated redshifts and we have corrected the counts for false positive detections using the rates estimated from simulations (see §4.4.1). The expected distributions, for a given passband, are computed from the expression

$$N_{exp}(z, k(z)) = V(z) \sum_{i=0}^{4} (\rho_i(z) P_i(z, k(z))) \qquad (32)$$

where $N_{exp}(z, k(z))$ is the expected number of clusters per square degree per unit redshift for an assumed k-correction model specified by $k(z)$, $V(z)$ is the volume of a comoving element from $z$ to $z + \delta z$ subtending a 1 square degree solid angle, $\rho_i(z)$ is the space density of clusters in this volume with richness class $i$, and $P_i(z, k(z))$ is the probability that a cluster of richness class $i$ at redshift $z$ would be detected by our cluster selection procedure for a given passband. Since the selection probability is a function of projected cluster profile, we need to assume something about the profiles of clusters in order to make our prediction. The projected profile of a typical $z \leq 0.2$ cluster is at least as steep as $r^{-1}$ but flatter than $r^{-2}$ over the range $100 \lesssim$ r $\lesssim 1000 h^{-1}$ kpc (Beers & Tonry 1986; Oegerle *et al.* 1987; Postman, Geller & Huchra 1988; Broadhurst 1995; Lubin & Postman 1995; Smail *et al.* 1995; Squires *et al.* 1995; Tyson & Fischer 1995). For this computation, we adopt the selection function for a $r^{-1.4}$ profile — the mean profile shape for the clusters in this catalog. Lastly, we assume a constant comoving density of clusters and a constant cluster mass function (*i.e.*, the relative abundances of clusters of different richness classes remains independent of redshift). The actual density of the various richness classes is derived from the cluster luminosity function given by Bahcall (1979), normalized to match the observed space density of low redshift clusters. We compute the predicted cluster counts for two different galaxy population scenarios – one where clusters are made up of nonevolving elliptical galaxies and one where clusters are made up of nonevolving Sbc galaxies (see Figure 3). The different scenarios account for the two substantially different expectations as a function of redshift and passband seen the Figures 22 and 23, although in the $I_4$ band cosmological effects are comparable with the effect of varying the k-correction. The estimated redshift distribution of the Palomar survey clusters is consistent with the hypothesis that the typical distant cluster galaxy is bluer than a nonevolving elliptical galaxy at the cluster redshift.

The predicted cluster counts depend critically on which data are used to determine the low redshift normalization. If we normalize to the observed space density of R $\geq$ 1 Abell clusters, we find that the expected number of such clusters with $z \leq 0.3$ per square degree is about a factor of 5 ($\pm 2$) smaller than the number in our Palomar catalog even after applying a statistical correction for spurious detections. This "discrepancy" is clearly demonstrated in Figure 21 — PDCS clusters with $N_R \geq 35$ or with $\Lambda_{cl} \gtrsim 40$ should correspond to Abell richness class 1 systems. The PDCS certainly has better completeness for poor clusters than either the Abell or GHO catalogs although assessing the exact incompleteness of these catalog is difficult given the visual selection procedures used in their construction. The space density of clusters in the PDCS is consistent with that previously found by GHO; however, a direct comparison is not very meaningful since there is no richness information for the GHO clusters and the GHO cluster selection function cannot be easily quantified. Extrapolating the density of $z \lesssim 0.1$, R $\geq$ 1 Abell clusters to $z \geq 0.3$ results in a substantial underestimate of the cluster space density inferred from the present survey. The uncertainty in the density of $z \leq 0.3$ clusters from the PDCS is large but improves significantly by $z = 0.6$ where R $\geq$ 1 PDCS clusters still outnumber the predicted R $\geq$ 1 Abell/ACO cluster density by a factor of 5 (after false positive correction). Hence, the expected counts shown in Figures 22 and 23 are based on a comoving density of R $\geq$ 1 clusters that is 5 times larger than that derived from the Abell and ACO catalogs. Our assumed Abell cluster space densities are $1.13 \times 10^{-5} h^3$ Mpc$^{-3}$, $4.04 \times 10^{-6} h^3$ Mpc$^{-3}$, $1.28 \times 10^{-6} h^3$ Mpc$^{-3}$, and $2.74 \times 10^{-7} h^3$ Mpc$^{-3}$ for richness classes 0, 1, 2, and $\geq$ 3, respectively. These densities are derived from Bahcall (1979) but are in good agreement with subsequent Abell cluster redshift surveys (Huchra *et al.* 1990; Postman *et al.* 1992; Lauer & Postman 1994) and southern hemisphere redshift surveys of systems comparable to R $\geq$ 0 Abell clusters (Dalton *et al.* 1992; Nichol *et al.* 1992).

At $z \leq 0.6$, the slope of the PDCS $N(< z)$ vs $z$ relation is consistent with a constant comoving density, albeit a density 5 times higher than the R $\geq$ 1 Abell cluster density. The flattening of the slope of the $N(< z)$ vs $z$ relation at $z \gtrsim 0.6$ is primarily due to the PDCS selection function (as demonstrated by Figure 23). Figure 22 shows that a significant constraint on the spectral evolution of cluster galaxies is the modal redshift of the cluster sample — for clusters dominated by non-evolving elliptical galaxies, the predicted mode of the PDCS cluster redshift distribution is between $0.3 - 0.45$ whereas for clusters dominated by evolving elliptical galaxies (or non-evolving Sbc galaxies), the predicted mode may be as high as 0.75. The uncertainties in the redshift estimates, however, prevent us from constraining the cluster space density at $z \sim 1$. Follow-up spectroscopic surveys should allow a reasonable determination of the behavior of the space density of clusters at $z > 0.6$. Postman *et al.* (1995) have completed a contiguous 13 square degree $I$ band survey which is approximately 1.5 mag deeper than the PDCS in an additional effort to address this issue.

There are several possible explanations for the difference between the abundances of Abell and PDCS clusters. The Abell catalog could easily be missing half the clusters (presumably mostly R $\leq$ 1). The objectively derived APM and EDM cluster catalogs yield a somewhat higher space density of clusters than that derived from the Abell catalog when cut at comparable richnesses although it is not inconsistent with the Abell value. However, nearly 60% of the EDM clusters with $m_{10} \leq 18.5$ are not detected in the ACO catalog. This is true at all richness cuts (Lumsden *et al.* 1992). Henry *et al.* (1992) produced an x-ray selected cluster sample to study cluster evolution. At $z \leq 0.08$, they detect 7 clusters but only 3 of these are in the Abell catalog. The 4 x-ray selected clusters not in the Abell catalog have x-ray luminosities that are comparable with the ones that are. X-ray luminosity is highly correlated with richness and central density (Edge & Stewart 1991).

Previous observations have shown that clusters at $z \sim 0.4$ have galaxy populations which are significantly bluer than their low redshift counterparts (Butcher & Oemler 1984; Oemler 1992). At higher redshifts, Aragón-Salamanca *et al.* (1993) report that "...by $z \sim 0.9$ there are no cluster galaxies as red as present day ellipticals." Such evolution would enhance the detection rate of high redshift clusters at a given threshold (particularly in the $V_4$ band). Recent HST image data (Dressler et al. 1994; Couch et al. 1994; Ellis 1994) reveal that many of the blue ($g-r < 1.2$) galaxies in $z \lesssim 0.5$ clusters are either "normal" spirals or have peculiar morphologies, producing non-elliptical fractions which are 3 to 5 times higher than the average current epoch cluster. A number of physical mechanisms have clearly reprocessed the morphological mix between epochs. The dense regions of clusters do provide unique sites where processes like ram-pressure stripping and galaxy-galaxy collisions can, in principle, become efficient means of altering the properties of a significant fraction of the galaxy distribution. One explanation could be spiral destruction (Dressler 1980; Postman & Geller 1984; Whitmore, Gilmore, & Jones 1993), although observations at redshifts approaching unity suggest that the fraction of excessively blue galaxies may jump to as high as 80% (Rakos & Schombert 1995). Fading through cessation of star formation may also play a role, but studies at low redshift indicate that this process alone cannot dominate the evolution of blue and/or spiral galaxies. Gunn & Dressler (1988) find that the spectra of cluster galaxies with $z \gtrsim 0.6$ show, on average, smaller 4000 Å decrements and a higher frequency of post-starburst features (the "E+A" spectral class) than those at $z < 0.6$. It is likely that in the cluster environment a combination of starbursts, tidal destruction, fading, and stripping conspire to produce the observed time variation of photometric and spectroscopic properties.

Some optically selected high redshift "clusters" may be larger scale structures (*e.g.*, "Great Wall"-like features) viewed at orientations that make them mimic clusters. Contamination by large scale structures would increase both the number of detections and the blue galaxy component (sheets of galaxies at low $z$ are composed largely of nonelliptical galaxies). Only an "edge-on" sheet would potentially be detected as cluster candidate since the galaxy surface density in a face-on sheet is too low (de Lapparent *et al.* 1986, 1989). If the fraction of sheets with near "edge-on" orientation is independent of direction or redshift then the probability of viewing a sheet of galaxies along its narrow dimension is just proportional to the volume of space surveyed. Hence, deep optical cluster surveys may be more contaminated by such structures than low $z$ cluster surveys. Lefevre *et al.* (1994) spectroscopically verified the existence of a galaxy structure of intermediate scale ($\sim 2-3h^{-1}$ Mpc) at $z = 0.985$.

Our estimated spurious detection rate due to alignments of field galaxies is $\lesssim 30\%$ (§4.4.1), and we are confident that field galaxy superposition is not the primary source of the factor of 5 discrepancy in cluster counts. Couch *et al.* (1991) claim that the higher cluster density in their $J$ band survey could, in part, be explained by foreground contamination from spiral rich groups. However, 88% of our clusters are detected in both the $V_4$ and $I_4$ passbands which argues against severe contamination from such groups. A related effect, which is probably important for our survey, is an enhanced detection rate due to the superposition of poor groups and/or clusters. Systems with redshifts that differ by $\Delta z \lesssim 0.2$ and with projected separations of $\lesssim 1h^{-1}$ Mpc would be detected as a single cluster with richness significantly higher than any of the individual superposed components. The Abell catalog contains a number of such cases (*e.g.*, A419, A1318, A1631). Spectroscopic follow-up is the only unambiguous way to measure the magnitude of this problem. We note, however, that the matched filter algorithm is able to resolve overlapping systems if the projected separation is $\gtrsim 1.25h^{-1}$ Mpc.

Finally, we address the question of how our measuring errors might contribute to this discrepancy. The space density of clusters in the Abell catalog is roughly exponential with Abell's richness parameter. If $f$ is the fraction of clusters at (or above) some richness $R$, $d \ln f / dR \approx -1.1$; that is, the space density of clusters falls by a factor of $\sim 3$ for each successive richness class. The richness class is roughly logarithmic with total galaxy population (which is proportional to $\Lambda_{cl}$), i.e. $d \ln \Lambda_{cl} / dR \approx 0.45$. Therefore, the fraction of clusters above a given value of $\Lambda_{cl}$ goes as a power law function of $\Lambda_{cl}$.

$$f \propto \Lambda_{cl}^{-2.4} \tag{33}$$

Thus, if we have overestimated the richnesses of all our clusters by a factor of $\sim 2$, the discrepancy can be explained. We have examined this possibility through extensive simulations of clusters with a broad range of profile shapes and color evolution (Lubin 1995). Firstly, we examine the relation between the *actual* richness and $N_R$. For most of the very distant clusters, we severely underestimate their richness; however, for some of the nearby clusters ($z_{est} \leq 0.4$), we can misclassify R = 0 clusters as R = 1 clusters and vice versa due to uncertainties in background subtraction and redshift errors. In the worse case, these misclassifications could cause a factor of $\sim 2$ overestimate in the space density of clusters at $z \leq 0.4$. Secondly, we have examined the effect of the profile shape on the filter richness parameter $\Lambda_{cl}$. The magnitude of $\Lambda_{cl}$ is proportional to the steepness of the cluster profile. Complete resolution of the discrepancy would require Abell R $\geq 1$ clusters to have $\Lambda_{cl} \geq 85$ and $\Lambda_{cl} \geq 70$ in the $V_4$ and $I_4$ bands, respectively. These values would correspond to an average cluster surface density profile of $r^{-2}$ or steeper. This profile is not consistent with the PDCS clusters. The average profile shape is $r^{-1.4}$ (Lubin & Postman 1995). Lastly, we have examined the effect of the shape of the cluster luminosity function on the relation between $\Lambda_{cl}$ and the Abell richness estimator. The relationship depends on the faint end slope, $\alpha$, but not on the k-correction. For values in the range $-0.6 \geq \alpha \geq -1.3$, the relationship varies by only $\pm 30\%$. The trend is such that if the cluster LF slope is greater (less) than the nominal flux filter value, a R = 1 cluster would yield a lower (higher) $\Lambda_{cl}$ value.

A more subtle question is the effect of systematic or random redshift errors. One might naively think that the error could be explained by a *relatively* small systematic underestimate of redshift as the volume, and hence the number, increases dramatically with distance. In the Euclidean case, $d \ln V/d \ln z = 3.0$ (where $V$ is the comoving volume), so a factor of five in space density would require a systematic underestimate of the distance by a factor of 0.6. At a redshift of 0.5 in an $\Omega = 1$ Einstein-deSitter universe, the value of $d \ln V/d \ln z$ is smaller, about 2.2; to explain the discrepancy, we would have to underestimate the distances by slightly more than a factor of 2. This seems hardly possible; however, the situation may be more complex as our estimate of richness is also distance-dependent. If we underestimate the distance, we may underestimate the richness as we have assumed a brighter characteristic $m^*$. The richness measured relative to this $m^*$ would be smaller than that measured relative to the actual, fainter $m^*$. The result is that we would confuse very rich clusters at great distance, of which there are few, with poorer ones nearby. We can estimate the magnitude of the effect as follows. From the simulations, we find that in the vicinity of the correct redshift, the inferred richness changes with the assumed redshift like $d \ln \Lambda_{cl}/d \ln z_a \approx 2.1$. Now let us focus on some assumed redshift $z_a$ and richness $\Lambda_{cl}^a$ and suppose that the clusters that we think are at that redshift are really at some other (say larger) redshift $z_t$. Let us then ask how the total measured number of clusters at $z < z_a$ and $\Lambda_{cl} > \Lambda_{cl}^a$ changes as the real redshift $z_t$ changes. The number of clusters that we infer to be in our sample as we change $z_t$ is the product of the total number of clusters, which we assume is proportional to the comoving volume, and the fraction of clusters at $z_t$ which are sufficiently rich to have measured richness $\Lambda_{cl}^a$ at the inferred redshift $z_a$. Thus

$$
\begin{aligned}
\frac{d \ln N}{d \ln z} & \propto \frac{d \ln V}{d \ln z} + \frac{d \ln f}{d \ln z} \\
& \propto \frac{d \ln V}{d \ln z} + \frac{d \ln f}{d \ln \Lambda_{cl}} \frac{d \ln \Lambda_{cl}}{d \ln z} \\
& \approx 2.2 - (2.4)(2.1) \\
& \approx -2.8
\end{aligned}
\tag{34}
$$

This richness effect is opposite in sign to the volume effect and more than cancels the former; the numbers change with almost the inverse third power of the redshift. We need to have *overestimated* the redshift by a factor of about 1.8 to explain the discrepancy. This is extremely unlikely. The validity of this richness correction calculation depends on the origin of the redshift errors. As we have discussed in Sect. 4.3 and above, any systematic redshift error is most likely due to our assumption of no evolution, rather than the choice of an inappropriate luminosity function. First, there is no evidence of gross variations in the luminosity function among clusters; second, our simulations, which are all based on a single Schechter luminosity function, display errors similar to the real sample and no substantial redshift biases. Therefore, we see no evidence to suggest that there are large luminosity function variations which would invalidate the notion of "richness" as measured by either the matched filter ($\Lambda_{cl}$) or the Abell technique ($N_R$) and the calculation for the richness correction (Eq. 34).

Let us now consider the ("Malmquist") bias due to *random* errors. As long as the number–distance relation is concave up (which is the case for both the richness-uncorrected and the richness-corrected relation in Eq. 34), the effect of random errors in the distances is to cause one to overestimate the galaxy density in a cluster. If the distribution of errors in $\ln z$ is assumed to be Gaussian with standard deviation $\sigma$, the density overestimate is the factor

$$F = \frac{\Lambda_{cl}^{inferred}}{\Lambda_{cl}^{true}} = \exp(\frac{\gamma^2 \sigma^2}{2}) \tag{35}$$

Here $\gamma = d\ln N/d\ln z \approx -2.8$, so the factor is approximately $F = \exp(3.9\sigma^2)$, which is 5 for $\sigma \approx 0.64$ and 2 for $\sigma \approx 0.42$. At our mean cluster redshift of $\sim 0.6$, the former corresponds to a mean redshift error of $\sim 0.4$, the latter to an error of $\sim 0.25$. We believe the errors in our estimated redshifts to be $\sim 0.2$ (see section 4.2.1). Therefore, perhaps one-half of the discrepancy can be accounted for by the Malmquist effect, and it seems likely that evolution and incompleteness in the Abell sample can account for the rest. However, we must wait for a substantial number of real redshifts for this sample before it is possible to clarify this issue satisfactorily. We have already begun a number of related spectroscopic surveys. Such programs are time consuming and one goal of publishing this catalog is to provide the community with a starting point for further exploration. Over the next five years, results from these and other ground-based spectroscopic surveys plus imaging with HST and x-ray satellites should yield a substantial improvement in our understanding of the formation and evolution of galaxy clusters.

These observations were made possible by the efforts of R. Lucinio, B. Zimmerman, M. Carr, E. Danielson, E. Lorenz, V. Nenow, J.D. Smith, and J. Westphal, who helped design, build, and test the 4-Shooter, and J. Carrasco, D. Tennant, M. Doyle, and J. Henning, members of the Palomar staff. We thank Jill Knapp for important discussions regarding star/galaxy classification, Robert Lupton for his excellent advice in statistical matters concerning the matched filter, and Tony Tyson for providing information about the response function of the $B_J$ filter. We thank Bob Nichol, Michael Strauss, and the referee for their thoughtful and constructive comments on this manuscript. Robert Deverill and David Saxe created some of the software used for the data processing and analysis. This work was supported in part by NASA contracts NAGW-2166 (MP), NAG5-1618 (DPS), NAS7-918 (JGH), NGT-51295 (LML) and by NSF grant AST86-18257A02 (JEG).

## Figure Captions

Fig.1 — A schematic diagram of the configuration of the 4-Shooter focal plane during the gathering of the survey data. The four CCDs are denoted as 1, 2, 3, and 4. The arrow on the left is the CCD readout direction (along the columns) and is the motion of the sky across the detectors.

Fig.2 — The response curves of the 4-Shooter $F555W$ ($V_4$) and $F785LP$ ($I_4$) filters used in the survey. The response includes the telescope, CCD, 4-Shooter camera, and filter quantum efficiencies. The upper curve for each filter is the response with no atmosphere; the lower curve is the response at an airmass of 1.2 at Palomar.

Fig.3 — (a) The k-correction as a function of redshift and morphology for the $V_4$ filter. (b) The k-correction as a function of redshift and morphology for the $I_4$ filter. (c) The $V_4-I_4$ color as a function of redshift and morphology.

Fig.4 — The photometric offset as a function of time for four pairs of adjacent scans. The upper two panels show offsets for scans taken on the same night; the bottom two panels show offsets for scans taken more than a month apart. The time is derived from the object declination, which is linearly related to the time elapsed since the start of the scan.

Fig.5 — (a) The differential galaxy surface density as a function of $V_4$ isophotal magnitude for the 5 survey fields. (b) The same for the $I_4$ band.

Fig.6 — Residuals between the mean galaxy surface density for all five fields and the galaxy surface density in each individual field as functions of $V_4$ and $I_4$ isophotal magnitude. Solid line represents mean residual level, dashed line gives the prediction based on the observed angular two point correlation function (Neushaefer *et al.* 1991).

Fig.7 — (a) The galaxy-to-star count ratio as a function of $V_4$ isophotal magnitude for all objects in the 13h field (solid dots) and for those objects which lie within $200 \leq$ CCD Column $\leq 600$. The results for the other survey fields are similar. (b) The same for the $I_4$ images.

Fig.8 — (a) $V_4$ band star counts as a function of CCD column number. The $I_4$ band results are similar. (b) $V_4$ band galaxy counts as a function of CCD column number. The $I_4$ band results are similar.

Fig.9 — (a) The $V_4$ flux filter as a function of redshift shown in intervals of 0.1 over the $z$ range [0.2,1.2]. The ordinate is the amplitude of the filter. Numbers indicate the filter redshift. No color evolution is assumed. (b) The $I_4$ flux filter.

Fig.10 — The radial filter as a function of redshift. The ordinate is the amplitude of the filter. A core radius of $100h^{-1}$ kpc and cutoff radius of $1h^{-1}$ Mpc are used. The 5 radial filters shown, from broadest to narrowest, are for $z = 0.2$, 0.3, 0.4, 0.6, and 1.0.

Fig.11 — The histogram of $S(i,j)$ values generated using a matched filter tuned to detect clusters at $z = 0.6$ in the $0^{\mathrm{h}}$ field.

Fig.12 — The cluster signal, $S_{cl}$, as a function of characteristic magnitude $m^*$ and passband. The corresponding filter redshift is shown along the upper abscissa. Points for a cluster at a given redshift are connect by solid lines. The solid point for each cluster denotes the redshift estimate that the matched filter algorithm would assign to the cluster. The estimates shown here are derived prior to the application of the cluster signal correction. The vertical dashed lines indicate the survey magnitude limits.

Fig.13 — The corrected cluster signal, $S_{cl}^{cor}$, as a function of characteristic magnitude $m^*$ and passband. The corresponding filter redshift is shown along the upper abscissa. Points for a cluster at a given redshift are connect by solid lines. The solid point for each cluster denotes the redshift estimate that the matched filter algorithm would assign to the cluster. Note the improved redshift estimate accuracy provided by the application of the $CSC$.

Fig.14 — The offset between estimated redshift and true redshift, $z_{est} - z_{true}$, as a function of true redshift for $\sim 400$ simulated clusters. Solid points represent clusters which are detected at the $\geq 4\sigma$ level. Crosses represent clusters detected below the $4\sigma$ level.

Fig.15 — (a) The offset between estimated redshift and true redshift, $z_{est} - z_{true}$, as a function of the corrected $V_4$ band matched filter signal, $S_{cl}^{cor} = (S(i,j) - S_{bg})/CSC$, for 6 real clusters. (b) The same for the $I_4$ band.

Fig.16 — The match-filtered $I_4$ band galaxy catalog as a function of filter redshift for the region centered on the GHO cluster 1322+3028 ($z_{obs} = 0.70$; cluster 063 in Table 4). The filter redshift is listed at the bottom of each $14.5' \times 14.5'$ subimage; north is at the top and west to the right. The GHO cluster 1322+3027 ($z_{obs} = 0.75$; cluster 059 in Table 4) is the peak to the east of the central cluster.

Fig.17 — The Abell richness $N_R(\leq 1.0h^{-1}$ Mpc) as a function of $\Lambda_{cl}$ for the PDCS clusters and for simulated clusters in the $V_4$ and $I_4$ passbands. The solid dots, open squares, and stars indicate cluster redshift estimates in the range $0.2 \leq z_{est} \leq 0.4$, $0.5 \leq z_{est} \leq 0.7$, and $z_{est} \geq 0.8$, respectively.

Fig.18 — Left: the mean spurious detection surface density in 10 simulated galaxy catalogs as a function of size and peak significance. Right: the mean surface density of observed cluster candidates as a function of size and peak significance.

Fig.19 — The matched filter richness, $\Lambda_{cl}$, as a function of redshift, richness, k-correction, and passband for the 3520 simulated clusters with $r^{-1.4}$ profiles. The dashed lines show the relationship for clusters with richness class 1–4 and with elliptical galaxy k-corrections. The solid lines show the relationship for clusters with Sbc k-corrections. The points are the actual detected clusters (open circles are supplemental clusters). The shaded area represents region below the typical $3\sigma$ detection threshold. Some matched cluster candidates lie within the shaded area in one passband but the significance of the detection in the other passband is above the $3\sigma$ limit.

Fig.20 — The selection function for clusters with $r^{-1.4}$ profiles. The dashed lines show the function for clusters with richness class 1–4 and with elliptical galaxy k-corrections. The solid lines show the function for clusters with Sbc k-corrections. If 4 separate solid lines are not seen it is because the selection probabilities for richer clusters are identical.

Fig.21 — The cumulative number of clusters per square degree as a function of redshift and richness for the PDCS, the GHO catalog, the Abell/ACO catalog, the APM cluster catalog, and the Edinburgh-Durham-Milano (EDM) cluster redshift catalog. The upper plot shows PDCS cluster counts as a function of the Abell richness estimator, $N_R$. The lower plot shows PDCS cluster counts as a function of the $\Lambda_{cl}$ richness estimator. For clarity, the APM and EDM values are shown on separate plots.

Fig.22 — The differential redshift distributions for clusters detected in the $V_4$ and $I_4$ bands. The curves show the expected distributions for clusters consisting of non-evolving elliptical (solid line) and Sbc (dashed line) galaxies for $\Omega_o = 0.2$ and $\Omega_o = 1$. The $\Omega_o = 1$ curves always predict fewer clusters at a given redshift. See text for further details.

Fig.23 — The cumulative redshift distributions for clusters detected in the $V_4$ and $I_4$ bands. The curves show the expected distributions for clusters consisting of non-evolving elliptical (solid line) and Sbc (dashed line) galaxies for $\Omega_o = 0.2$ and $\Omega_o = 1$. The $\Omega_o = 1$ curves always predict fewer clusters at a given redshift. See text for further details.

## Plate Captions

Plate 1 — Finding charts for PDCS clusters $001 - 020$. The images are of the central $134'' \times 134''$ regions; north is up and west is to the right. The cluster ID is given at the top of each picture. The clusters are shown in the passband in which the significance of the peak signal is highest. Occasionally there are narrow gaps in the images – these are just the unusable areas between 4-Shooter CCDs.

Plate 2 — Finding charts for PDCS clusters $021 - 040$.

Plate 3 — Finding charts for PDCS clusters $041 - 060$.

Plate 4 — Finding charts for PDCS clusters $061 - 079$ and supplemental PDCS cluster 001 (supplemental clusters have an S after the ID number).

Plate 5 — Finding charts for supplemental PDCS clusters $002 - 021$.

Plate 6 — Finding charts for supplemental PDCS clusters $022 - 028$.

TABLE 1. PDCS Fields

| Center Coordinates | | | $l$ | $b$ | $A_{V_4}$ | $A_{I_4}$ | Eff. Area (deg$^2$) | |
|---|---|---|---|---|---|---|---|---|
| R.A. | (J2000) | Dec. | | | | | $V_4$ | $I_4$ |
| 00$^h$ 29$^m$ 11$^s$ | +05° 31′ 55″ | | 112.76 | −56.9 | 0.055 | 0.027 | 1.074 | 0.970 |
| 02$^h$ 28$^m$ 33$^s$ | +00° 55′ 45″ | | 166.64 | −53.5 | 0.028 | 0.014 | 1.120 | 0.927 |
| 09$^h$ 52$^m$ 57$^s$ | +47° 33′ 20″ | | 170.21 | +49.8 | 0.004 | 0.002 | 1.052 | 0.995 |
| 13$^h$ 26$^m$ 01$^s$ | +29° 52′ 55″ | | 54.89 | +81.9 | 0.031 | 0.015 | 1.062 | 1.014 |
| 16$^h$ 06$^m$ 17$^s$ | +41° 32′ 40″ | | 65.92 | +47.9 | 0.019 | 0.009 | 1.088 | 1.026 |

TABLE 2. PDCS Photometric Standards

| Standard | $V_4$ | $I_4$ |
|---|---|---|
| HD 19445 | 8.04 | 7.77 |
| Ross 34 | 11.10 | 9.70 |
| Feige 67 | 11.71 | 12.70 |
| Ross 484 | 10.81 | 9.93 |
| BD +26 2606 | 9.71 | 9.46 |
| BD +28 4211 | 10.40 | 11.39 |
| BD +17 4708 | 9.45 | 9.20 |

TABLE 3. Matched Filter and Cluster Detection Parameters

| Parameter | Value |
|---|---|
| $M^*_{V_4}$ ($h = 0.75$, $q_o = 0.5$) | $-21.00$ |
| $M^*_{I_4}$ ($h = 0.75$, $q_o = 0.5$) | $-21.90$ |
| $\alpha$ (LF slope) | $-1.10$ |
| $r_c$ (Core Radius) | $100h^{-1}$ kpc |
| $r_{co}$ (Cutoff Radius) | $1.0h^{-1}$ Mpc |
| Detection Threshold | $95^{\text{th}}$ pctl |
| Peak Signal ($S^{peak}$) (dual band detection) | $\max(S^{peak}_{V_4}, S^{peak}_{I_4}) \geq 3\sigma_{bg}$ |
| Peak Signal (single band detection) | $\geq 4\sigma_{bg}$ |
| $V_4$ magnitude limit | 23.8 |
| $I_4$ magnitude limit | 22.5 |

TABLE 4. The Palomar Distant Cluster Catalog

| ID | R.A. (J2000) | Dec. (J2000) | $V_4$ Rad | $z_{est}$ | Peak | $\sigma$ | $\Lambda_{cl}$ | $N_o$ | $N_R$ | $I_4$ Rad | $z_{est}$ | Peak | $\sigma$ | $\Lambda_{cl}$ | $N_o$ | $N_R$ |
|---|---|---|---|---|---|---|---|---|---|---|---|---|---|---|---|---|
| 001 | 00 29 03.0 | +05 01 24 | 69 | 0.60 | 0.35 | 2.98 | 70.6 | 19± 3 | 50± 7 | 114 | 0.60 | 0.73 | 3.07 | 61.5 | 19± 5 | 39±18 |
| 002 | 00 29 28.4 | +05 04 03 | 120 | 0.40 | 0.55 | 3.12 | 44.4 | 20± 3 | 60± 8 | 138 | 0.40 | 1.23 | 3.98 | 52.1 | 25± 5 | 94±20 |
| 003 | 00 28 36.6 | +05 07 44 | 151 | 0.60 | 0.44 | 3.73 | 88.5 | 27± 4 | 138± 9 | 33 | 0.60 | 0.61 | 2.58 | 51.7 | 10± 3 | 20±13 |
| 004 | 00 29 11.1 | +05 08 55 | 164 | 0.50 | 0.66 | 4.67 | 87.1 | 24± 4 | 91± 8 | 199 | 0.60 | 1.12 | 4.74 | 94.9 | 32± 6 | 121±21 |
| 005 | 00 27 38.6 | +05 09 46 | 88 | 0.20 | 1.23 | 4.05 | 31.8 | 13± 3 | 8± 6 | 77 | 0.20 | 1.52 | 3.95 | 27.7 | 11± 3 | 16±14 |
| 006 | 00 29 52.2 | +05 12 36 | 50 | 0.40 | 0.54 | 3.07 | 43.5 | 22± 4 | 44± 8 | 109 | 0.50 | 1.04 | 3.71 | 62.9 | 28± 6 | 55±19 |
| 007 | 00 30 56.2 | +05 14 57 | 209 | 0.20 | 1.37 | 4.50 | 35.3 | 12± 3 | 40± 6 | 162 | 0.30 | 1.30 | 4.01 | 36.7 | 15± 3 | 55±14 |
| 008 | 00 28 31.4 | +05 18 01 | 68 | 0.60 | 0.38 | 3.18 | 75.3 | 27± 4 | 53± 9 | 30 | 0.50 | 0.70 | 2.50 | 42.4 | 22± 5 | 44±21 |
| 009 | 00 30 46.3 | +05 37 24 | 68 | 0.30 | 0.67 | 3.03 | 30.9 | 12± 3 | 32± 7 | 101 | 0.40 | 1.37 | 4.44 | 58.1 | 18± 4 | 25±13 |
| 010 | 00 28 52.1 | +05 49 12 | 187 | 0.30 | 1.27 | 5.78 | 58.8 | 24± 4 | 73± 8 | 176 | 0.40 | 1.15 | 3.72 | 48.7 | 14± 4 | 38±16 |
| 011 | 00 29 44.9 | +05 51 43 | 267 | 0.40 | 1.29 | 7.29 | 104.5 | 28± 4 | 96± 8 | 258 | 0.30 | 1.91 | 5.90 | 54.1 | 27± 5 | 94±19 |
| 012 | 00 27 59.8 | +05 56 27 | 247 | 0.30 | 1.63 | 7.42 | 75.5 | 26± 4 | 65± 7 | 223 | 0.40 | 1.83 | 5.92 | 77.5 | 12± 4 | 73±16 |
| 013 | 00 31 20.0 | +05 33 29 | 145 | 0.20 | 1.27 | 4.17 | 32.8 | 8± 2 | 18± 5 | –– | 0.20 | 1.60 | 4.17 | 29.3 | 11± 3 | 18±13 |
| 014 | 02 27 24.7 | +00 25 49 | 20 | 0.30 | 0.66 | 2.49 | 25.1 | 18± 3 | 40± 7 | 103 | 0.40 | 1.16 | 3.39 | 46.4 | 27± 6 | 71±17 |
| 015 | 02 30 21.0 | +00 26 21 | 41 | 0.60 | 0.35 | 2.41 | 62.5 | 17± 4 | 22± 7 | 98 | 1.10 | 0.27 | 3.13 | 78.5 | 18± 5 | 78±20 |
| 016 | 02 28 26.5 | +00 32 20 | 119 | 0.50 | 0.75 | 4.26 | 87.8 | 32± 4 | 90± 9 | 178 | 0.50 | 1.94 | 6.48 | 108.6 | 35± 6 | 110±16 |
| 017 | 02 30 22.8 | +00 36 40 | 110 | 0.70 | 0.44 | 3.67 | 115.0 | 8± 2 | 49± 7 | 15 | 0.70 | 0.45 | 2.17 | 46.7 | 28± 6 | 120±23 |
| 018 | 02 27 25.5 | +00 40 04 | 98 | 0.40 | 0.77 | 3.82 | 54.0 | 20± 4 | 82± 8 | 158 | 0.40 | 1.31 | 3.80 | 52.1 | 23± 5 | 118±20 |
| 019 | 02 27 19.5 | +00 46 30 | 95 | 0.70 | 0.32 | 2.71 | 85.0 | 23± 4 | 93± 8 | 78 | 0.60 | 0.86 | 3.43 | 66.3 | 19± 4 | 86±17 |
| 020 | 02 28 32.7 | +00 57 26 | 69 | 0.40 | 0.72 | 3.55 | 50.2 | 27± 4 | 56± 8 | 38 | 0.50 | 0.90 | 3.01 | 50.4 | 22± 4 | 60±13 |
| 021 | 02 26 43.6 | +00 58 12 | 204 | 0.30 | 1.18 | 4.49 | 45.2 | 17± 3 | 66± 7 | 185 | 0.40 | 1.42 | 4.13 | 56.6 | 19± 4 | 71±13 |
| 022 | 02 26 51.5 | +01 05 42 | 304 | 0.40 | 0.93 | 4.58 | 64.7 | 18± 3 | 115± 8 | 205 | 0.60 | 1.08 | 4.30 | 82.9 | 26± 5 | 109±18 |
| 023 | 02 27 56.1 | +01 05 14 | 299 | 0.20 | 2.14 | 7.49 | 46.6 | 12± 3 | 19± 5 | 355 | 0.20 | 2.42 | 7.05 | 44.6 | 13± 3 | 38± 9 |
| 024 | 02 30 27.4 | +01 09 04 | 165 | 0.40 | 0.96 | 4.72 | 66.8 | 19± 3 | 46± 6 | 118 | 0.40 | 1.34 | 3.90 | 53.4 | 21± 4 | 66±13 |
| 025 | 02 29 42.5 | +01 08 50 | 129 | 0.30 | 1.30 | 4.92 | 49.5 | 19± 3 | 59± 7 | 62 | 0.30 | 1.27 | 3.48 | 35.2 | 14± 3 | 59±13 |
| 026 | 02 29 47.8 | +00 39 12 | 180 | 0.20 | 1.23 | 4.32 | 25.2 | 6± 2 | 16± 6 | –– | 0.20 | 1.03 | 3.01 | 19.0 | 8± 2 | 15±10 |
| 027 | 02 29 37.9 | +01 21 55 | 45 | 1.10 | 0.07 | 4.08 | 101.1 | 31± 8 | 11±10 | –– | 1.10 | 0.03 | 0.38 | 10.6 | 12± 4 | 25±18 |
| 028 | 02 28 48.0 | +01 23 24 | 255 | 0.20 | 1.57 | 5.50 | 32.0 | 18± 3 | 55± 6 | –– | –– | –– | –– | –– | –– | –– |
| 029 | 09 53 12.1 | +47 08 58 | 41 | 0.40 | 0.51 | 2.47 | 32.5 | 9± 2 | 22± 6 | 68 | 0.40 | 1.04 | 3.32 | 36.9 | 20± 5 | 52±17 |
| 030 | 09 54 46.3 | +47 10 48 | 116 | 0.30 | 1.40 | 5.07 | 46.5 | 9± 3 | 28± 6 | 105 | 0.30 | 1.46 | 3.93 | 33.7 | 10± 3 | 20±14 |
| 031 | 09 53 39.5 | +47 12 58 | 41 | 1.10 | 0.08 | 3.71 | 120.0 | 19± 8 | 55±14 | 57 | 1.00 | 0.21 | 2.55 | 54.6 | 19± 5 | 14±19 |
| 032 | 09 52 29.0 | +47 17 49 | 122 | 0.30 | 0.97 | 3.51 | 32.2 | 10± 3 | 23± 7 | 32 | 0.40 | 0.89 | 2.85 | 31.7 | 9± 3 | 19±13 |
| 033 | 09 52 13.1 | +47 16 48 | 37 | 0.50 | 0.38 | 2.43 | 41.3 | 22± 4 | 31± 8 | 64 | 0.50 | 0.81 | 3.16 | 42.5 | 23± 5 | 46±18 |
| 034 | 09 55 09.1 | +47 29 55 | 282 | 0.30 | 1.87 | 6.76 | 62.0 | 22± 3 | 74± 7 | 211 | 0.40 | 1.79 | 5.74 | 63.8 | 31± 6 | 67±19 |
| 035 | 09 52 31.2 | +47 36 27 | 112 | 0.60 | 0.40 | 3.32 | 67.5 | 12± 3 | 37± 7 | 53 | 0.60 | 0.56 | 2.72 | 42.2 | 26± 6 | 44±21 |
| 036 | 09 53 53.7 | +47 40 15 | 255 | 0.30 | 1.65 | 5.97 | 54.7 | 15± 3 | 56± 6 | 271 | 0.30 | 2.22 | 5.97 | 51.1 | 14± 3 | 62±12 |
| 037 | 09 51 41.5 | +47 41 30 | 66 | 0.60 | 0.31 | 2.63 | 53.5 | 20± 4 | 38± 8 | 83 | 0.60 | 0.80 | 3.88 | 60.0 | 26± 6 | 26±22 |
| 038 | 09 51 09.9 | +47 43 54 | 79 | 0.30 | 1.44 | 5.20 | 47.7 | 13± 3 | 28± 7 | 68 | 0.30 | 1.68 | 4.53 | 38.8 | 15± 4 | 45±16 |
| 039 | 09 51 25.2 | +47 49 50 | 91 | 0.60 | 0.37 | 3.07 | 62.3 | 25± 4 | 32± 8 | 89 | 0.60 | 0.84 | 4.09 | 63.3 | 25± 5 | 46±17 |
| 040 | 09 53 25.6 | +47 58 55 | 226 | 0.20 | 1.68 | 5.03 | 28.3 | 14± 3 | 39± 6 | 252 | 0.20 | 2.51 | 6.23 | 35.3 | 15± 3 | 56±12 |
| 041 | 09 54 16.9 | +47 58 41 | 22 | 0.70 | 0.17 | 1.94 | 43.5 | 20± 7 | 46±14 | 80 | 0.60 | 0.80 | 3.85 | 59.6 | 19± 4 | 40±14 |
| 042 | 09 53 54.3 | +48 00 04 | 31 | 0.90 | 0.11 | 2.32 | 61.0 | 10± 3 | 31± 7 | 146 | 0.60 | 1.03 | 5.00 | 77.4 | 25± 5 | 69±20 |
| 043 | 09 52 15.1 | +47 57 44 | 200 | 0.20 | 1.75 | 5.23 | 29.5 | 6± 2 | 13± 4 | –– | 0.20 | 1.41 | 4.41 | 25.0 | 7± 2 | 6±10 |
| 044 | 09 52 18.6 | +48 02 32 | 55 | 1.10 | 0.12 | 5.35 | 173.1 | 21± 9 | 2±12 | –– | 1.10 | 0.00 | -0.06 | -1.4 | 8± 5 | 4±19 |
| 045 | 09 54 38.8 | +47 15 59 | –– | 0.40 | 0.58 | 2.78 | 36.6 | 18± 4 | 80± 8 | 129 | 0.40 | 1.27 | 4.06 | 45.2 | 18± 5 | 88±19 |

TABLE 4. The Palomar Distant Cluster Catalog

| | R.A. | Dec. | $V_4$ | | | | | | | $I_4$ | | | | | | |
|---|---|---|---|---|---|---|---|---|---|---|---|---|---|---|---|---|
| ID | (J2000) | (J2000) | Rad | $z_{est}$ | Peak | $\sigma$ | $\Lambda_{cl}$ | $N_o$ | $N_R$ | Rad | $z_{est}$ | Peak | $\sigma$ | $\Lambda_{cl}$ | $N_o$ | $N_R$ |
| 046 | 13 27 20.5 | +29 22 15 | 30 | 0.30 | 0.80 | 2.48 | 28.8 | 16±3 | 12±6 | 39 | 0.30 | 1.07 | 3.26 | 29.9 | 10±3 | 22±11 |
| 047 | 13 23 45.4 | +29 23 05 | 147 | 0.30 | 1.24 | 3.84 | 44.6 | 14±3 | 47±6 | 93 | 0.30 | 1.38 | 4.22 | 38.5 | 12±4 | 32±12 |
| 048 | 13 25 03.4 | +29 24 23 | 60 | 0.80 | 0.28 | 2.96 | 104.7 | 20±4 | 44±8 | 27 | 0.30 | 1.02 | 3.11 | 28.4 | 11±4 | 8±13 |
| 049 | 13 26 15.7 | +29 25 11 | 353 | 0.20 | 2.88 | 8.10 | 53.0 | 15±3 | 38±5 | 359 | 0.20 | 2.84 | 8.72 | 51.2 | 14±3 | 41±9 |
| 050 | 13 26 24.1 | +29 32 59 | 211 | 0.40 | 1.20 | 5.11 | 80.8 | 25±4 | 68±7 | 199 | 0.50 | 1.65 | 6.69 | 96.8 | 40±6 | 109±19 |
| 051 | 13 23 44.8 | +29 34 13 | 197 | 0.40 | 1.14 | 4.83 | 76.5 | 23±3 | 80±7 | 151 | 0.40 | 1.21 | 4.14 | 50.0 | 27±5 | 61±16 |
| 052 | 13 25 32.9 | +29 38 25 | 193 | 0.40 | 1.57 | 6.67 | 105.5 | 37±4 | 87±8 | 172 | 0.50 | 1.95 | 7.90 | 114.3 | 33±6 | 67±17 |
| 053 | 13 25 01.1 | +29 43 06 | 58 | 0.60 | 0.46 | 3.07 | 79.5 | 9±2 | 22±5 | 89 | 0.20 | 1.08 | 3.31 | 19.4 | 15±4 | 27±16 |
| 054 | 13 25 34.7 | +29 46 14 | 67 | 0.50 | 0.62 | 3.17 | 70.5 | 21±3 | 27±7 | 53 | 0.50 | 0.74 | 2.99 | 43.2 | 15±4 | 20±14 |
| 055 | 13 26 25.7 | +29 58 20 | 210 | 0.30 | 2.08 | 6.45 | 74.9 | 19±3 | 57±7 | 214 | 0.30 | 2.59 | 7.90 | 72.2 | 21±5 | 62±15 |
| 056 | 13 27 08.3 | +30 00 18 | 79 | 0.50 | 0.63 | 3.21 | 71.4 | 20±4 | 42±8 | 107 | 0.50 | 1.05 | 4.24 | 61.4 | 20±5 | 54±18 |
| 057 | 13 23 47.3 | +30 03 31 | 79 | 0.50 | 0.65 | 3.32 | 73.7 | 14±3 | 19±6 | 108 | 0.80 | 0.50 | 3.84 | 72.7 | 23±5 | 65±16 |
| 058 | 13 27 39.7 | +30 06 12 | 61 | 0.30 | 0.90 | 2.78 | 32.3 | 18±3 | 58±7 | 108 | 0.30 | 1.28 | 3.90 | 35.7 | 8±3 | 48±15 |
| 059 | 13 24 48.8 | +30 11 36 | 72 | 0.80 | 0.31 | 3.26 | 115.3 | 37±9 | 88±15 | 49 | 0.60 | 0.63 | 3.15 | 51.3 | 21±5 | 45±20 |
| 060 | 13 23 39.0 | +30 12 12 | 42 | 0.30 | 0.95 | 2.94 | 34.1 | 16±3 | 36±6 | 48 | 0.30 | 1.11 | 3.39 | 31.0 | 13±4 | 35±14 |
| 061 | 13 27 07.4 | +30 18 01 | 45 | 0.30 | 0.92 | 2.85 | 33.2 | 9±3 | 26±7 | 132 | 0.30 | 1.71 | 5.21 | 47.7 | 15±3 | 29±10 |
| 062 | 13 23 39.0 | +30 22 26 | 111 | 0.40 | 1.43 | 6.08 | 96.2 | 26±3 | 44±6 | 119 | 0.40 | 1.94 | 6.64 | 80.3 | 31±5 | 59±12 |
| 063 | 13 24 20.6 | +30 12 52 | 168 | 0.60 | 0.60 | 4.01 | 103.7 | 23±4 | 59±8 | 44 | 0.50 | 0.75 | 3.04 | 44.0 | 20±5 | 64±19 |
| 064 | 13 26 22.3 | +30 15 20 | –– | 1.00 | 0.09 | 1.45 | 81.6 | 9±3 | −11±7 | 105 | 1.00 | 0.39 | 4.35 | 103.8 | 40±10 | 153±33 |
| 065 | 16 03 51.8 | +41 01 53 | 89 | 0.40 | 0.83 | 4.32 | 61.0 | 24±4 | 39±8 | 51 | 0.40 | 0.98 | 3.08 | 33.8 | 19±5 | 58±16 |
| 066 | 16 04 48.6 | +41 05 06 | 22 | 0.90 | 0.17 | 2.05 | 87.7 | 11±3 | 10±6 | 52 | 0.40 | 1.00 | 3.15 | 34.5 | 13±3 | 22±13 |
| 067 | 16 03 49.4 | +41 11 13 | 54 | 0.50 | 0.50 | 3.06 | 59.9 | 26±4 | 39±8 | 89 | 0.50 | 0.94 | 3.44 | 45.7 | 21±5 | 78±20 |
| 068 | 16 06 03.3 | +41 15 37 | 111 | 0.50 | 0.49 | 3.03 | 59.4 | 23±4 | 62±9 | 86 | 0.60 | 0.67 | 3.02 | 49.8 | 14±5 | 45±20 |
| 069 | 16 07 39.2 | +41 19 17 | 171 | 0.30 | 1.08 | 4.93 | 43.5 | 11±2 | 26±6 | 64 | 0.70 | 0.46 | 2.59 | 43.9 | 14±5 | 64±22 |
| 070 | 16 09 06.7 | +41 21 32 | 80 | 0.20 | 0.83 | 3.35 | 17.7 | 14±3 | −10±9 | 65 | 0.20 | 1.30 | 3.04 | 16.7 | 17±6 | 28±19 |
| 071 | 16 04 07.2 | +41 27 18 | 81 | 0.70 | 0.36 | 3.12 | 95.5 | 21±4 | 58±9 | 97 | 0.50 | 0.80 | 2.92 | 38.4 | 12±4 | 57±16 |
| 072 | 16 07 16.7 | +41 27 01 | 99 | 0.60 | 0.46 | 3.33 | 85.1 | 34±5 | 61±10 | 36 | 0.40 | 0.77 | 2.42 | 26.5 | 14±4 | 30±16 |
| 073 | 16 06 19.6 | +41 35 54 | 48 | 0.30 | 0.75 | 3.43 | 30.3 | 14±3 | 19±7 | 117 | 0.30 | 1.29 | 3.53 | 28.2 | 28±6 | 77±20 |
| 074 | 16 04 45.6 | +41 38 50 | 150 | 0.30 | 0.96 | 4.37 | 38.6 | 18±3 | 43±6 | 185 | 0.30 | 1.49 | 4.10 | 32.7 | 21±5 | 107±19 |
| 075 | 16 06 53.3 | +41 39 09 | 88 | 0.90 | 0.27 | 3.26 | 147.7 | 12±3 | 42±8 | 49 | 1.10 | 2.02 | 2.67 | 71.0 | 16±4 | 46±18 |
| 076 | 16 04 10.8 | +41 49 60 | 157 | 0.30 | 1.34 | 6.12 | 54.0 | 15±3 | 34±6 | 185 | 0.30 | 1.99 | 5.46 | 43.7 | 14±3 | 66±14 |
| 077 | 16 08 34.9 | +41 52 43 | 336 | 0.20 | 1.83 | 7.38 | 39.0 | 11±3 | 35±5 | 266 | 0.20 | 2.65 | 6.21 | 34.3 | 15±3 | 54±11 |
| 078 | 16 08 37.6 | +41 20 08 | 145 | 1.00 | 0.32 | 4.50 | 304.9 | >50±10 | >80±20 | –– | 1.00 | 0.00 | 0.05 | 1.0 | 12±5 | 9±23 |
| 079 | 16 05 41.4 | +41 57 33 | –– | 0.40 | 0.33 | 1.72 | 24.2 | 12±4 | −6±8 | 96 | 0.40 | 1.30 | 4.08 | 44.7 | 16±4 | 37±15 |

TABLE 5. The Supplemental Palomar Distant Cluster Catalog

| ID | R.A. (J2000) | Dec. (J2000) | $V_4$ Rad | $z_{est}$ | Peak | $\sigma$ | $\Lambda_{cl}$ | $N_o$ | $N_R$ | $I_4$ Rad | $z_{est}$ | Peak | $\sigma$ | $\Lambda_{cl}$ | $N_o$ | $N_R$ |
|---|---|---|---|---|---|---|---|---|---|---|---|---|---|---|---|---|
| 001 | 00 30 18.9 | +05 46 30 | –– | 0.50 | 0.26 | 1.86 | 34.7 | 22± 4 | 63± 9 | 144 | 0.50 | 0.82 | 2.92 | 49.5 | 31± 6 | 91±23 |
| 002 | 00 30 56.9 | +05 55 05 | –– | 0.40 | 0.39 | 2.21 | 31.9 | 14± 3 | 34± 8 | 55 | 0.40 | 1.02 | 3.32 | 43.4 | 16± 4 | 33±17 |
| 003 | 02 26 48.7 | +01 18 10 | 47 | 0.80 | 0.20 | 2.28 | 79.6 | 25± 7 | 141±17 | –– | 0.80 | 0.14 | 0.86 | 19.8 | 12± 5 | 80±21 |
| 004 | 09 50 02.2 | +47 20 13 | 49 | 1.00 | 0.11 | 3.25 | 103.8 | 9± 3 | 3± 7 | –– | 1.00 | 0.14 | 1.68 | 36.1 | 10± 4 | 18±15 |
| 005 | 09 55 05.3 | +47 50 19 | 117 | 0.90 | 0.15 | 3.18 | 85.5 | 23± 7 | 79±15 | –– | 0.90 | 0.07 | 0.71 | 13.8 | 13± 5 | 16±22 |
| 006 | 09 49 56.5 | +47 06 40 | –– | 0.40 | 0.53 | 2.53 | 33.2 | 15± 3 | 32± 6 | 78 | 0.40 | 1.10 | 3.50 | 39.0 | 15± 4 | 44±13 |
| 007 | 09 55 31.0 | +47 12 45 | –– | 1.00 | 0.03 | 1.00 | 32.1 | 17± 3 | 18± 8 | 80 | 1.00 | 0.24 | 2.97 | 63.7 | 23± 8 | 118±30 |
| 008 | 09 53 49.9 | +47 52 28 | –– | 0.60 | 0.36 | 3.05 | 61.9 | 21± 4 | 43± 8 | 92 | 0.60 | 0.77 | 3.72 | 57.6 | 23± 6 | 60±21 |
| 009 | 09 55 41.7 | +48 00 36 | –– | 1.20 | 0.06 | 3.56 | 121.1 | 11± 4 | 19± 8 | 123 | 1.20 | 0.22 | 3.84 | 105.1 | 15± 7 | 68±25 |
| 010 | 13 28 01.1 | +29 24 26 | 50 | 1.10 | 0.16 | 3.04 | 210.9 | 15± 6 | 24±11 | –– | 1.10 | 0.03 | 0.33 | 9.5 | 6± 4 | 44±20 |
| 011 | 13 25 02.0 | +29 35 47 | 112 | 0.90 | 0.22 | 3.02 | 130.7 | 17± 4 | 39± 8 | –– | 0.90 | 0.15 | 1.40 | 29.4 | 6± 2 | 17±13 |
| 012 | 13 25 12.9 | +29 48 55 | 82 | 0.80 | 0.27 | 2.86 | 101.4 | 9± 3 | 22± 8 | –– | 0.80 | 0.10 | 0.77 | 14.9 | 13± 5 | 33±20 |
| 013 | 13 25 01.3 | +30 05 52 | 116 | 0.60 | 0.44 | 2.97 | 76.9 | 21± 4 | 54± 8 | –– | 0.60 | 0.15 | 0.76 | 12.4 | 23± 6 | 57±23 |
| 014 | 13 25 21.0 | +30 09 55 | 49 | 0.90 | 0.24 | 3.22 | 139.8 | > 40± 8 | > 80±12 | –– | 0.90 | 0.01 | 0.13 | 2.6 | 14± 5 | 27±21 |
| 015 | 13 25 08.7 | +30 24 39 | 118 | 0.50 | 0.76 | 3.91 | 86.9 | 22± 4 | 58± 7 | –– | 0.50 | 0.09 | 0.36 | 5.2 | 14± 5 | 41±17 |
| 016 | 13 24 01.5 | +30 24 32 | 70 | 0.80 | 0.31 | 3.30 | 117.0 | 29± 7 | 86±13 | –– | 0.80 | 0.28 | 2.17 | 41.8 | 28± 6 | 58±19 |
| 017 | 13 25 45.2 | +29 25 59 | –– | 0.50 | 0.49 | 2.52 | 55.9 | 13± 4 | 48± 9 | 54 | 0.50 | 0.69 | 2.78 | 40.3 | 23± 6 | 64±22 |
| 018 | 13 27 53.2 | +29 28 37 | –– | 0.70 | -0.07 | -0.62 | -18.1 | 13± 4 | 36± 8 | 83 | 0.70 | 0.46 | 2.89 | 51.5 | 12± 5 | 66±22 |
| 019 | 13 25 58.8 | +29 50 16 | –– | 0.80 | 0.07 | 0.80 | 28.1 | 17± 4 | 13± 8 | 30 | 0.80 | 0.28 | 2.18 | 40.7 | 23± 6 | 52±22 |
| 020 | 13 24 32.9 | +29 59 35 | –– | 0.40 | 0.39 | 1.67 | 26.5 | 9± 3 | 10± 7 | 47 | 0.40 | 0.83 | 2.82 | 34.1 | 21± 5 | 52±19 |
| 021 | 13 25 25.4 | +30 18 28 | –– | 0.50 | 0.40 | 2.03 | 45.1 | 10± 2 | 35± 6 | 46 | 0.50 | 0.77 | 3.13 | 45.3 | 15± 4 | 37±18 |
| 022 | 13 28 17.1 | +30 18 34 | –– | 0.70 | 0.07 | 0.61 | 17.9 | 11± 4 | 18± 9 | 25 | 0.70 | 0.41 | 2.59 | 46.2 | 22± 6 | 65±20 |
| 023 | 16 07 27.9 | +41 09 26 | 68 | 0.80 | 0.27 | 2.79 | 105.9 | 19± 4 | -8± 8 | –– | 0.80 | 0.14 | 0.96 | 19.3 | 15± 5 | 30±21 |
| 024 | 16 09 01.6 | +41 09 25 | 105 | 1.10 | 0.22 | 3.62 | 302.8 | > 30± 8 | > 50±10 | –– | 1.10 | 0.02 | 0.22 | 5.8 | 6± 5 | 26±19 |
| 025 | 16 07 17.7 | +41 32 16 | 131 | 0.60 | 0.47 | 3.36 | 85.9 | 24± 5 | 86±10 | –– | 0.60 | 0.23 | 1.04 | 17.2 | 17± 6 | 68±25 |
| 026 | 16 07 35.6 | +41 36 36 | 67 | 1.10 | 0.21 | 3.35 | 280.7 | > 30± 8 | > 30± 8 | –– | 1.10 | -0.08 | -1.00 | -27.2 | 6± 5 | 9±23 |
| 027 | 16 04 13.2 | +41 35 53 | 93 | 0.70 | 0.39 | 3.39 | 103.8 | 8± 3 | 34± 7 | –– | 0.70 | 0.28 | 1.56 | 28.8 | 14± 4 | 54±15 |
| 028 | 16 03 35.4 | +41 52 34 | 52 | 1.00 | 0.23 | 3.19 | 215.8 | > 30± 8 | > 50±10 | –– | 1.00 | 0.06 | 0.64 | 15.5 | 9± 4 | 41±19 |

## NOTES TO TABLES 4 AND 5

<u>Cluster 005</u>: GHO cluster 0025+0455.

<u>Clusters 009 and 013</u>: These two detections may be part of the same system. Cluster 013 not detected as a separate system in the $I_4$ band.

<u>Cluster 011</u>: GHO cluster 0027+0535 (also known as 00 $\alpha\alpha$).

<u>Cluster 012</u>: GHO cluster 0025+0539.

<u>Cluster 017</u>: GHO cluster 0227+0024 (also known as 02 KP$\alpha$).

<u>Clusters 026 and 027</u>: These clusters are not detected in the $I_4$ band. Low $V_4$ richness values.

<u>Cluster 028</u>: $I_4$ band data for this cluster is not available. It would have been detected in the $I_4$ band otherwise.

<u>Cluster 029</u>: Cluster was not detected in the $I_4$ band. Low $V_4$ richness values.

<u>Clusters 030 and 045</u>: These two detections may be associated with one another.

<u>Cluster 032</u>: GHO cluster 0949+4732.

<u>Cluster 033</u>: This detection may be associated with cluster 032.

<u>Cluster 038</u>: GHO cluster 0947+4758.

<u>Cluster 041</u>: GHO cluster 0951+4813.

<u>Clusters 043</u>: Cluster was not detected in the $I_4$ band. Low $V_4$ richness values.

<u>Clusters 044</u>: Cluster was not detected in the $I_4$ band. Low $V_4$ richness values. Faint galaxies in $V_4$ data lie below $I_4$ completeness limit.

<u>Cluster 045</u>: Cluster was not detected in the $V_4$ band. Near bright star.

<u>Cluster 049</u>: Abell 1739.

<u>Clusters 050</u>: GHO cluster 1324+2948. This detection may be associated with Abell 1739 (Cluster 049).

<u>Cluster 051</u>: GHO cluster 1321+2949.

<u>Cluster 052</u>: GHO cluster 1323+2953 (also known as 13 KPN$\beta$).

<u>Cluster 055</u>: GHO cluster 1324+3014 (also known as 13 KPA).

<u>Cluster 056</u>: GHO cluster 1324+3015.

<u>Cluster 058</u>: GHO cluster 1325+3021.

<u>Cluster 059</u>: GHO cluster 1322+3027 (also known as 13 KPN$\alpha$).

<u>Cluster 063</u>: GHO cluster 1322+3028 (also known as 13 KPN$\gamma$).

<u>Cluster 064</u>: Cluster was not detected in the $V_4$ band. Faint galaxies near 4-Shooter pyramid shadow region.

<u>Cluster 073</u>: GHO cluster 1604+4144 (also known as 16 $\gamma$).

<u>Cluster 078</u>: Cluster was not detected in $I_4$ band. Faint galaxies in $V_4$ data lie below $I_4$

completeness limit. High $V_4$ richness values.

<u>Cluster 079</u>: Cluster was not detected in $V_4$ band.

<u>Supplemental Cluster 017</u>: This detection may be associated with Abell 1739 (cluster 049).

<u>Supplemental Cluster 020</u>: GHO cluster 1322+3014.

<u>Supplemental Cluster 023</u>: Background cluster behind GHO cluster 1605+4119.

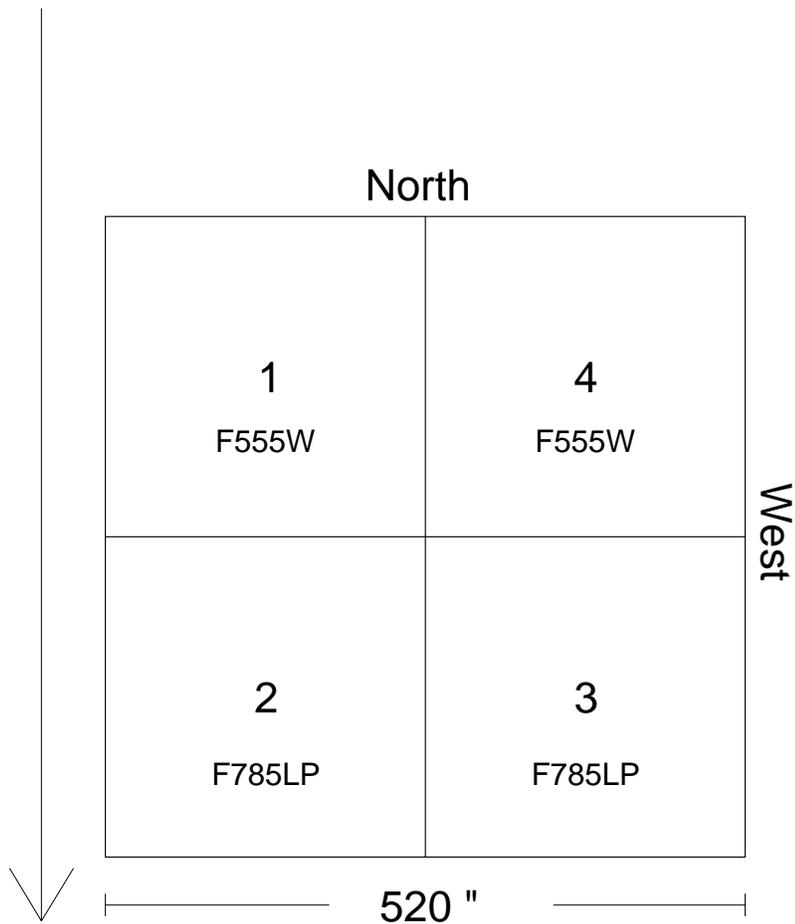

Fig. 1.— A schematic diagram of the configuration of the 4-Shooter focal plane during the gathering of the survey data. The four CCDs are denoted as 1, 2, 3, and 4. The arrow on the left is the CCD readout direction (along the columns) and is the motion of the sky across the detectors.

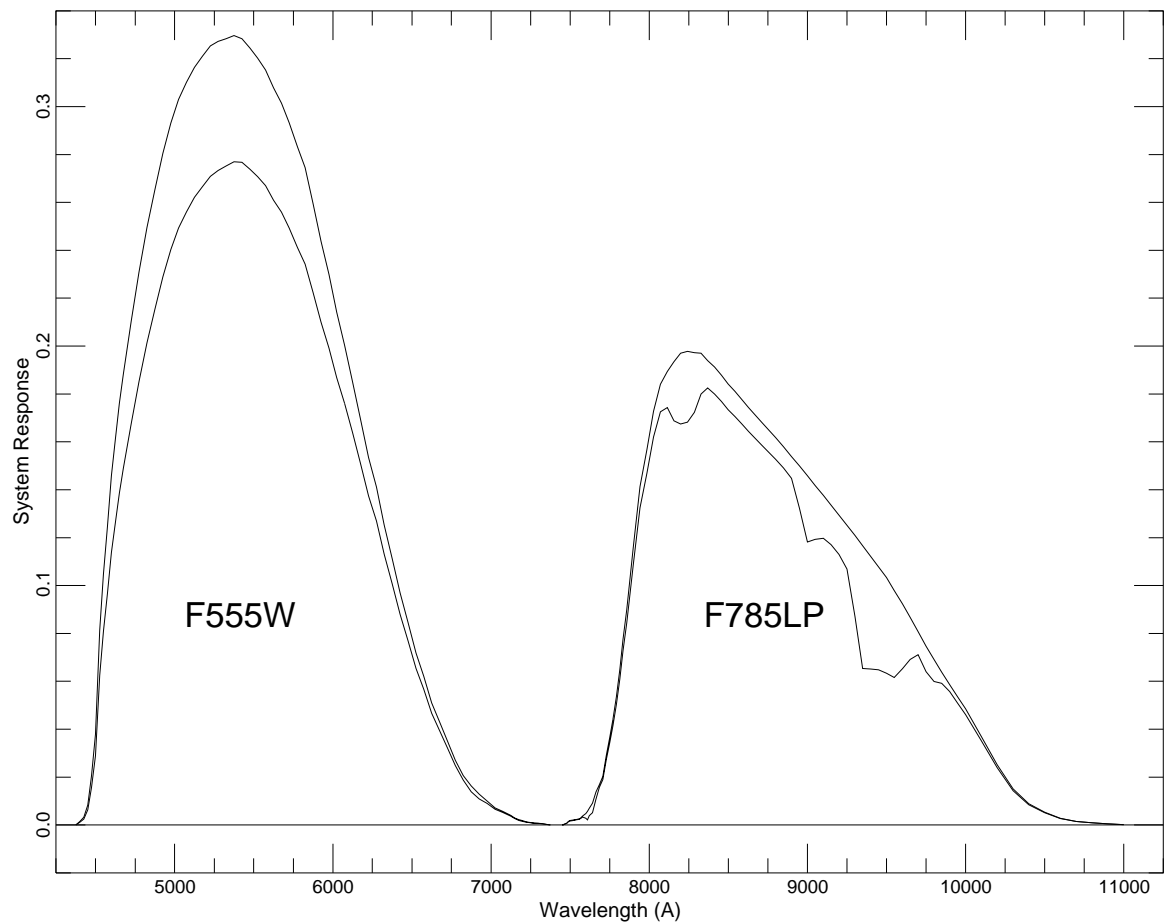

Fig. 2.— The response curves of the 4-Shooter $F555W$ ($V_4$) and $F785LP$ ($I_4$) filters used in the survey. The response includes the telescope, CCD, 4-Shooter camera, and filter quantum efficiencies. The upper curve for each filter is the response with no atmosphere; the lower curve is the response at an airmass of 1.2 at Palomar.

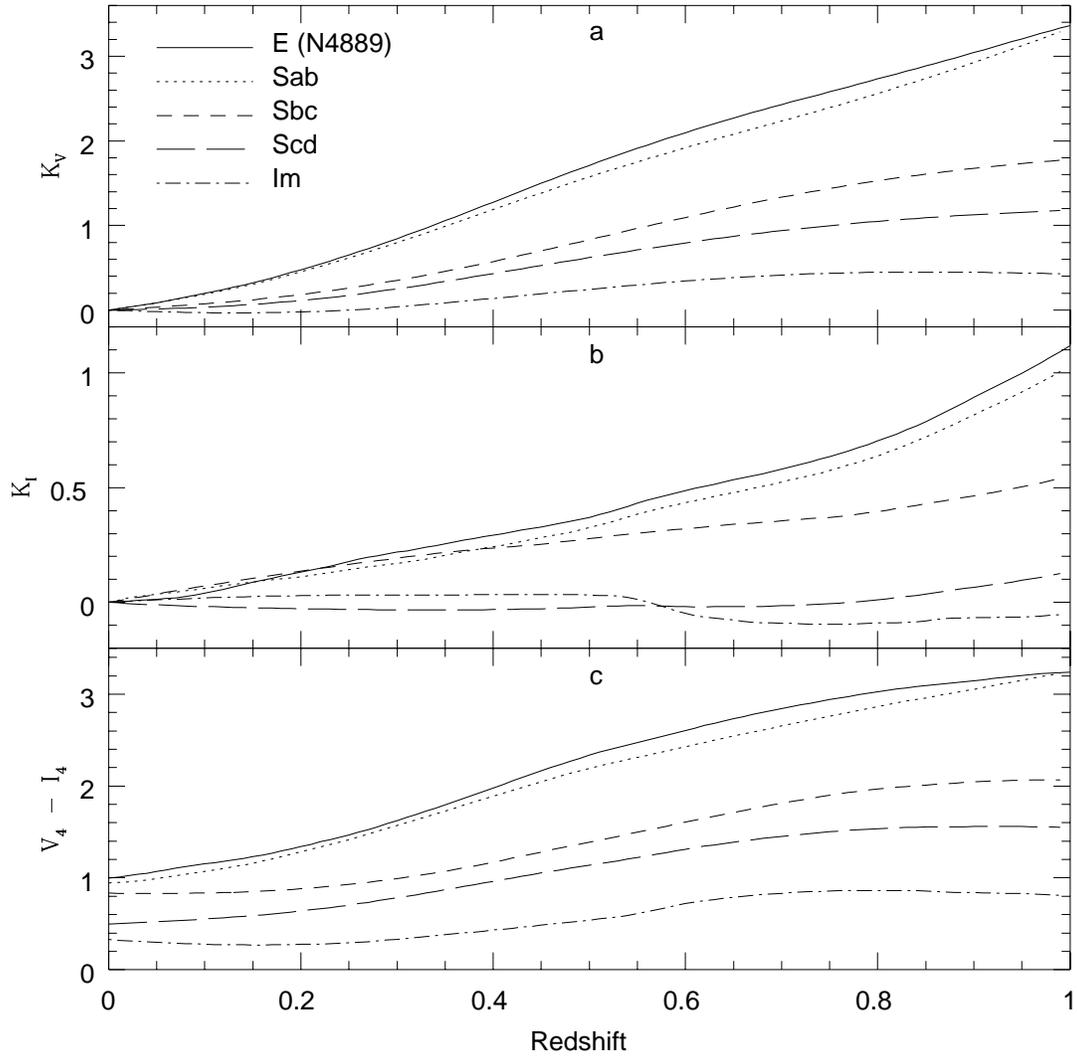

Fig. 3.— (a) The k-correction as a function of redshift and morphology for the $V_4$ filter. (b) The k-correction as a function of redshift and morphology for the $I_4$ filter. (c) The $V_4-I_4$ color as a function of redshift and morphology.

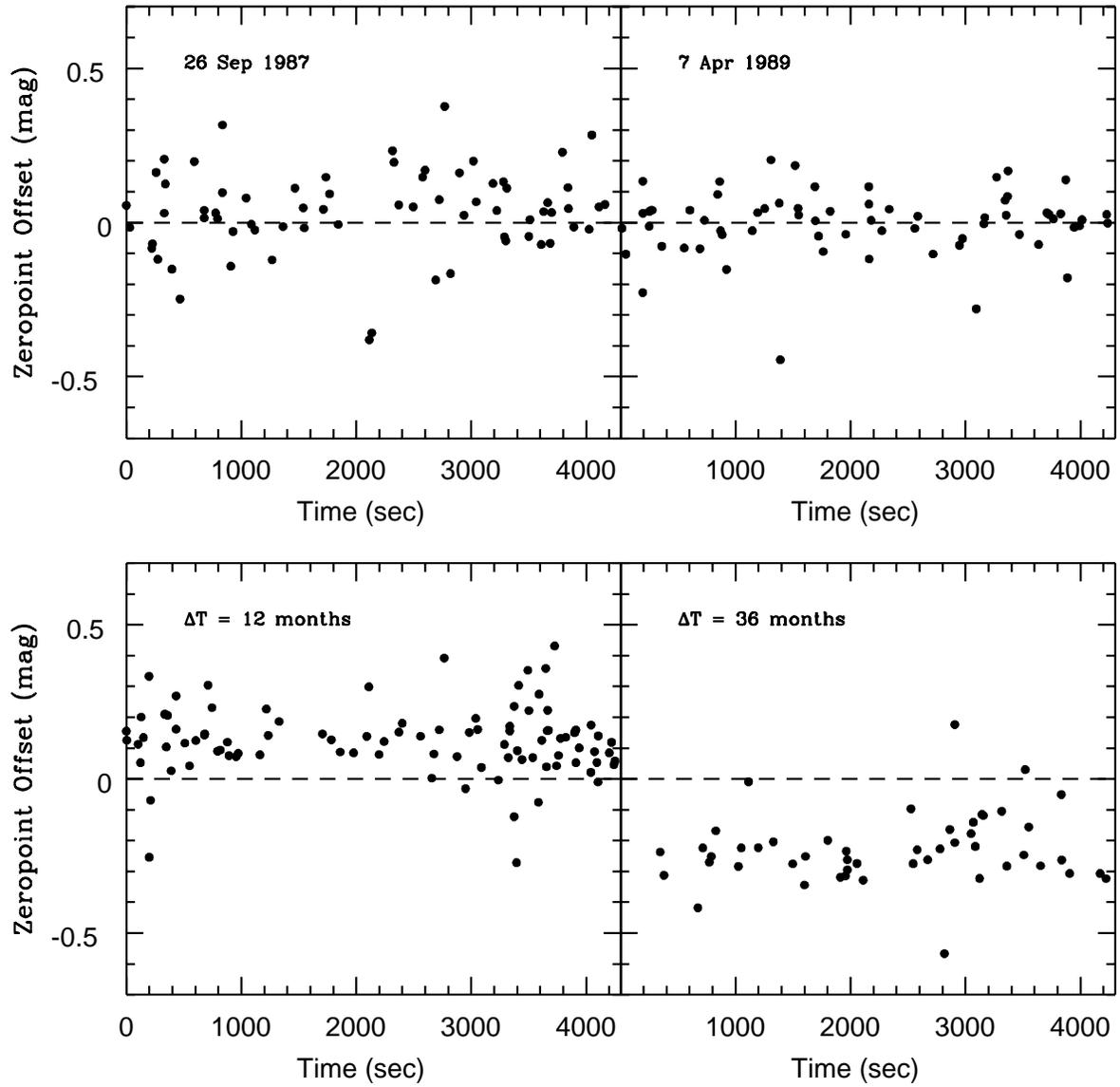

Fig. 4.— The photometric offset as a function of time for four pairs of adjacent scans. The upper two panels show offsets for scans taken on the same night; the bottom two panels show offsets for scans taken more than a month apart. The time is derived from the object declination, which is linearly related to the time elapsed since the start of the scan.

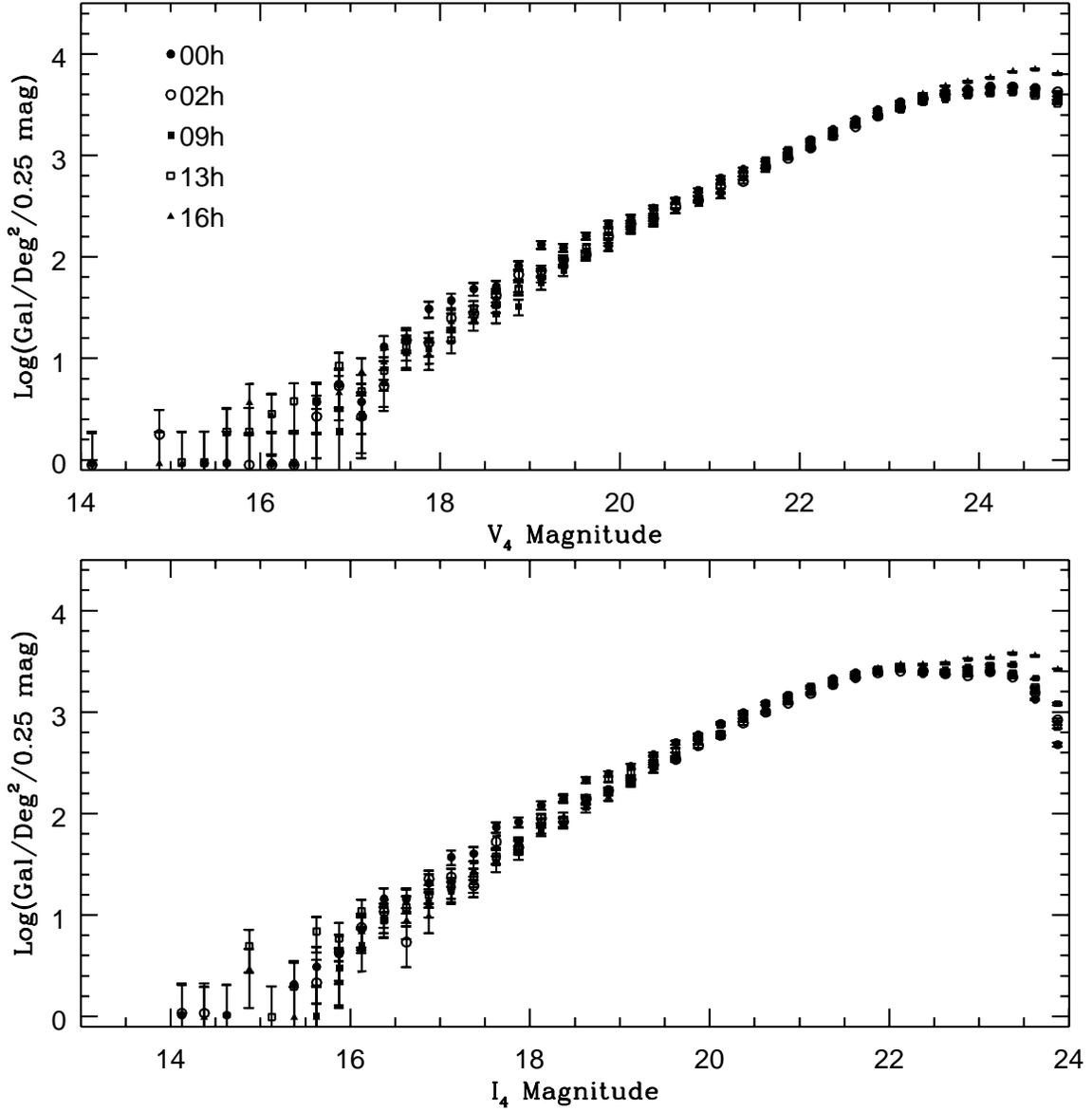

Fig. 5.— (a) The differential galaxy surface density as a function of $V_4$ isophotal magnitude for the 5 survey fields. (b) The same for the $I_4$ band.

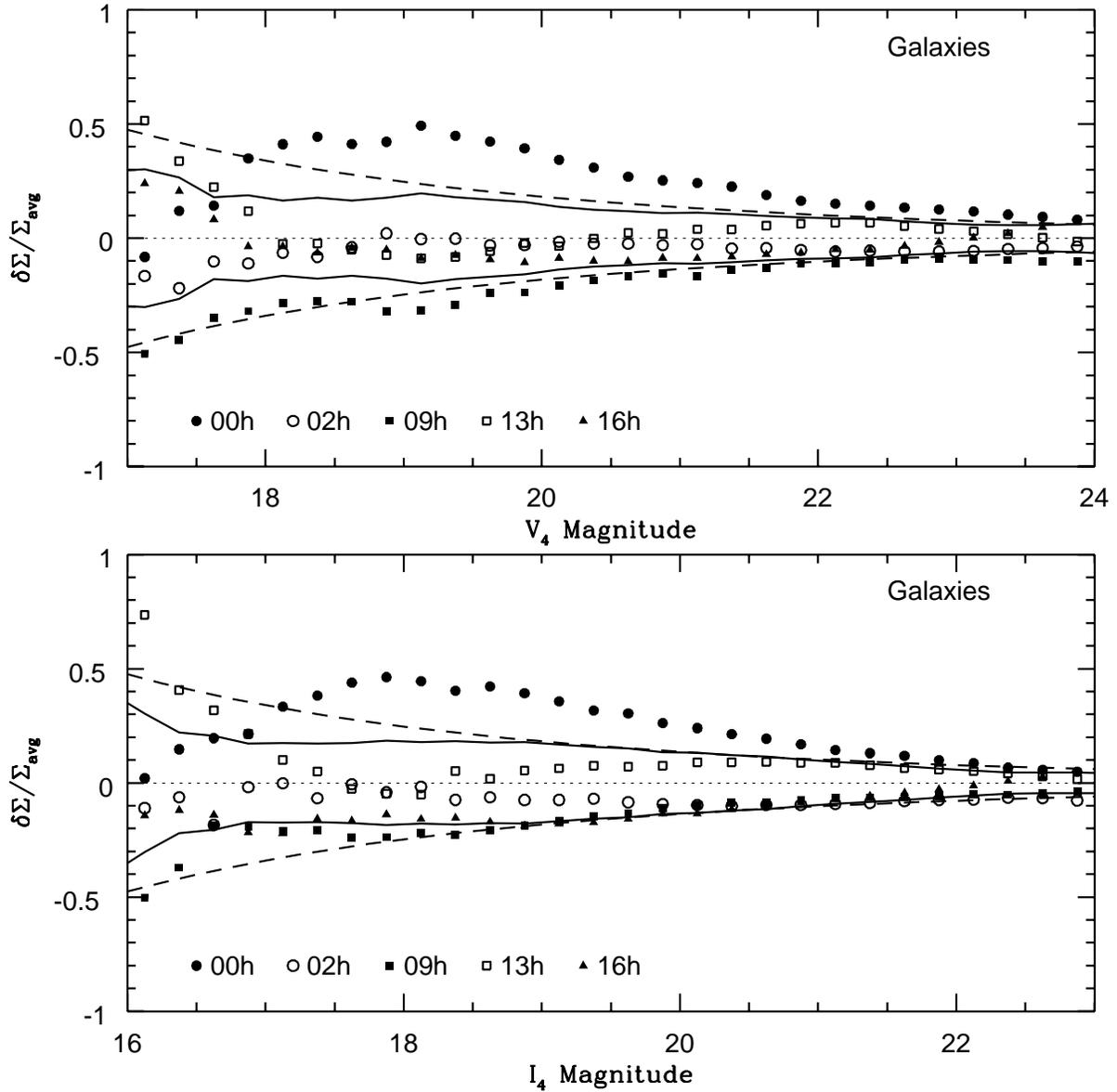

Fig. 6.— Residuals between the mean galaxy surface density for all five fields and the galaxy surface density in each individual field as functions of $V_4$ and $I_4$ isophotal magnitude. Solid line represents mean residual level, dashed line gives the prediction based on the observed angular two point correlation function (Neushaefer *et al.* 1991).

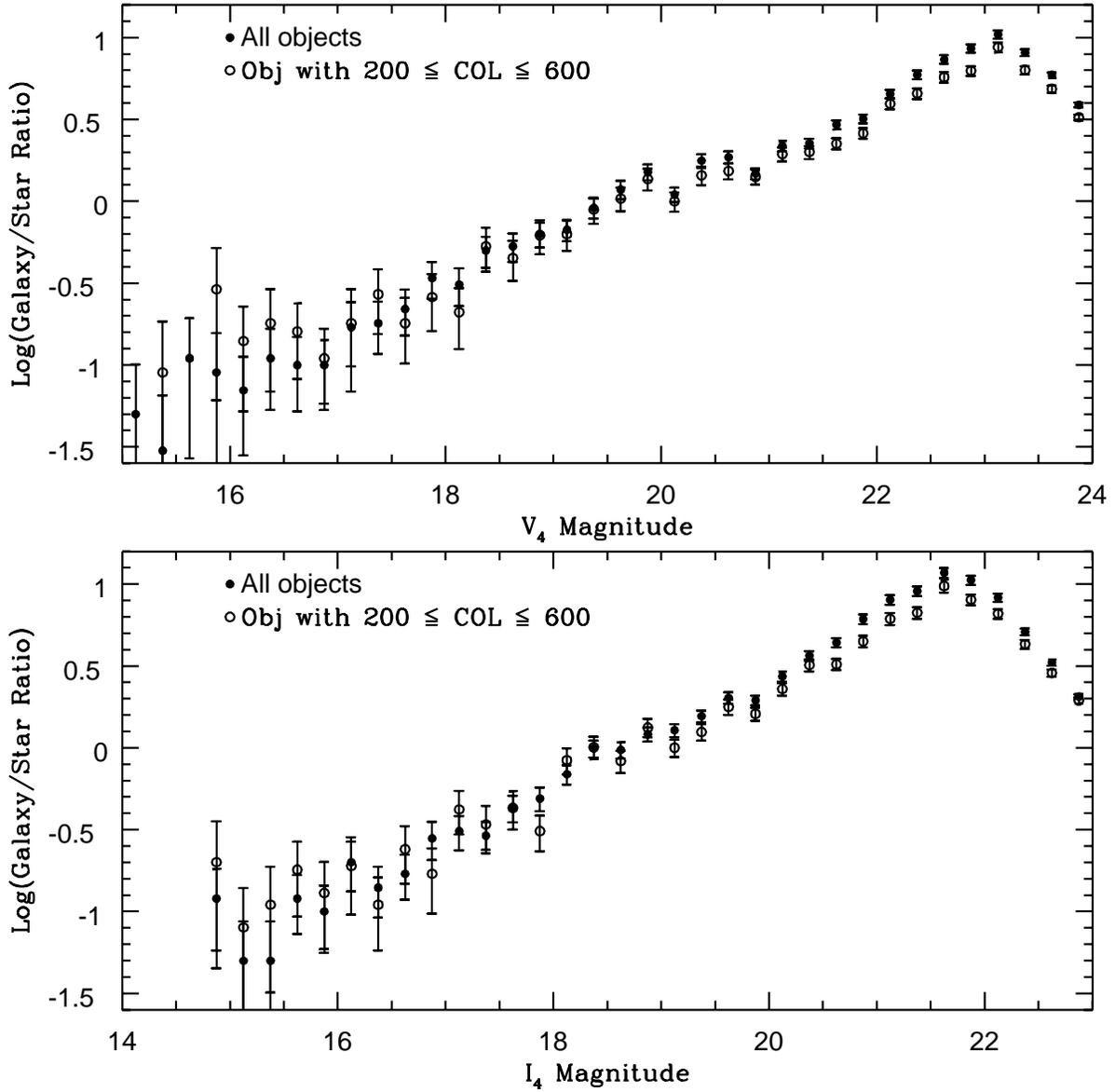

Fig. 7.— (a) The galaxy-to-star count ratio as a function of $V_4$ isophotal magnitude for all objects in the 13h field (solid dots) and for those objects which lie within $200 \leq$ CCD Column $\leq 600$. The results for the other survey fields are similar. (b) The same for the $I_4$ images.

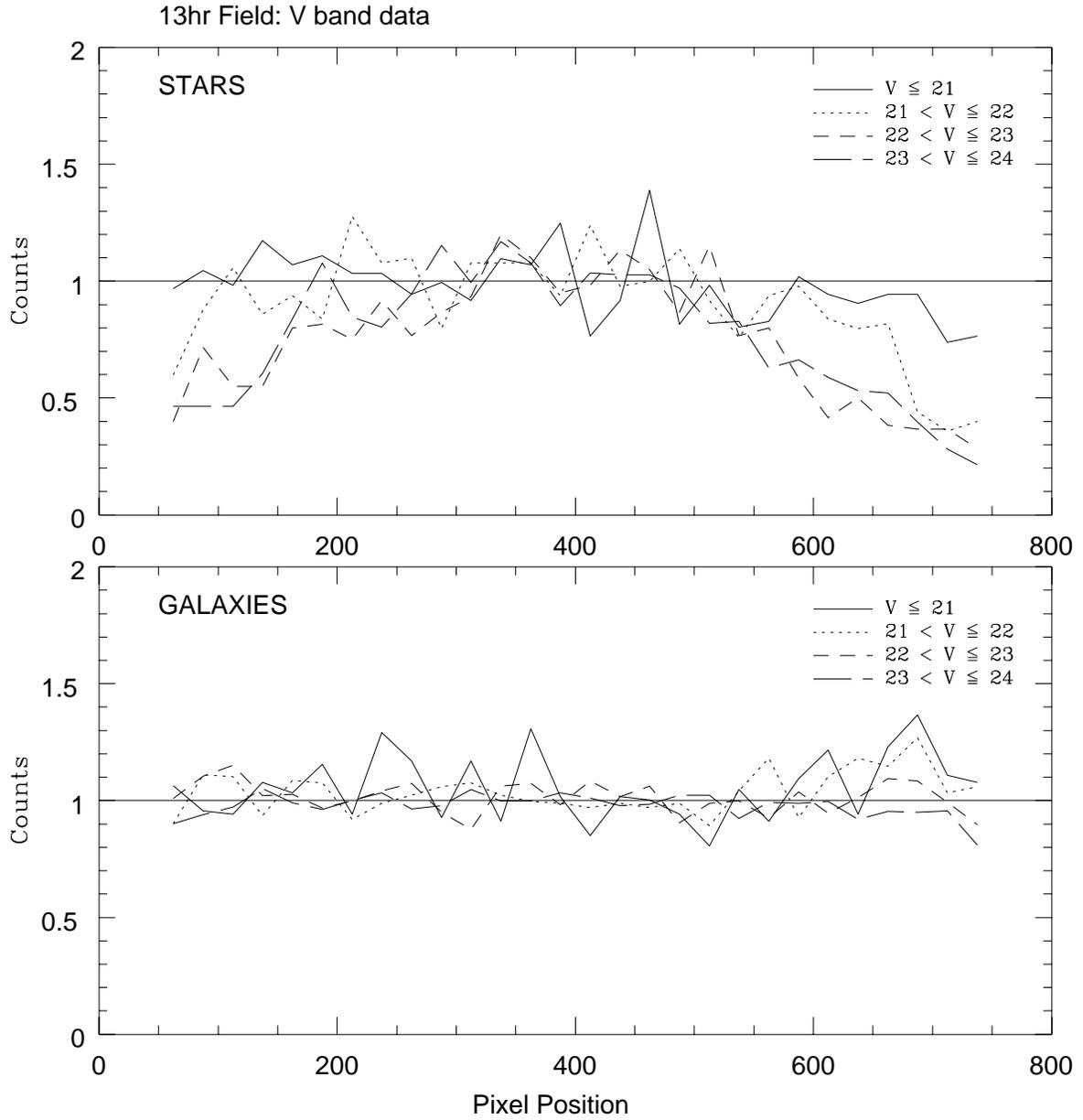

Fig. 8.— (a) $V_4$ band star counts as a function of CCD column number. The $I_4$ band results are similar. (b) $V_4$ band galaxy counts as a function of CCD column number. The $I_4$ band results are similar.

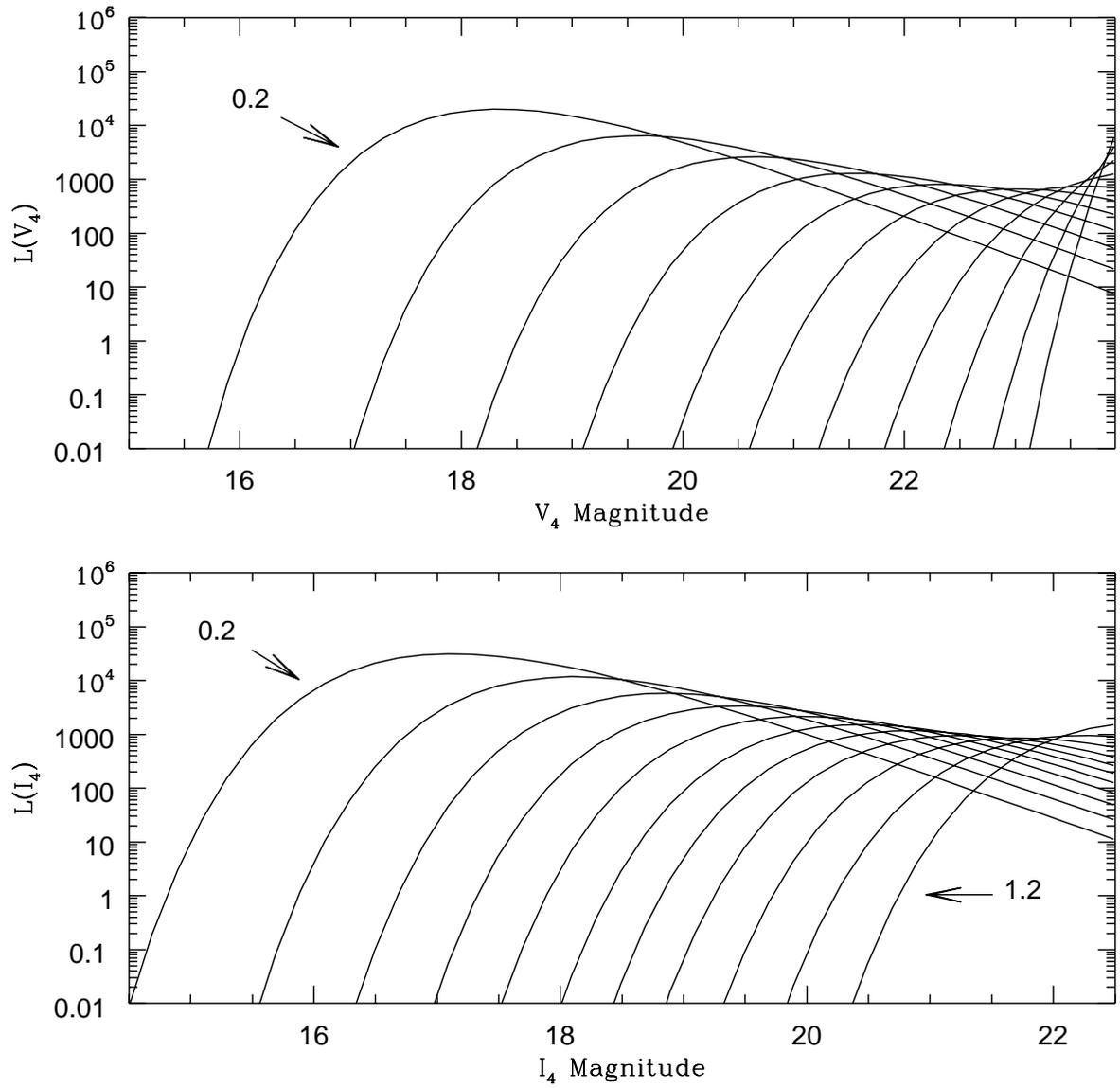

Fig. 9.— (a) The $V_4$ flux filter as a function of redshift shown in intervals of 0.1 over the $z$ range [0.2,1.2]. The ordinate is the amplitude of the filter. Numbers indicate the filter redshift. No color evolution is assumed. (b) The $I_4$ flux filter.

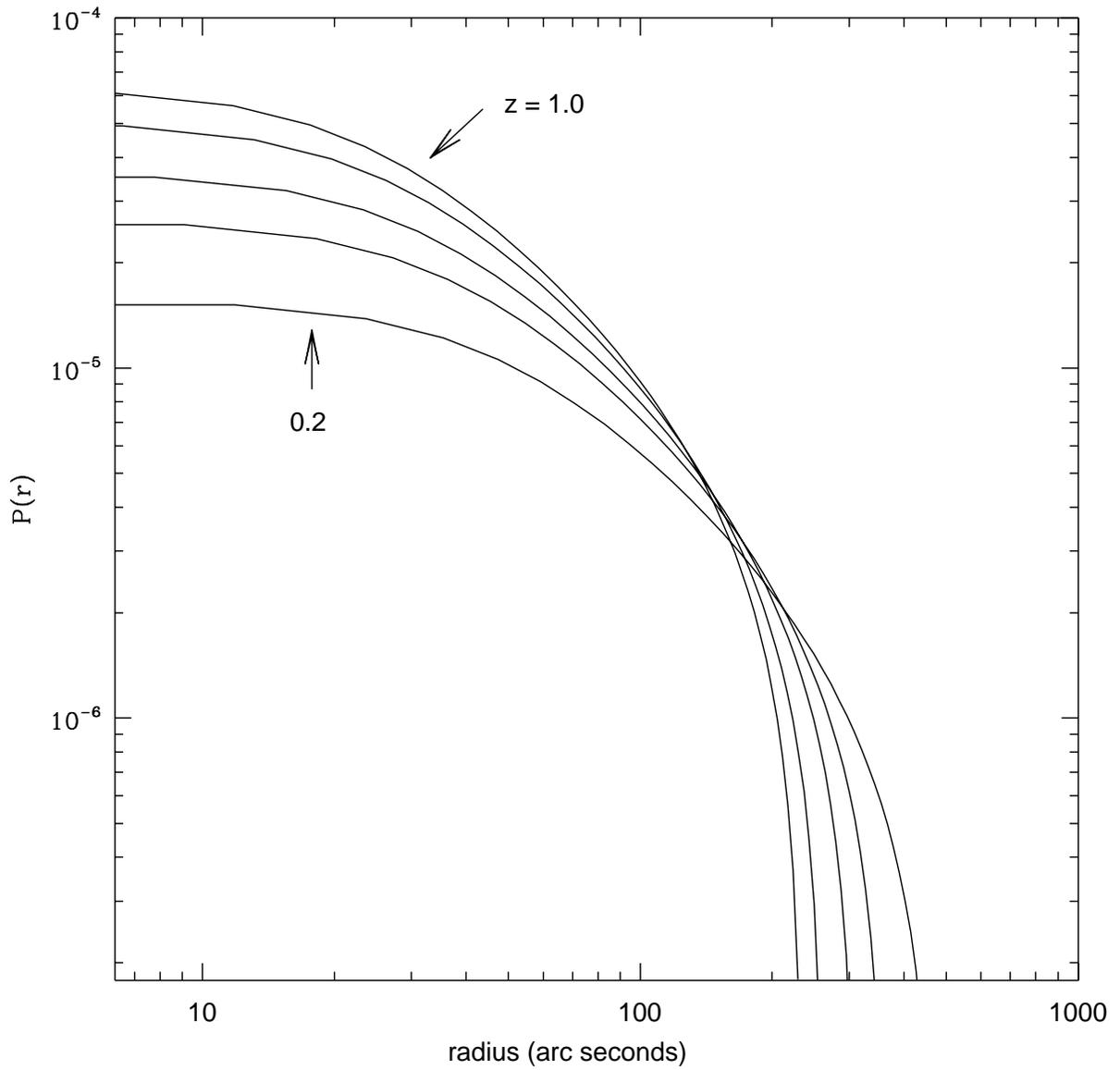

Fig. 10.— The radial filter as a function of redshift. The ordinate is the amplitude of the filter. A core radius of $100h^{-1}$ kpc and cutoff radius of $1h^{-1}$ Mpc are used. The 5 radial filters shown, from broadest to narrowest, are for $z = 0.2$, 0.3, 0.4, 0.6, and 1.0.

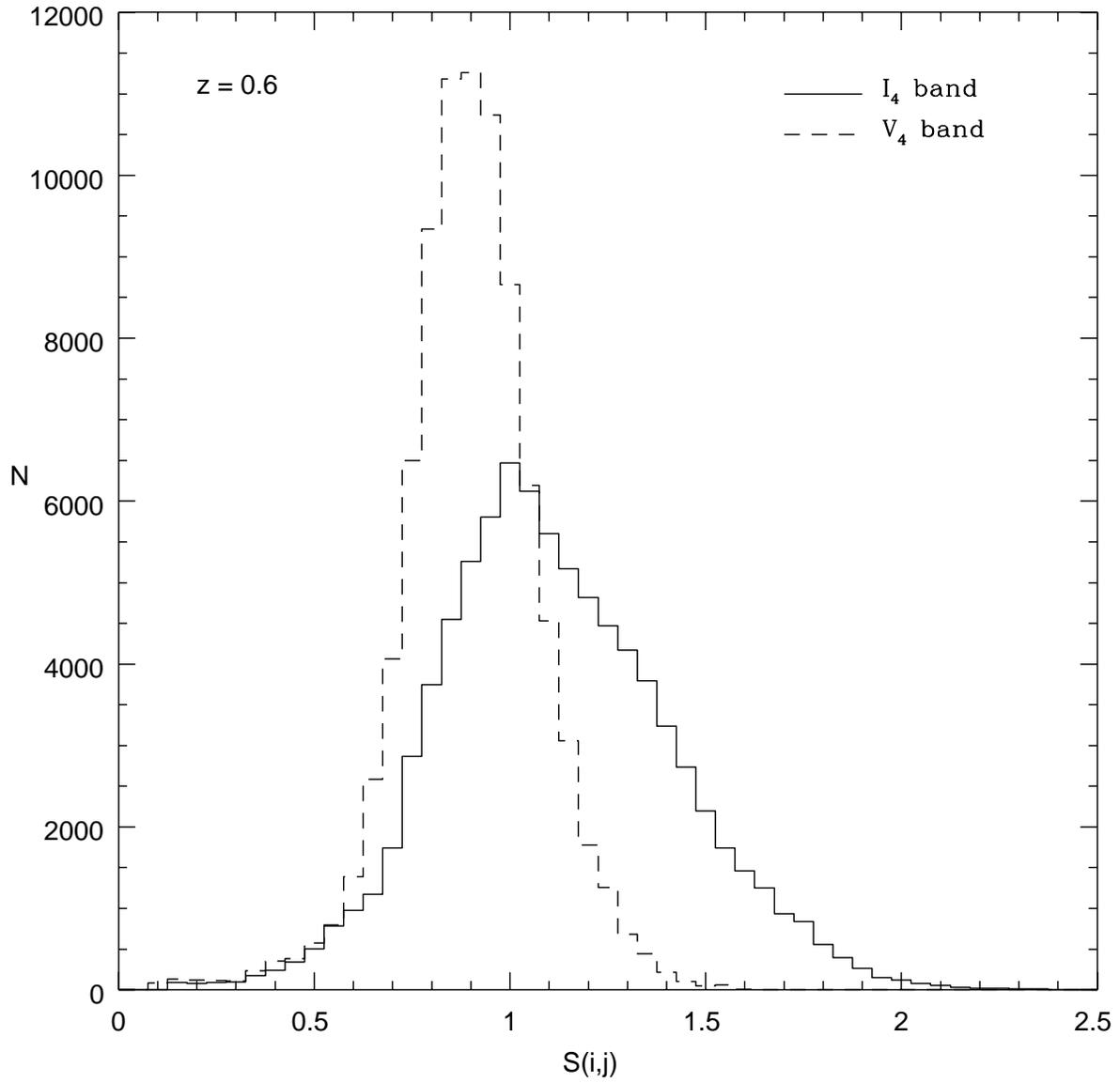

Fig. 11.— The histogram of $S(i,j)$ values generated using a matched filter tuned to detect clusters at $z = 0.6$ in the $0^{\mathrm{h}}$ field.

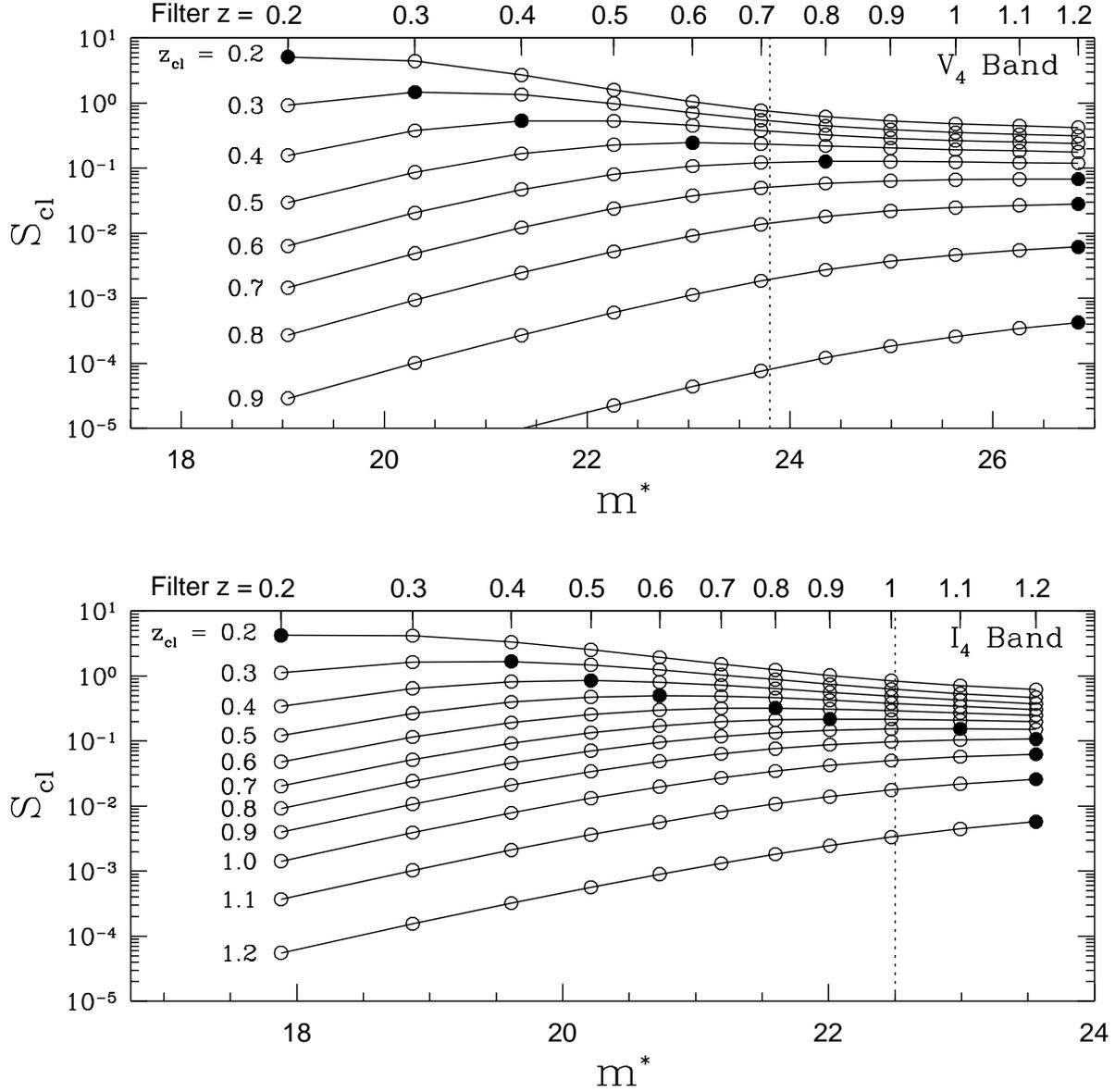

Fig. 12.— The cluster signal, $S_{cl}$, as a function of characteristic magnitude $m^*$ and passband. The corresponding filter redshift is shown along the upper abscissa. Points for a cluster at a given redshift are connect by solid lines. The solid point for each cluster denotes the redshift estimate that the matched filter algorithm would assign to the cluster. The estimates shown here are derived prior to the application of the cluster signal correction. The vertical dashed lines indicate the survey magnitude limits.

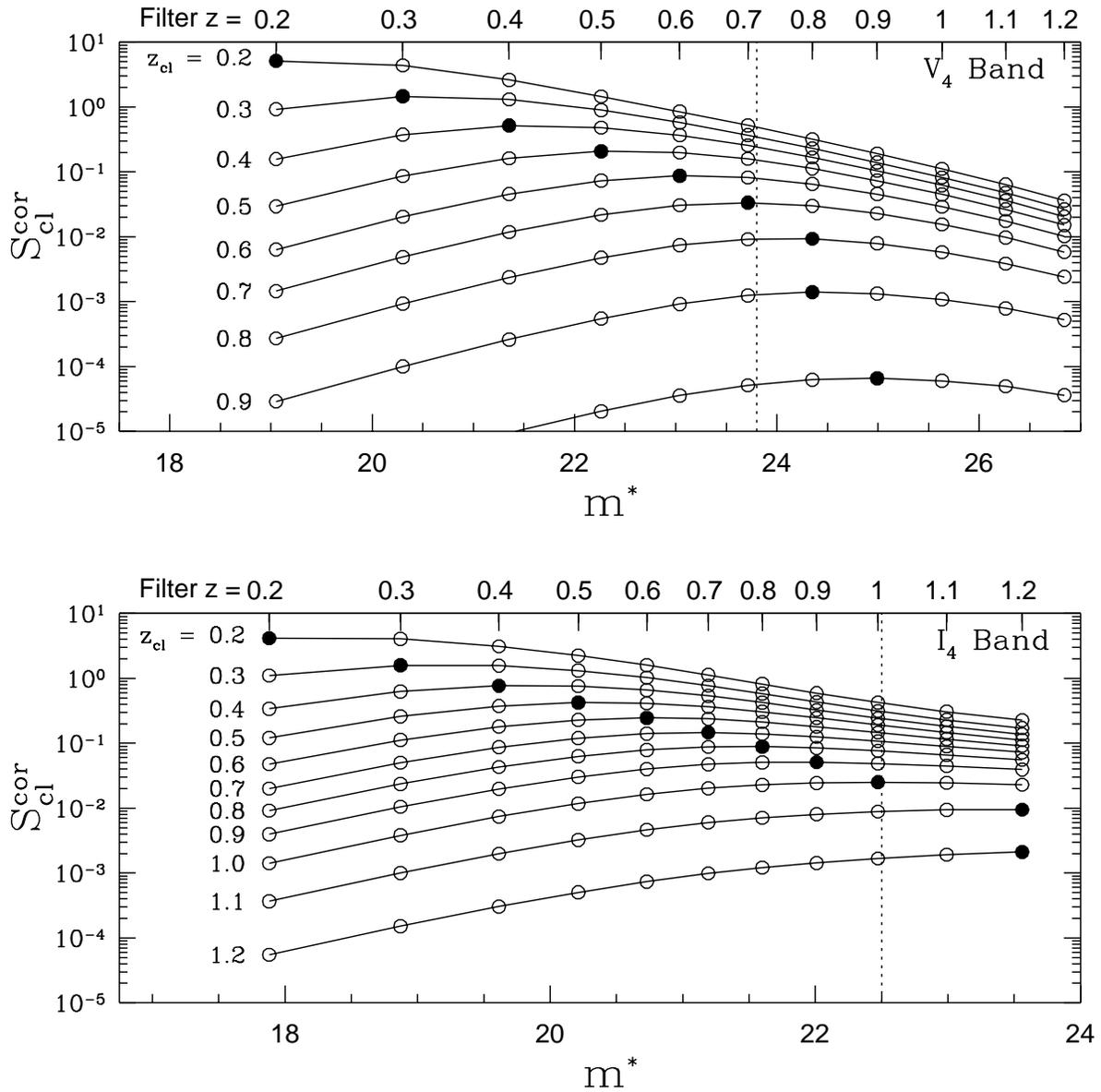

Fig. 13.— The corrected cluster signal, $S_{cl}^{cor}$, as a function of characteristic magnitude $m^*$ and passband. The corresponding filter redshift is shown along the upper abscissa. Points for a cluster at a given redshift are connect by solid lines. The solid point for each cluster denotes the redshift estimate that the matched filter algorithm would assign to the cluster. Note the improved redshift estimate accuracy provided by the application of the $CSC$.

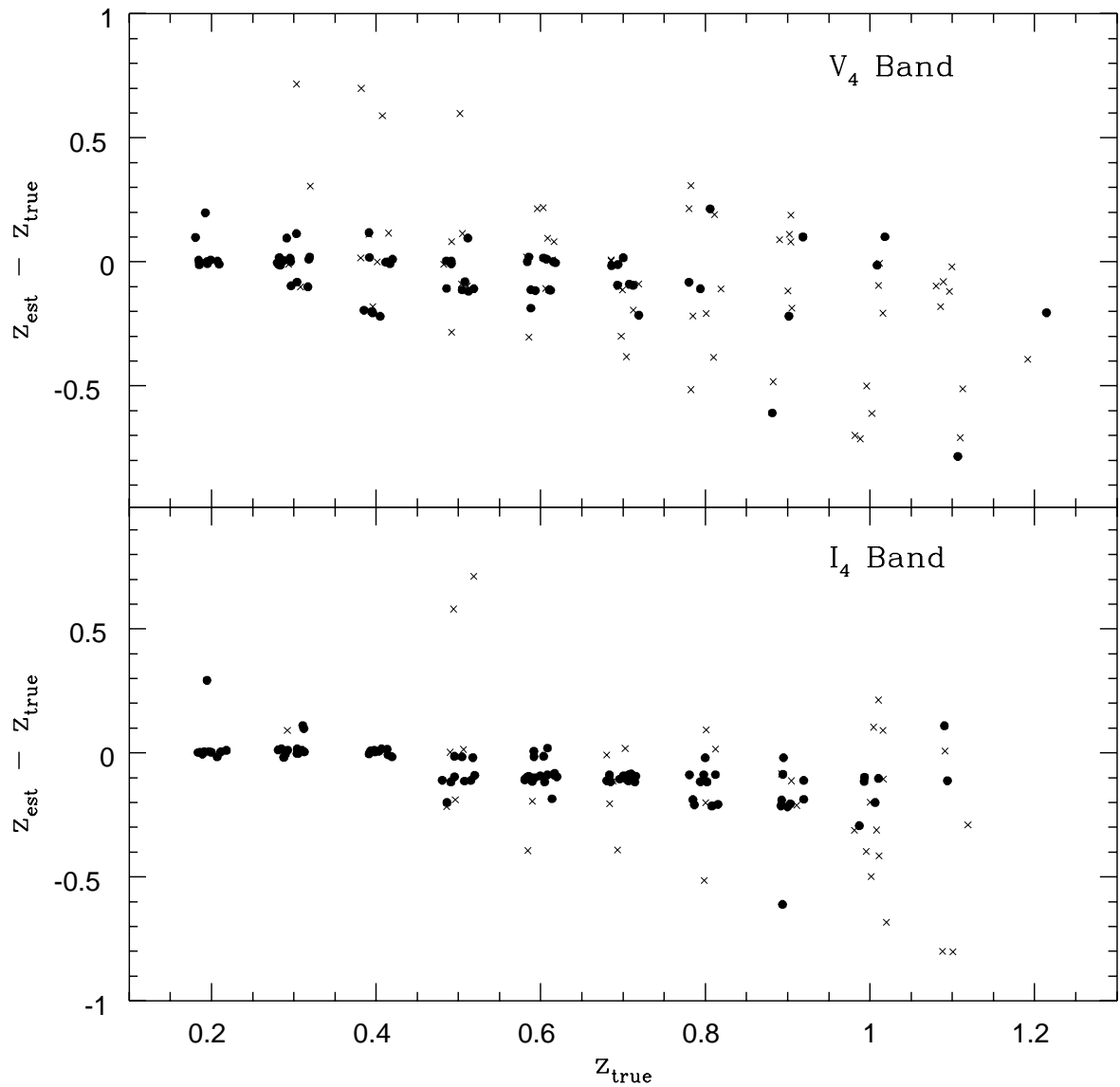

Fig. 14.— The offset between estimated redshift and true redshift, $z_{est} - z_{true}$, as a function of true redshift for $\sim 400$ simulated clusters. Solid points represent clusters which are detected at the $\geq 4\sigma$ level. Crosses represent clusters detected below the $4\sigma$ level.

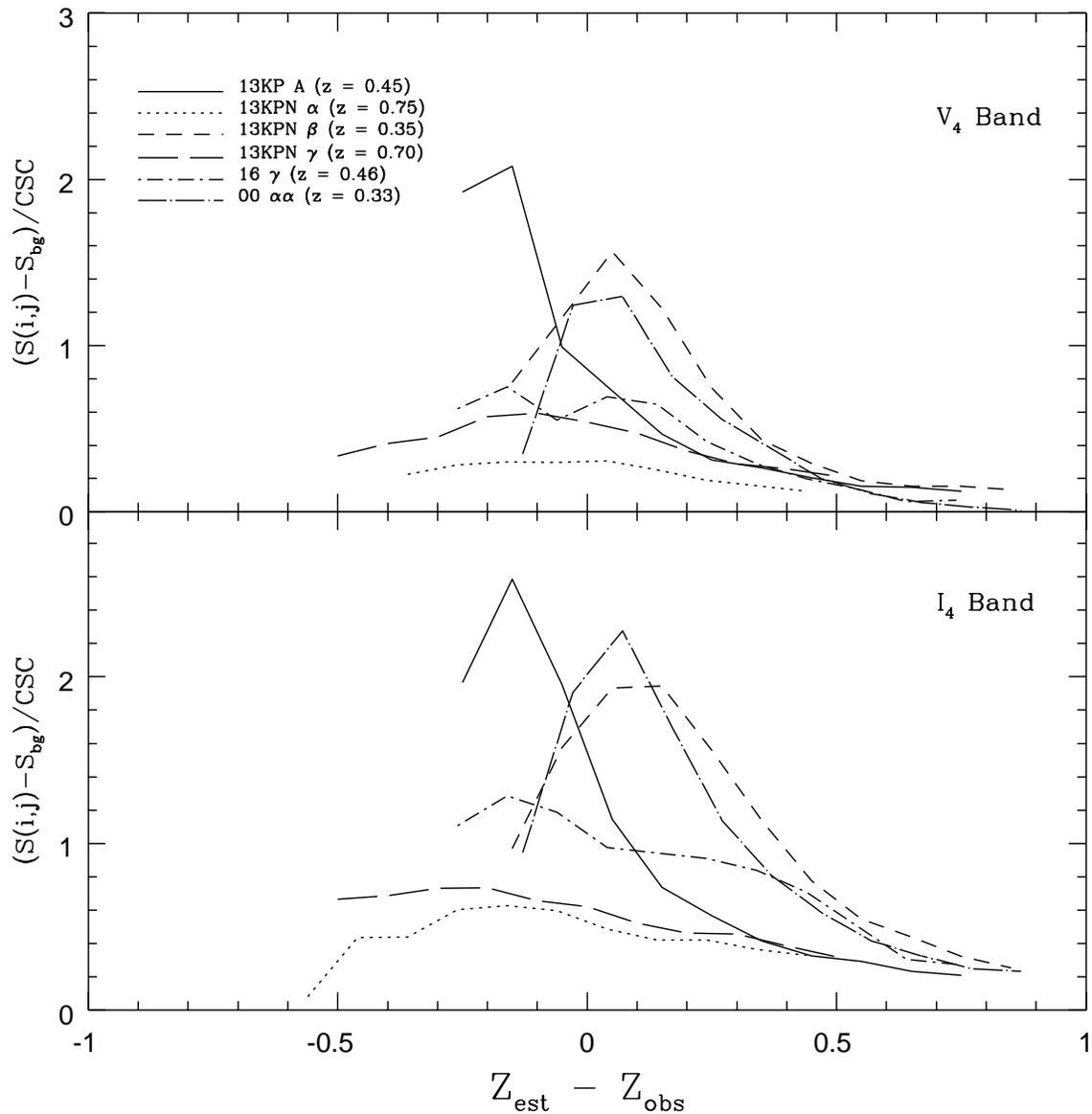

Fig. 15.— (a) The offset between estimated redshift and true redshift, $z_{est} - z_{true}$, as a function of the corrected $V_4$ band matched filter signal, $S_{cl}^{cor} = (S(i, j) - S_{bg})/CSC$, for 6 real clusters. (b) The same for the $I_4$ band.

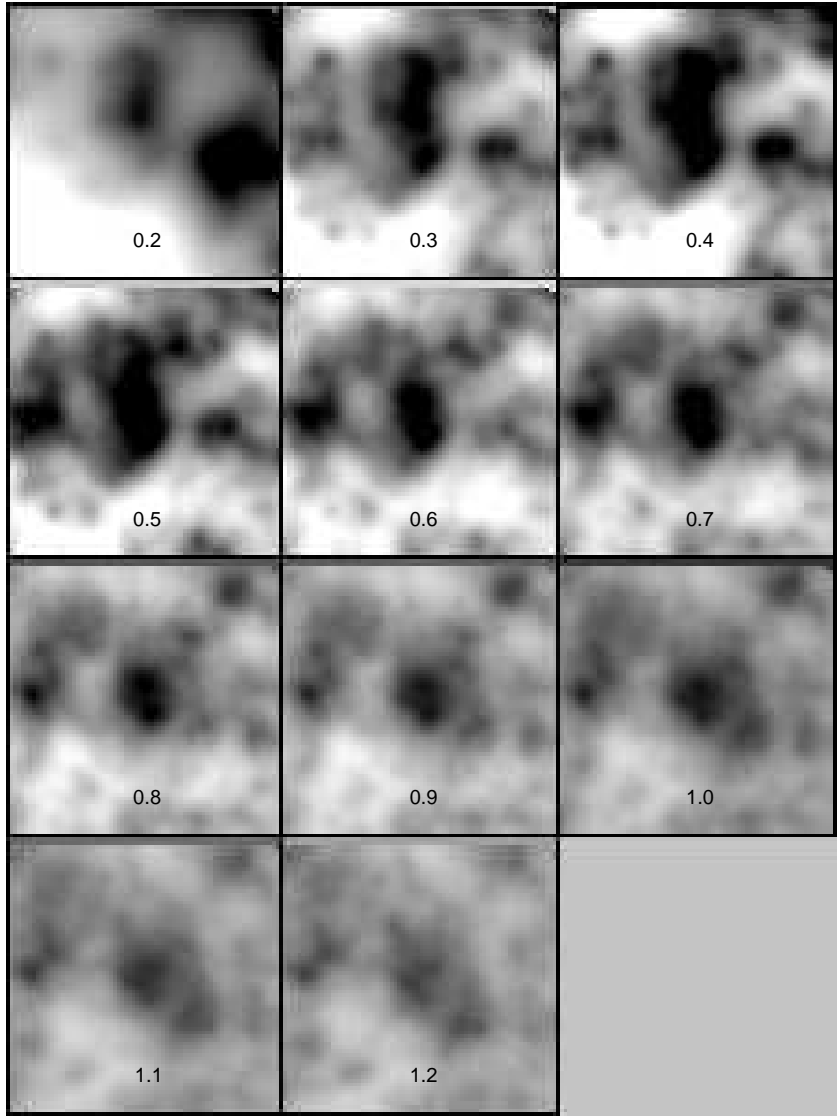

Fig. 16.— The match-filtered $I_4$ band galaxy catalog as a function of filter redshift for the region centered on the GHO cluster 1322+3028 ($z_{obs} = 0.70$; cluster 063 in Table 4). The filter redshift is listed at the bottom of each $14.5' \times 14.5'$ subimage; north is at the top and west to the right. The GHO cluster 1322+3027 ($z_{obs} = 0.75$; cluster 059 in Table 4) is the peak to the east of the central cluster.

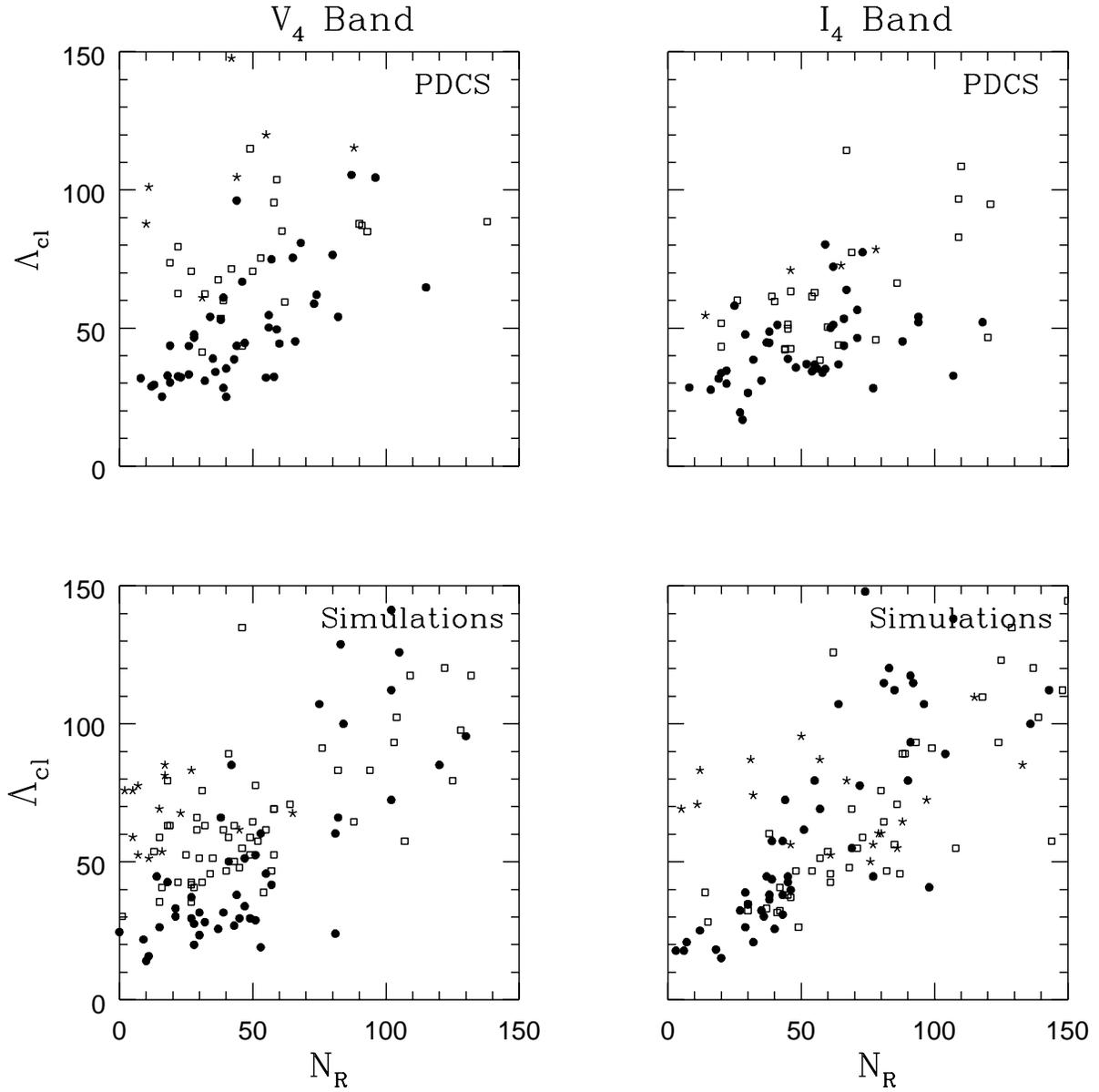

Fig. 17.— The Abell richness $N_R (\leq 1.0 h^{-1}$ Mpc$)$ as a function of $\Lambda_{cl}$ for the PDCS clusters and for simulated clusters in the $V_4$ and $I_4$ passbands. The solid dots, open squares, and stars indicate cluster redshift estimates in the range $0.2 \leq z_{est} \leq 0.4$, $0.5 \leq z_{est} \leq 0.7$, and $z_{est} \geq 0.8$, respectively.

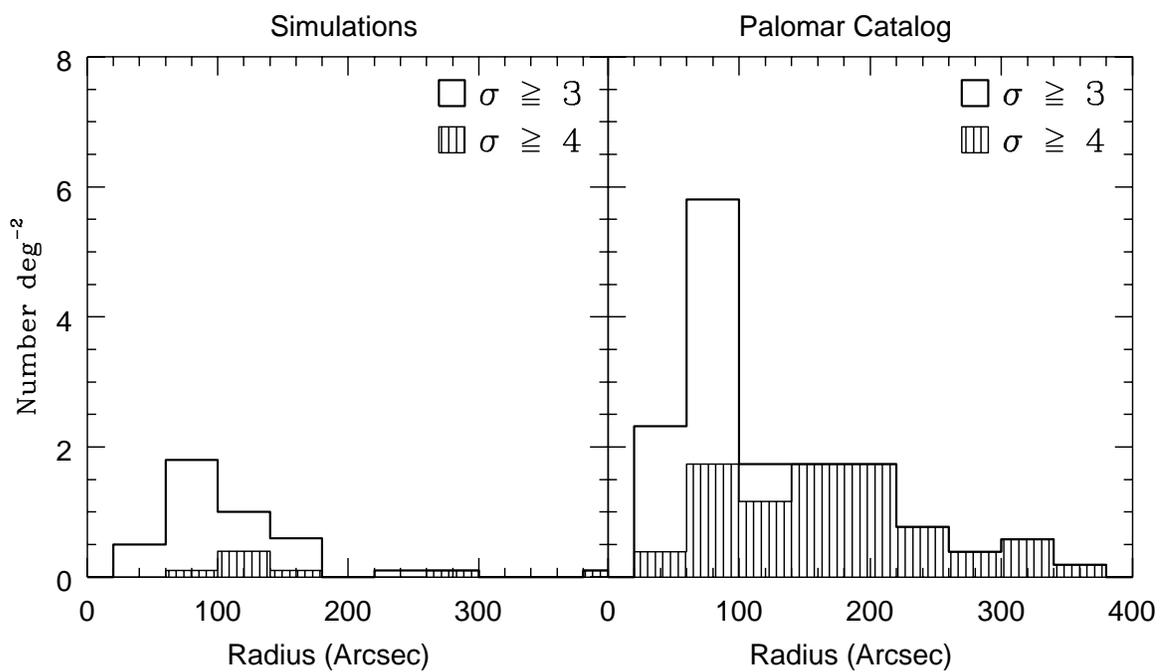

Fig. 18.— Left: the mean spurious detection surface density in 10 simulated galaxy catalogs as a function of size and peak significance. Right: the mean surface density of observed cluster candidates as a function of size and peak significance.

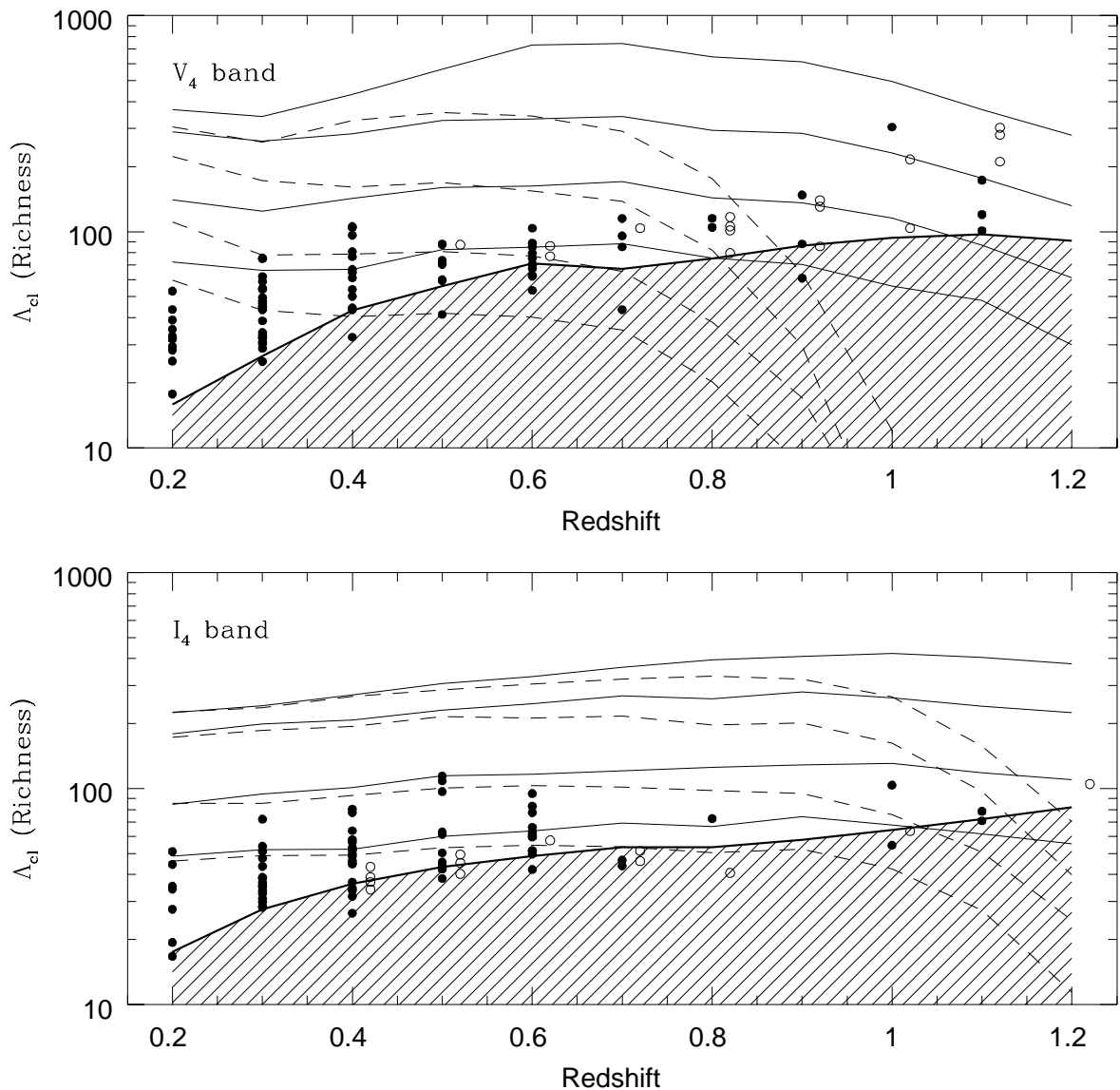

Fig. 19.— The matched filter richness, $\Lambda_{cl}$, as a function of redshift, richness, k-correction, and passband for the 3520 simulated clusters with $r^{-1.4}$ profiles. The dashed lines show the relationship for clusters with richness class 1–4 and with elliptical galaxy k-corrections. The solid lines show the relationship for clusters with Sbc k-corrections. The points are the actual detected clusters (open circles are supplemental clusters). The shaded area represents region below the typical $3\sigma$ detection threshold. Some matched cluster candidates lie within the shaded area in one passband but the significance of the detection in the other passband is above the $3\sigma$ limit.

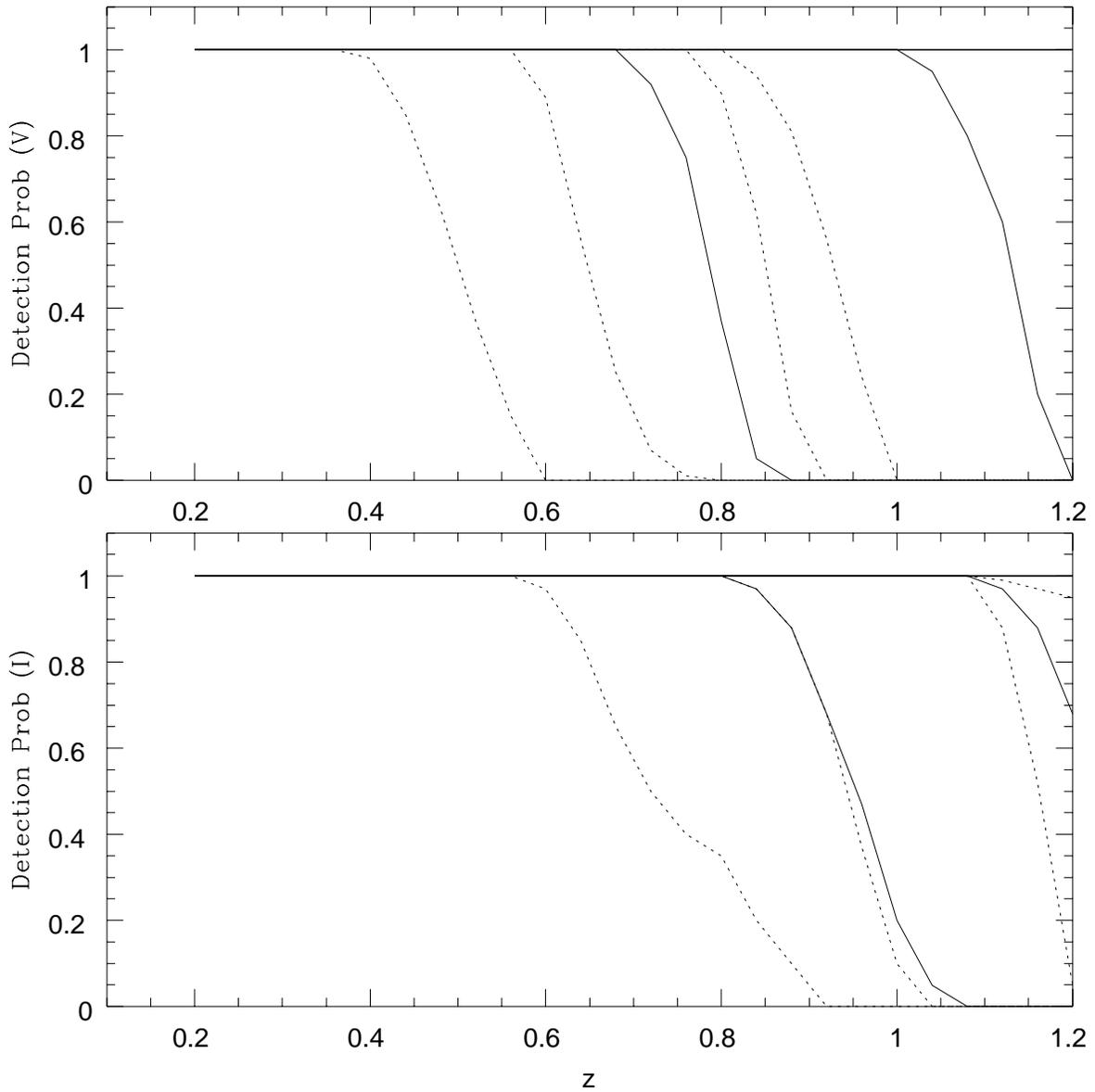

Fig. 20.— The selection function for clusters with $r^{-1.4}$ profiles. The dashed lines show the function for clusters with richness class 1–4 and with elliptical galaxy k-corrections. The solid lines show the function for clusters with Sbc k-corrections. If 4 separate solid lines are not seen it is because the selection probabilities for richer clusters are identical.

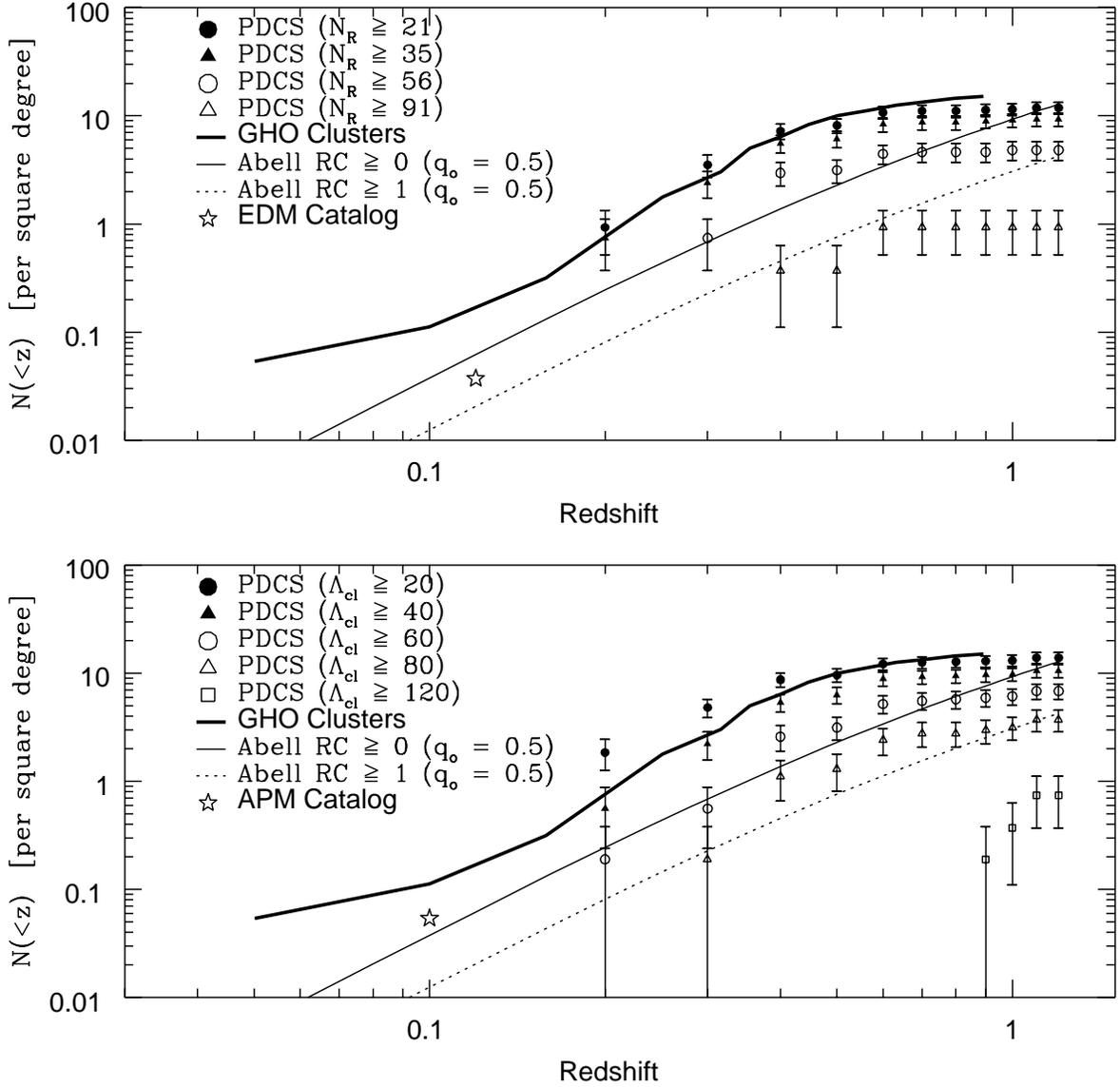

Fig. 21.— The cumulative number of clusters per square degree as a function of redshift and richness for the PDCS, the GHO catalog, the Abell/ACO catalog, the APM cluster catalog, and the Edinburgh-Durham-Milano (EDM) cluster redshift catalog. The upper plot shows PDCS cluster counts as a function of the Abell richness estimator, $N_R$. The lower plot shows PDCS cluster counts as a function of the $\Lambda_{cl}$ richness estimator. For clarity, the APM and EDM values are shown on separate plots.

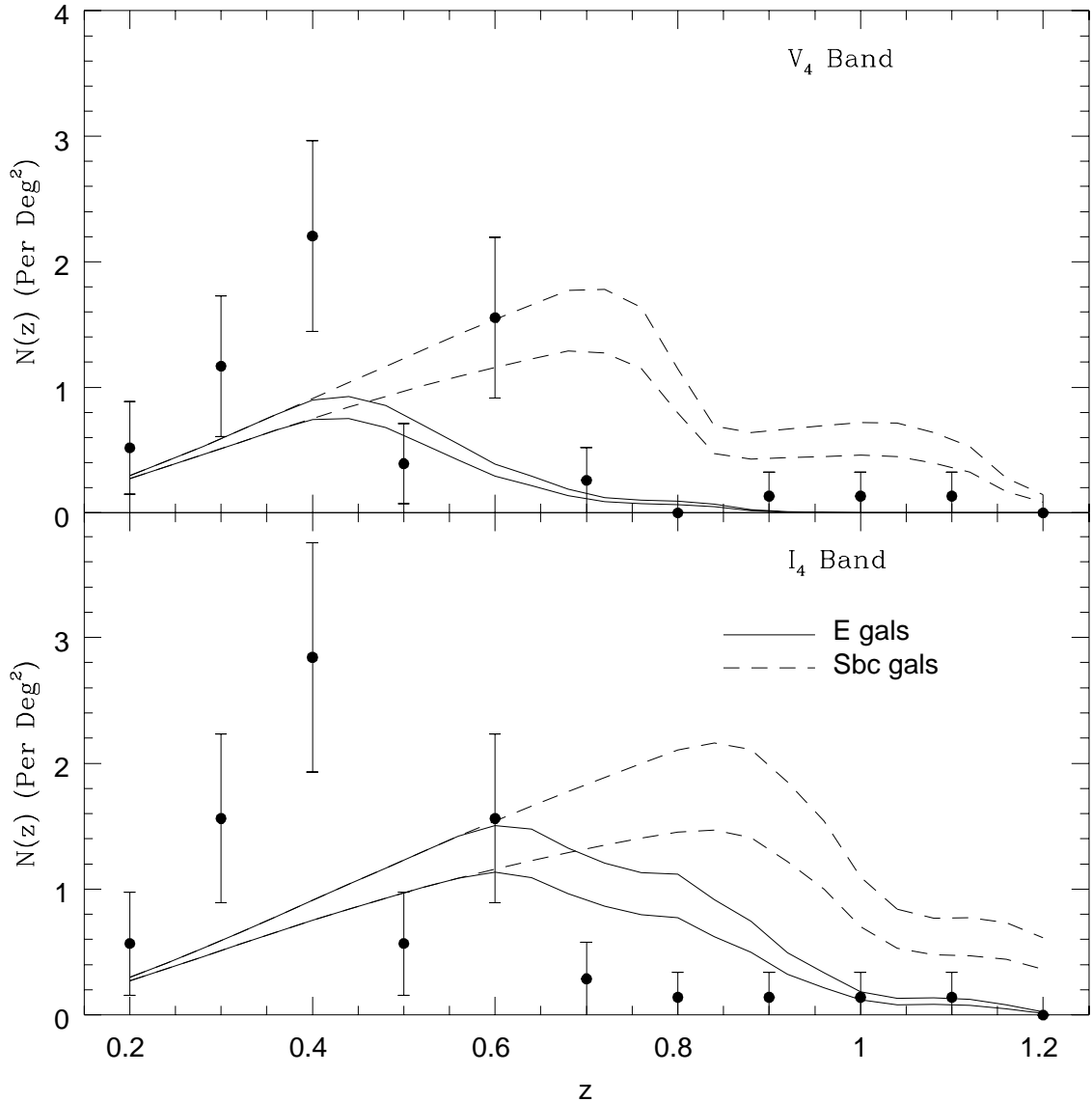

Fig. 22.— The differential redshift distributions for clusters detected in the $V_4$ and $I_4$ bands. The curves show the expected distributions for clusters consisting of non-evolving elliptical (solid line) and Sbc (dashed line) galaxies for $\Omega_o = 0.2$ and $\Omega_o = 1$. The $\Omega_o = 1$ curves always predict fewer clusters at a given redshift. See text for further details.

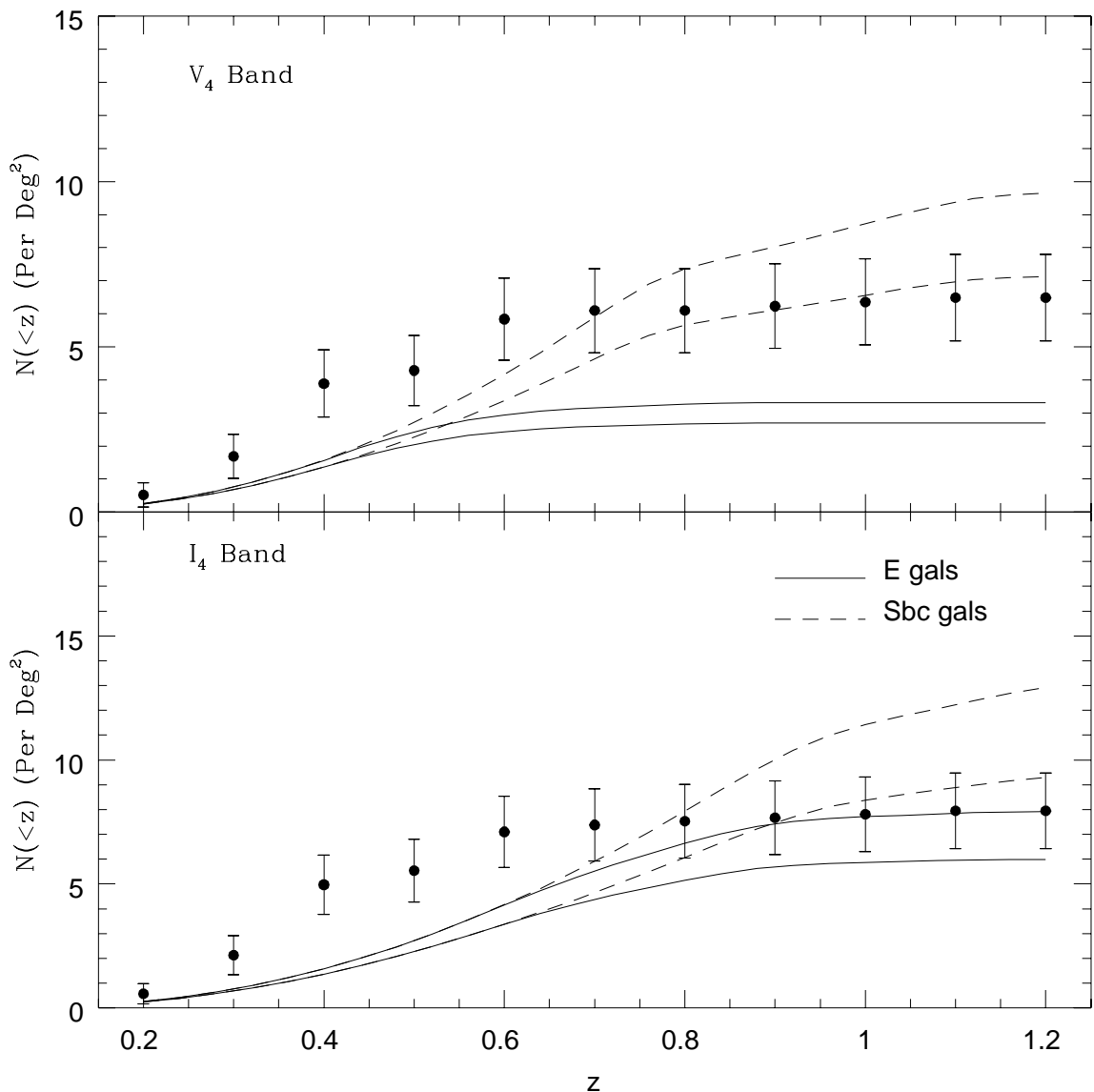

Fig. 23.— The cumulative redshift distributions for clusters detected in the $V_4$ and $I_4$ bands. The curves show the expected distributions for clusters consisting of non-evolving elliptical (solid line) and Sbc (dashed line) galaxies for $\Omega_o = 0.2$ and $\Omega_o = 1$. The $\Omega_o = 1$ curves always predict fewer clusters at a given redshift. See text for further details.